\documentclass[12pt]{article}
\usepackage{epsfig,amssymb}
\usepackage{latexsym}

\newcommand{\bq}{\begin{equation}}
\newcommand{\eq}{\end{equation}}
\newcommand{\bqa}{\begin{eqnarray}}
\newcommand{\eqa}{\end{eqnarray}}
\newcommand{\ben}{\begin{enumerate}}
\newcommand{\een}{\end{enumerate}}
\newcommand{\bc}{\begin{center}}
\newcommand{\ec}{\end{center}}
\newcommand{\bqb}{\begin{eqnarray*}}
\newcommand{\eqb}{\end{eqnarray*}}

\def\gsim{\gtrsim}
\def\lsim{\lesssim}

\def\pr#1#2#3{ Phys. Rev. ${\bf{#1}}$:#2 (#3)}
\def\prl#1#2#3{ Phys. Rev. Lett. ${\bf{#1}}$:#2 (#3)}

\def\prep#1#2#3{ Phys. Rep. ${\bf{#1}}$:#2 (#3)}

\def\np#1#2#3{ Nucl. Phys. ${\bf{#1}}$:#2 (#3)}
\def\zp#1#2#3{ Z. f. Phys. ${\bf{#1}}$:#2 (#3)}

\def\epj#1#2#3{ Eur. Phys. J. ${\bf{#1}}$:#2 (#3)}
\def\ijmp#1#2#3{ Int. J. Mod. Phys. ${\bf{#1}}$:#2 (#3)}

\def\jphys#1#2#3{ J. Phys. ${\bf{#1}}$,#2 (#3)}

\def\polon#1#2#3{Acta Phys. Polon. ${\bf{#1}}$:#2 (#3) }
\global\nulldelimiterspace = 0pt

\def\mw{m_W}

\def\tchi{\tilde \chi}

\begin{document}
\pagenumbering{arabic}
\thispagestyle{empty}
\def\thefootnote{\fnsymbol{footnote}}
\setcounter{footnote}{1}

\begin{flushright}
July 7,  2009\\
corrected version\\
arXiv: 0903.4532  [hep-ph]\\

 \end{flushright}
\vspace{2cm}
%---------------------titre ---------------------------------------
\begin{center}
{\Large {\bf A new class of SUSY signatures in  the processes  $gg  \to HH', ~VH$.    } } \\
 \vspace{1.5cm}
%-----------------------------------------------------------------
{\large G.J. Gounaris$^a$, J. Layssac$^b$,
and F.M. Renard$^b$}\\
\vspace{0.2cm}
$^a$Department of Theoretical Physics, Aristotle
University of Thessaloniki,\\
Gr-54124, Thessaloniki, Greece.\\
\vspace{0.2cm}
$^b$Laboratoire de Physique Th\'{e}orique et Astroparticules,
UMR 5207\\
Universit\'{e} Montpellier II,
 F-34095 Montpellier Cedex 5.\\
\end{center}

\vspace*{1.cm}
\begin{center}
{\bf Abstract}
\end{center}
Within the MSSM and  SM frameworks, we analyze the 1loop electroweak (EW) predictions for
the helicity amplitudes describing   the 17 processes $gg\to HH'$, and the 9 processes $gg\to VH$;
where $H,H'$ denote  Higgs   or Goldstone bosons,
while $V= Z, ~W^\pm$. Concentrating on MSSM, we then investigate how the asymptotic
  helicity conservation (HCns) property of SUSY,
affects the amplitudes   at the LHC energy range;
and what is the corresponding situation in SM,
where no HCns theorem exists.
HCns  is subsequently used to construct
many relations among  the  cross sections of  the  above MSSM processes,
 depending only on the standard MSSM angles $\alpha$ and $\beta$ characterizing the two Higgs doublets. These relations  should be asymptotically exact; but as the energy decreases towards the LHC range,  mass-depending deviations  should start appearing. Provided the SUSY scale is not too high,  these relations   may  remain roughly correct,  even at the LHC energy range.

\vspace{0.7cm}

\def\thefootnote{\arabic{footnote}}
\setcounter{footnote}{0}
\clearpage

\section{Introduction}

The fact that Supersymmetry confers remarkable properties to scattering amplitudes at high
energy, has already been noticed  in the literature. One aspect of it
emphasized some time ago, is that  in processes involving standard external particles
and non-vanishing Born contributions, the coefficients of the 1loop linear logarithmic
corrections at  high energy differ  strikingly, between
the minimal supersymmetric model (MSSM) and the standard model (SM),
reflecting  the differences in the    gauge and Yukawa interactions
\cite{MSSMrules, SMrules, eeV1V2}.

Another aspect concerns the  important  helicity conservation (HCns)  theorem
  established in  supersymmetry (SUSY) \cite{heli}.
  This property demands that for any 2-to-2 process,
all amplitudes that violate HCns, \underline{exactly} vanish, at   energies
 much higher than all masses,   and fixed  angles.
More explicitly this theorem states that for any process
\bq
a_{\lambda_a}+b_{\lambda_b} \to c_{\lambda_c}+d_{\lambda_d} ~~, \label{gen-process}
\eq
with  $\lambda_j$ denoting  the particle helicity, all amplitudes satisfying
\bq
\lambda_a+\lambda_b -\lambda_c-\lambda_d \neq 0 ~~, \label{HV-constraint}
\eq
vanish exactly at asymptotic energies. The amplitudes obeying  (\ref{HV-constraint}),
are called below  helicity violating (HV) amplitudes; while those
satisfying  $\lambda_a+\lambda_b -\lambda_c-\lambda_d= 0$,
are termed as helicity conserving (HC) amplitudes. HCns should be true
to all orders in the SUSY couplings, drastically reducing  the number
of the asymptotically non-vanishing amplitudes \cite{heli}.

This  HCns  theorem  is particularly non-trivial for processes
involving external gauge bosons,  where   huge  cancelations among the various  diagrams
 conspire for its realization \cite{heli}. Moreover, the theorem      crucially depends
on the renormalizability  of the  model;
any anomalous  coupling will violate it \cite{Kasimierz}.\\

In SM there is no general all-order proof for HCns.
Nevertheless, in several processes, it has been found to be
approximately correct.
Thus, if the Born contribution is non-vanishing,
then at the tree level,  the HV amplitudes for any 2-to-2 processes
always vanish asymptotically, while the HC ones tend to usually
non vanishing constants \cite{heli}.
If 1loop corrections are included to  such processes, then HCns remains
approximately correct; in the sense that the
HC amplitudes receive considerable  $\ln$- and $\ln^2$-corrections  at high energies,
and are always   much larger than the HV amplitudes, which however
do not necessarily vanish asymptotically \cite{MSSMrules}.

Concerning processes with vanishing Born contributions, we mention
 $\gamma \gamma \to ZZ, ~\gamma Z, ~ \gamma \gamma $, studied some time ago,
    at  the complete 1loop EW order, in both SM and MSSM \cite{gamgamV10V20}.
    In these  cases, it   has  then been  seen explicitly in both, SM and MSSM,
     that the HC amplitudes   rise logarithmically,  due to the gauge
     (and gaugino in MSSM) loop contributions, and are predominantly
     imaginary \cite{gamgamV10V20}.
    On the contrary, the HV amplitudes  tend to angle-dependent small constants in SM,
 but vanish  in MSSM \cite{gamgamV10V20}.

Thus, in all SM cases studied so far,
HCns is approximately valid;  in the sense that the HC amplitudes
dominate the HV ones, but the HV amplitudes do not necessarily vanish asymptotically.
 For SM processes with vanishing   Born contributions though, no general statement on,
 even the approximate validity of  HCns  exists in the literature.   \\

Coming back to the supersymmetric case, where  HCns
has been proved to all orders for  asymptotic energies  \cite{heli};
we remark that  its relevance for realistic energies is  process-dependent and
 needs to be separately investigated.

 To this aim, the complete 1loop
 electroweak (EW) corrections were calculated for
 $ug\to dW^+$, which determines $W$+jet production at LHC \cite{ugdW}.
 Assuming that   the SUSY masses are in the range set
by the $SPS1a'$ benchmark  of the SPA convention\footnote{This model is very close
to the best fit of the precision data in \cite{OBuchmueller}.} \cite{SPA},
it has  been found that the HC amplitudes are much larger than  the  HV ones,
for energies $\gsim 0.5~{\rm TeV}$, and  a wide range of angles   \cite{ugdW}.
Similar results are expected   for benchmarks with somewhat heavier SUSY masses, like those
 in Table 1.
\begin{table}[h]
\begin{center}
{ Table 1: Input  parameters at the grand scale,
for three  constrained MSSM benchmark models with  $\mu>0$;
  dimensional parameters  in GeV. }\\
  \vspace*{0.3cm}
\begin{small}
\begin{tabular}{||c|c|c|c||}
\hline \hline
  & $SPS1a'$ \cite{SPA} & BBSSW \cite{Baer}  & FLN mSP4 \cite{Nath}      \\ \hline
 $m_{1/2}$ &250  & 900   & 137   \\
 $m_0$ & 70  & 4716 & 1674   \\
 $A_0$ &-300  &  0 & 1985   \\
$\tan\beta$ & 10  & 30  & 18.6  \\
  \hline \hline
\end{tabular}
 \end{small}
\end{center}
\end{table}

Furthermore, to the 1loop EW order in MSSM, HCns was used
to derive  relations between the  differential  cross sections
 for the subprocess $ug\to dW^+$ and $ug \to \tilde d_L  \tilde \chi_i^+$,
where $\tilde d_L $ denotes an L-down-squark and $ \tilde \chi_i^+$ describes any
of the two charginos \cite{ugdW-ugsdWino}. The  derivation of these relations was based
on the asymptotic properties of the helicity  amplitudes.
But for benchmarks  like those in Table 1, the relations remained
approximately  correct,  even  at  LHC energies;
where the  HCns validity for the   $ug \to \tilde d_L \tilde \chi_i^+$
amplitudes, is not yet reached \cite{ugdW-ugsdWino}.
Similar  relations should be true for many other analogous pairs of processes. \\

In the present work  we propose to study  more stringently the helicity conservation
property; i.e. to study the energies needed for establishing HCns in MSSM,
 and possibly identify cases where it is strongly  violated in SM.

We therefore  look at processes   where the dominant HC amplitudes
do not increase logarithmically at high energies,
but rather tend to  angular dependent, "constants".
Our previous experience implies that in such cases there should
be no Born contribution \cite{MSSMrules}; and moreover, that there should not be any
gauge exchange contributions, like those  in $\gamma \gamma \to ZZ, ~\gamma Z, ~ \gamma \gamma $
  \cite{gamgamV10V20}.

In the  MSSM case,  where HCns is obeyed, we could then  also derive asymptotic
relations analogous to those in \cite{ugdW-ugsdWino}; hoping that they may again be
useful, even  at  the LHC range.

Thus, we  study here the gluon-gluon fusion  to gauge or Higgs bosons,
 at the complete 1loop EW order,  in either SM or MSSM.
 For simplicity, we assume a CP invariant framework, where all soft breaking terms
 and    superpotential  and Yukawa couplings  are real.
  More explicitly the processes we study are
\bq
g(l, \mu ) g(\l', \mu')\to H(p)H'(p') ~~~, ~~~
g(l,\mu ) g(l',\mu')\to V(p,\tau ) H(p')  ~~ ~,
\label{gluon-HH-VH}
\eq
where
\bq
H,~H'  ~ \Rightarrow ~ H^{\pm}, ~G^{\pm},~ H_{SM}, ~H^0, ~h^0, ~G^0, ~A^0 ~~, \label{ggHH-proc1}
\eq
 denote the   Higgs or Goldstone bosons in MSSM or SM , and\footnote{$V=\gamma$
 is impossible due to  CP invariance.}
\bq
V ~ \Rightarrow ~  W^{\pm}, ~ Z~~ . \label{ggVH-proc1}
\eq
In (\ref{gluon-HH-VH}),  $(\mu, \mu', \tau)$
describe the helicities of the two incoming gluons
and the final vector boson respectively,
while $(l,l')$ are the incoming momenta, and $(p,p')$ the outgoing.

Concerning $gg\to HH'$, we consider the   17 processes
\bqa
{\rm 4~~ SM ~ processes} &\to & HH, ~ G^0H,~ G^+G^-, ~G^0G^0, ~~~ \nonumber \\
{\rm 13 ~~ MSSM ~ processes}  &\to &  H^+H^-,~ H^0H^0,~ h^0h^0, ~H^0h^0,
~ A^0h^0, ~ A^0H^0, ~ A^0A^0,
\nonumber \\
&&   G^0h^0,~ G^0H^0, ~ G^+H^-, ~ G^0A^0, ~ G^+G^-, ~G^0G^0, ~  \label{ggHH-proc2}
\eqa
calculated from the general graphs of Fig.\ref{ggHH-diag-fig}.
For   each of these process, we study  the energy and angular behaviour
of the four helicity amplitudes corresponding to  $\mu=\pm 1$ and $\mu'=\pm 1$,
emphasizing    the difference between the HC and HV amplitudes.

Turning next to $gg\to VH$, we consider   the 9  processes
\bqa
{\rm  3~~ SM ~ processes} &\to &  ZH, ~W^+G^-, ~ ZG^0,  ~~ \nonumber \\
{\rm 6 ~~ MSSM~ processes }  &\to &   W^+H^-,~ ZH^0,~ Zh^0,~ ZA^0,~ W^+G^-,~ ZG^0,~
 \label{ggVH-proc2}
\eqa
calculated from the diagrams in Fig.\ref{ggVH-diag-fig}.
In these cases, we have a richer helicity structure with $\mu=\pm1$, $\mu'=\pm1$ and
$\tau=\pm 1,0$.

FORTRAN codes calculating the  helicity amplitudes for all these processes are
constructed, which are  released in \cite{code}.\\

We indeed find that the  HC amplitudes dominate at high energies in MSSM,
behaving  like  angular dependent "constants", for  both groups of processes in
(\ref{ggHH-proc2}) and (\ref{ggVH-proc2}).  Several relations among the dominant
HC amplitudes for such  processes
   are established. These are used to derive  asymptotic relations among
 various cross sections, which  may lead to interesting tests
 of the underlying supersymmetric structure, even at non-asymptotic energies.

In SM, the HC amplitudes of (\ref{gluon-HH-VH})  are again  found
to behave asymptotically like angular dependent "constants".
But the HCns picture is   distorted, and
some  helicity violating (HV) amplitudes
also tend to  "constants", comparable in magnitude to those of the HC ones.
There exist processes though, where in SM also, the HV amplitudes  vanish at high energies.

Cross sections for the 1loop EW contributions
to many such processes  exist in the literature \cite{Zerwas,Hollik, Han, Melles, Kniehl};
but a detail amplitude analysis  studying  the helicity
conservation property, has not yet been done. \\

The contents of the paper are: In Section 2 we present the
general structure of the   $gg\to HH'$ and $gg\to VH$
amplitudes. In  Section 3,  the high energy behaviours of the helicity
amplitudes for the various processes, are  analyzed; and
the   asymptotic  relations among several cross
sections are  derived. In  Section 4, we introduce the aforementioned FORTRAN codes,
which calculate the 1loop EW  helicity amplitude;
and we give  our   numerical results.  Particular attention is
payed  towards  investigating the behaviour of the above asymptotic cross section relations,
 as the energy decreases. Finally, Section 5 contains the  summary and an outlook.

We  we do  not make any detail proposal for an LHC observable, in this paper.
Applications to  LHC would require
additional  work including QED   and (most importantly) QCD corrections \cite{Dawson},
as well as the final state identification
and background analysis, which are beyond the scope of this paper. \\

\section{The  $gg\to HH'$ and $gg\to VH$ amplitudes}

{\bf  The $gg\to HH'$  case. }\\
 Defining the kinematics  for the process $g g \to HH'$
 as in (\ref{gluon-HH-VH}),
 the corresponding helicity amplitudes are  written as
$F^{HH'}_{\mu \mu'}(s,  \theta)$, where $s$ is the square of the c.m. energy, and
$\theta$ is the corresponding scattering angle; ($0<\theta<\pi$).
 A  color factor $\delta^{ab}$ has always been removed from the amplitudes,
where $(a,b)$ describe the color indices of the two incoming gluons.
The  phase of $F_{\mu \mu'}(s,  \theta)$, is related
 to the phase of the $S$-matrix by $S=iF\delta^{ab}$.

Bose statistics for the initial gluons and
 CP invariance  imply
\bqa
{\rm Bose} & \Rightarrow & F_{\mu \mu'}(\theta)= F_{\mu' \mu  }(\pi-\theta)
~~, \nonumber \\
{\rm CP }  & \Rightarrow &  F^{H_a^0H_{a'}^0}_{\mu \mu' }(\theta)=
F^{H_a^0H_{a'}^0}_{-\mu -\mu'}(\theta) ~~, ~~
 F^{H_b^0H_{b'}^0}_{\mu \mu'}(\theta)=
F^{H_b^0H_{b'}^0}_{-\mu -\mu'}(\theta) ~~, \nonumber \\
&  &  F^{H_a^0H_{b}^0}_{\mu \mu'}(\theta)=
- F^{H_a^0H_{b}^0}_{-\mu -\mu'}(\theta) ~~,
\nonumber \\
&&  F^{H^\pm H^\mp}_{\mu \mu'}(\theta)
=F^{H^\mp H^\pm}_{-\mu -\mu'}(\theta) ~~,
 \label{HH-Bose-CP}
\eqa
where the charged final state  relations  also  apply for
the $H^\pm G^\mp $ and $G^\pm G^\mp $ amplitudes.
In MSSM we use the notation $H^0_a=(H^0,h^0)$ and $H^0_b=(A^0,G^0)$, while in SM
 we identify  $H^0_a=H$ and $H^0_b=G^0$.

 Relations  (\ref{HH-Bose-CP})  constrain the four $g g \to HH'$
 amplitudes
 \bq
 F_{++}, ~ F_{+-}, ~F_{-+} , ~F_{--}~,  \label{independent-F-HH}
 \eq
 so that the first two may be  considered as independent.
  According to  HCns,    only $F_{\pm\mp}$  survive asymptotically  in MSSM \cite{heli}.
The corresponding  cross section  is
\bq
{d\sigma (gg\to HH')\over d\cos\theta}
={|\vec{p}|\over512\pi s\sqrt{s}}
\sum_{\mu,\mu'} |F_{\mu \mu'}|^2~~, \label{dsigmaHH1 }
\eq
where the summation is  over all possible  $(\mu=\pm1,~\mu'=\pm1)$, and
 $|\vec{p}|$ denotes the absolute value of the 3-momentum in the c.m. of the $HH'$ pair.\\

\begin{figure}[p]
\vspace{-1cm}
\[
\hspace{-4.5cm}\epsfig{file=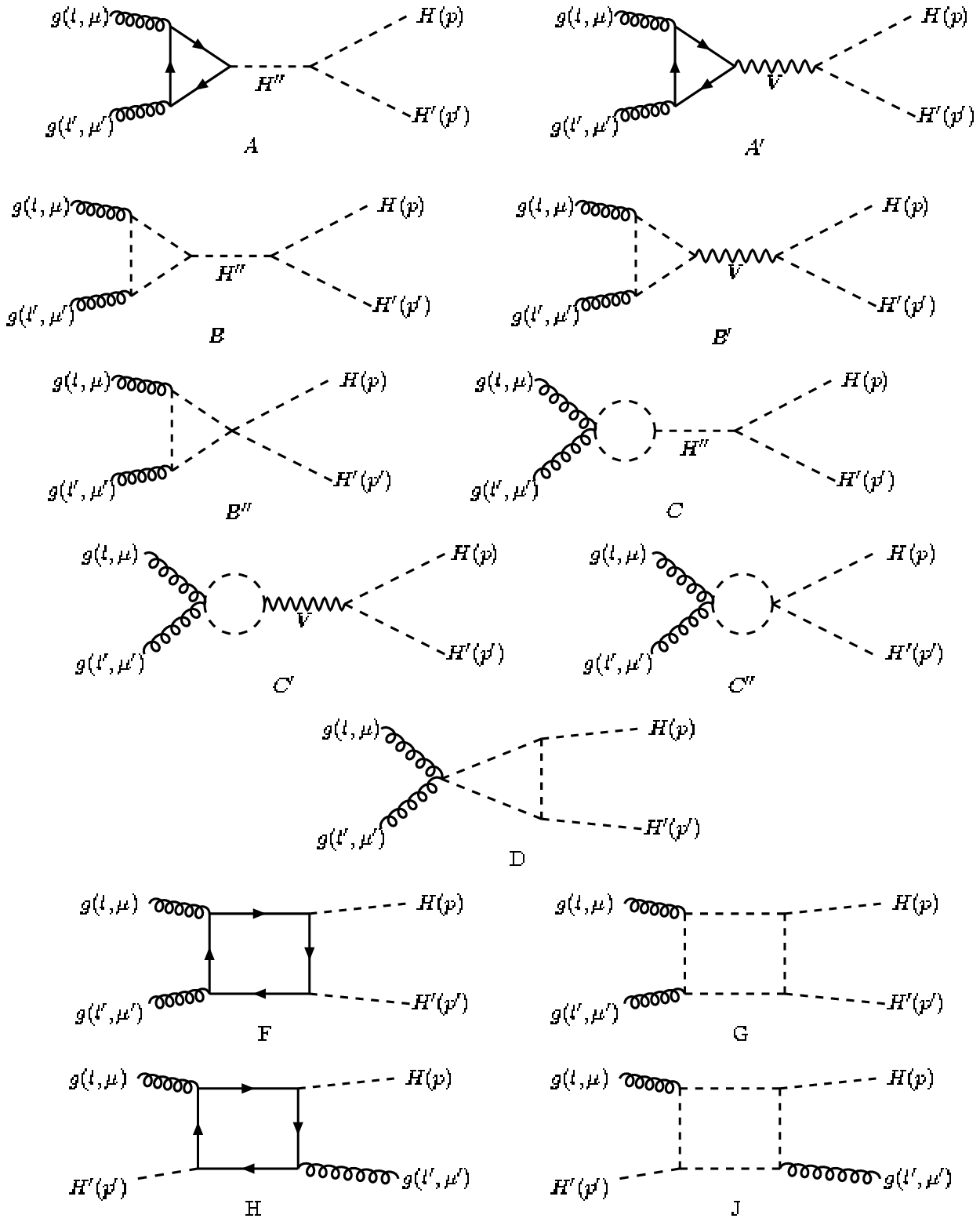, height=21cm, width=17cm}
\]
\vspace{-7cm}
\caption[1]{Independent diagrams for calculating $gg\to H H'$  in  MSSM and SM;
with $(H, H')$ denoting scalar Higgs particles or Goldstone bosons.
The diagrams are named as $A,~A',~B,~B',~B'',~ C,~C',~C'',~D,~ F,~G,~H,~J $.
The $s$-channel scalar or vector exchanges in some of the triangular and bubble graphs are
named as $H''$ and $V$.
Full, broken and wavy lines describe respectively fermionic, scalar and vector particles.
The incoming and outgoing momenta
and helicities are indicated in parentheses.}
\label{ggHH-diag-fig}
\end{figure}

The generic set of the 1loop EW diagrams for  $gg\to H H'$  in  MSSM and SM is presented in
Fig.\ref{ggHH-diag-fig}, where full, broken and wavy lines describe respectively
fermionic, scalar and vector particles.
The  contributions from interchanging the two gluons should be added for the
 diagrams\footnote{The diagram-names are indicated in Fig.\ref{ggHH-diag-fig},
 as well as the  definitions of $H''$ and $V$ used below.}
  $A, A', B, B', B'', F, G, H, J$;  on the contrary, for  the diagrams $C,C',C'',D$,  the
gluon-symmetrization is automatically included.

No $(H,H')$ symmetrization is assumed. Consequently,
for the  $F$ and $G$ boxes, the  respective   quark- and squark-loops are independent of
the  corresponding antiquark- and antisquark-loops, which should therefore
be added respectively.
For the rest of the graphs, only the quark or squark loops are needed.

The specific graphs of Fig.\ref{ggHH-diag-fig} contributing to
each of the 17 processes in (\ref{ggHH-proc2}), are

\begin{itemize}

\item
In SM, the only relevant  boxes are    F and H, which  contribute to all possible
processes in (\ref{ggHH-proc2}).

In  MSSM,  all F, G, H, J boxes contribute to the processes in (\ref{ggHH-proc2}).

\item
 Triangle and bubble contributions in SM arise   as follows:

\begin{itemize}

\item
 for $gg\to H H, ~ G^0G^0$,   they come from  graph $A $ with $H''=H$;

\item
for $gg\to G^0H$,  they come from graph $A$ with    $H''=G^0$, and graph $A'$ with $V=Z$;

\item
for $gg\to G^+G^- $, they come from graph $A $ with $H''=H$.

\end{itemize}

\item
 Triangle and bubble contributions in MSSM arise  as follows:

\begin{itemize}

\item
for $gg\to H^0H^0, ~h^0h^0, ~H^0h^0, ~ A^0A^0, ~G^0G^0, ~ A^0G^0 $,
they come  from graphs  $A,~B,  ~ C, $  with $H''=H^0,~h^0$;
and from graphs $ B'', ~C'', ~ D $;

\item
for $gg\to A^0H^0,~A^0h^0, G^0H^0,~ G^0h^0 $,
they come from graph  $A $  with $(H''=A^0,~G^0)$; and graphs  $A', ~ B',~ C'~$ ~ with $V=Z$ ;

\item
for $gg\to H^+H^-,~ G^+G^-   $,
they come from graphs  $A,~B,  ~ C, $ with $(H''=H^0,~h^0)$; and from graphs $B'', ~C'', ~ D$;

\item
for $gg\to G^+H^- $,
they come from graph  $A $ with $(H''=H^0,~h^0,~ A^0 )$; from graphs $ B,  ~ C, $
with $(H''=H^0,~h^0 )$;  and from graphs  $B'', ~C'', ~ D$.\\

\end{itemize}

\end{itemize}

\noindent
{\bf The $gg\to VH$  amplitudes. }\\
Using again the notation   (\ref{gluon-HH-VH}), we  describe  the helicity amplitudes
as $F^{VH}_{\mu \mu'\tau}(s,  \theta)$.
 The same phase conventions as in the previous subsection  are used,
 and a color factor $\delta^{ab}$  is again removed.

Bose statistics for the initial gluons and
 CP invariance  imply
 \bqa
{\rm Bose} & \Rightarrow & F_{\mu\mu'\tau}(\theta)= (-1)^\tau F_{\mu' \mu  \tau}(\pi-\theta)
~~, \nonumber \\
{\rm CP }  & \Rightarrow &  F^{ZH_a^0}_{\mu\mu'\tau}(\theta)=(-1)^{(1-\tau)}
F^{ZH_a^0}_{-\mu -\mu'-\tau}(\theta) ~~,  \nonumber \\
 & \Rightarrow &  F^{ZH_b^0}_{\mu\mu'\tau}(\theta)=-(-1)^{(1-\tau)}
 F^{ZH_b^0}_{-\mu -\mu'-\tau}(\theta) ~~,  \nonumber \\
 & \Rightarrow &  F^{W^+H^-}_{\mu\mu'\tau}(\theta)=(-1)^{(1-\tau)}
 F^{W^-H^+}_{-\mu -\mu'-\tau}(\theta) ~~, \nonumber \\
 && F^{W^+G^-}_{\mu\mu'\tau}(\theta)=(-1)^{(1-\tau)}
 F^{W^-G^+}_{-\mu -\mu'-\tau}(\theta) ~~,  \label{VH-Bose-CP}
 \eqa
 where $H_a^0,~H_b^0$ are defined immediately after  (\ref{HH-Bose-CP}).

Relations  (\ref{VH-Bose-CP})  constrain the 12 possible helicity amplitudes
\bqa
&& F_{+++}~, ~ F_{++-}~ ,~ F_{++0}~, ~F_{+-+}~,~F_{+--},F_{+-0}~,~ \nonumber \\
&& F_{---}~,~ F_{--0} ~,~ F_{--+}~, F_{-++}~,~ F_{-+-}~,~ F_{-+0}~,~
\label{independent-F-VH}
\eqa
so that  the first six may  be considered as the independent for neutral final states,
while for charged final states we take the first nine as independent.
According to the HCns theorem, only
 $F_{\pm\mp0}$ may survive  at asymptotic energies in MSSM \cite{heli}.
The corresponding  cross section  is given by
\bq
{d\sigma(gg\to VH) \over d\cos\theta}
={|\vec{p}|\over512\pi s\sqrt{s}}
\sum_{\mu,\mu',\tau} |F_{\mu \mu'\tau}|^2~~, \label{dsigmaVH1 }
\eq
where the summation is done over all possible  $(\mu=\pm1,~\mu'=\pm1)$
and $(\tau=\pm 1, 0)$.
In (\ref{dsigmaVH1 }),  $|\vec{p}|$ denotes the absolute value of the 3-momentum
in the c.m. of the final $VH$ pair.\\

\begin{figure}[p]
\vspace{-1cm}
\[
\hspace{-4.5cm}\epsfig{file=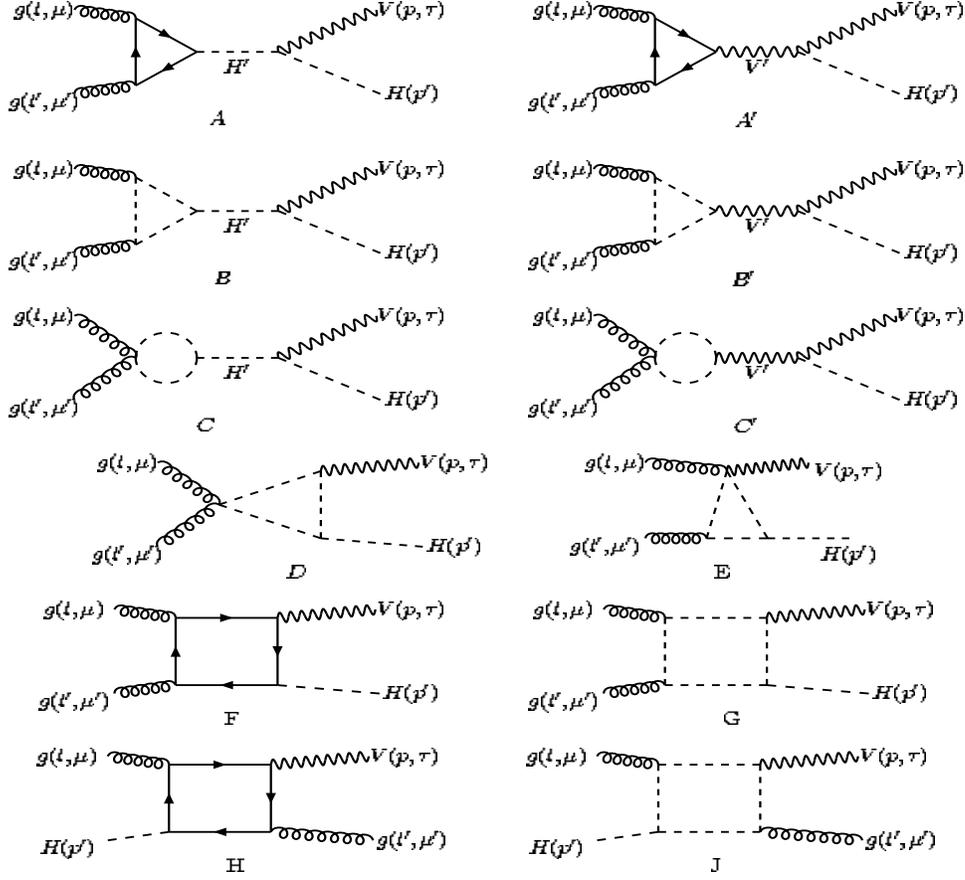, height=23cm, width=17cm}
\]
\vspace{-11cm}
\caption[1]{Independent diagrams for calculating $gg\to V H $  in  MSSM and SM,
with $V$ denoting a vector particle, and $H$ describing  a scalar Higgs-type particle
or Goldstone boson. The diagrams are named as $A,~A',~B,~B',~ C,~C',~D,~ E,~F,~G,~H,~J $.
The $s$-channel scalar or vector exchanges in some of the triangular and bubble graphs are
named as $H'$ and $V'$.
Full, broken and wavy lines describe respectively fermionic,
scalar and vector particles. The incoming and outgoing momenta
and helicities are indicated in parentheses.}
\label{ggVH-diag-fig}
\end{figure}

The generic set of the 1loop EW diagrams for  $gg\to VH $  in  MSSM and SM is presented in
Fig.\ref{ggVH-diag-fig}; where full, broken and wavy lines again describe respectively
the fermionic, scalar and vector particles. As before, the  contributions from
interchanging the two gluons should be added for the diagrams\footnote{The names of the diagrams are defined in Fig.\ref{ggVH-diag-fig}.}
 $A, A', B, B', E, F, G, H, J$; while for  $C,C',D$,   the
gluon-symmetrization is automatically included. For the  $F$ and $G$ boxes we should add
to the respective quark- and squark-loop
contributions, the corresponding antiquark-and antisquark-loops. For the rest
of the graphs, only the quark or squark loops are needed.

The specific graphs of Fig.\ref{ggVH-diag-fig} contributing to
each of the 9 processes in (\ref{ggVH-proc2}), are\footnote{The definitions of $H'$ and $V'$
for the items below are given in Fig.\ref{ggVH-diag-fig}.}:

\begin{itemize}

\item
In SM, the relevant boxes  F and H  contribute to all  processes in (\ref{ggVH-proc2}).

In  MSSM,  all F, G, H, J boxes contribute to the    processes in (\ref{ggVH-proc2}).

\item
Triangle and bubble contributions in SM appear  as follows:

\begin{itemize}

\item
 for $gg\to ZH$,  they come from  graph $A $ with $H'=G^0$, and from graph $A'$ with $V'=Z$ ~~;

\item
for $gg\to ZG^0$, they come from graph $A$ with    $H'=H$ ~~;

\item
for $gg\to W^+G^- $,  they come from graph $A $ with $(H'=H, G^0)$, and from
graph $A'$ with $(V'=\gamma, Z)$ ~~.\\

\end{itemize}

\item
Triangle and bubble contributions in MSSM appear  as follows:

\begin{itemize}

\item
for $gg\to ZH^0,~Zh^0 $, they come from graph $A$ with $(H'=A^0, G^0)$, and from
graphs  $A', B', C'$  with $V'=Z$;

\item
for $gg\to ZA^0,~ZG^0 $, they come from graphs $A, B, C$ with $(H'=h^0, H^0)$;

\item
for $gg \to W^+H^-$, they come from graph $A$ with $(H'=h^0, H^0, A^0)$, from graphs $B, C$ with
$(H'=h^0, H^0)$, and from graphs  $D, E$;

\item
for $gg \to W^+G^-$, they come from graph $A$ with $(H'=h^0, H^0, G^0)$, from graphs $B, C$ with
$(H'=h^0, H^0)$, from graphs $A', B', C'$ with $(V'=\gamma, Z)$, and from graphs  $D, E$.\\

\end{itemize}

\end{itemize}

Finally we note that the processes in (\ref{ggHH-proc2},\ref{ggVH-proc2})
which involve  final Goldstone bosons,
provide  a useful test of the validity of our
calculations at high energies. This comes from the equivalence theorem which
 states that at high energies we should have \cite{equiv}
 \bqa
  F_{\mu\mu'0}(g g \to W^\pm H^\mp) &  \simeq &  \mp \xi_W
  F_{\mu\mu'}(gg \to G^\pm H^\mp) ~~ ,  \nonumber \\
-i  F_{\mu\mu'0}(g g \to Z H^0_{a,b}) & \simeq & \xi_Z F_{\mu\mu'}(gg \to G^0 H^0_{a,b}) ~~ .
\label{VH-GH-equivalence}
 \eqa
 We have checked that these relations are satisfied by the results of
 our codes, where $\xi_W=\xi_Z=1$ is always used.

 Similarly, the processes in (\ref{ggHH-proc2}) involving two final Goldstones,
 determine   the high energy behavior of  $gg\to V_1V_2$,
 for two longitudinal vector bosons \cite{equiv}.\\

\section{High energy properties}

\subsection{Analytical results for $gg \to HH'$ in MSSM}

The high energy behaviour of the amplitudes for the $gg \to HH'$ processes in (\ref{ggHH-proc2}),
may be analytically obtained   from the  diagrams in Fig.\ref{ggHH-diag-fig} and the
asymptotic  expressions given e.g. in  \cite{techpaper}.

The only diagrams of  Fig.\ref{ggHH-diag-fig},
which are not suppressed at high energies, are  $B'',C'',F,H$.
Considered separately,  the $B''$ and $C''$ contributions  go to constants;
while   the  $F$ and $H$ boxes    have     linear ln(s) behaviours, which
 cancel out in their sum. Thus, the complete contribution  behaves like an
 angle-dependent, but energy independent   "constant", for all $gg\to HH'$  processes.
Subtleties arise  in specific processes though, depending on the relative importance
of these diagrams.\\

Thus, in SM, where the squark diagrams are absent, the available
 processes $HH$, $G^0G^0$, $G^0H$, $G^+G^-$ receive their complete
 asymptotic "constant" contribution solely  from the F and H boxes.\\

 For MSSM, we first concentrate on  the   $G^0G^0$ and $G^+G^-$ processes,
 where the constants from the $F+H$ boxes are  canceled  in   $F_{\pm\pm}$,
 by  opposite constants  coming from the squark diagrams $(B'',C'')$, leaving
 mass-suppressed contributions that vanish at  high energies and fixed angles.
   Only the HC amplitudes
  $F_{\pm\mp}$ survive asymptotically,
 characterized by  energy  independent, but angle  dependent "constants".
 Similar situations arise also for all
 other  MSSM processes, in agreement with  HCns  \cite{heli}.

To describe in more detail  the HC  asymptotic amplitudes,
 it is convenient to divide these  MSSM processes
 into three classes, as follows:

\begin{itemize}

\item
Class a: ~ It contains the 6 processes (k=1,6)
\[
 G^0G^0 ~,~ G^0A^0 ~,~  A^0A^0 ~,~  H^0H^0 ~,~ h^0h^0 ~,~ H^0h^0 ~,~
 \]
  characterized by neutral final bosons carrying identical
CP eigenvalues. The corresponding asymptotic limits for $F^k_{\pm\mp}$ may then be expressed
as\footnote{We use the same conventions  as in \cite{Gunion}.}
\bqa
F^k_{\pm\mp} & \to &  R_{ak} C^I_{\pm\mp}(\theta)~~, ~~~~~{\rm with} \nonumber \\
R_{a1} & = & m^2_t+m^2_b ~~,\nonumber \\
R_{a2} & = & m^2_t\cot\beta-m^2_b\tan\beta ~~, \nonumber \\
R_{a3} & = & m^2_t\cot^2\beta+m^2_b\tan^2\beta ~~, \nonumber \\
R_{a4} & = & {m^2_t\sin^2\alpha\over\sin^2\beta}+{m^2_b\cos^2\alpha\over\cos^2\beta} ~~,
\nonumber \\
R_{a5} &= & {m^2_t\cos^2\alpha\over\sin^2\beta}+ {m^2_b\sin^2\alpha\over\cos^2\beta} ~~,
\nonumber \\
R_{a6} & = & {m^2_t\sin\alpha\cos\alpha\over\sin^2\beta}-
{m^2_b\cos\alpha\sin\alpha\over\cos^2\beta}~~, \label{Class-a}
\eqa
where   $C^I_{\pm\mp}(\theta)$ describe the process-independent part of  these limits,
while the real quantities  $R_{ak}$ describe  the    process-dependent part.
The later solely depend on the MSSM angles $\alpha$, (describing  the standard  two-Higgs-doublet mixing angle),  and $\beta$ (related to the ratio of the Higgs vacuum expectation values) \cite{Gunion}.  \\

\item
Class b: ~ It contains the 4 processes (k=1,4)
\[
G^0H^0 ~,~  G^0h^0 ~,~ A^0H^0 ~,~ A^0h^0 ~,~
\]
 characterized by neutral final bosons carrying opposite
CP eigenvalues. The corresponding asymptotic limits for $F^k_{\pm\mp}$ then become
\bqa
F^k_{\pm\mp} & \to &  R_{bk} C^J_{\pm\mp}(\theta)~~, ~~~~~ {\rm with}  \nonumber  \\
R_{b1} & = & {m^2_t\sin\alpha\over\sin\beta}-{m^2_b\cos\alpha\over\cos\beta}~~, \nonumber \\
R_{b2} & = &  {m^2_t\cos\alpha\over\sin\beta}+{m^2_b\sin\alpha\over\cos\beta}~~, \nonumber \\
R_{b3} & = &  {m^2_t\sin\alpha\cot\beta\over\sin\beta}+
{m^2_b\cos\alpha\tan\beta\over\cos\beta}~~, \nonumber \\
R_{b4} & = &  {m^2_t\cos\alpha\cot\beta\over\sin\beta}-
{m^2_b\sin\alpha\tan\beta\over\cos\beta}~~, \label{Class-b}
\eqa
where $C^J_{\pm\mp}(\theta)$ describe the process-independent
part of  these limits, while $R_{bk}$ are again real and depend on the process.

It is important to remark that the relative phase
of $C^I_{\pm\mp}(\theta)$ and $C^J_{\pm\mp}(\theta)$, defined by the asymptotic
limits  in (\ref{Class-a}) and (\ref{Class-b}),
is always $\pi/2$. This is due to CP invariance in our model,
  and  the  fact that the product of the CP eigenvalues in   each  pair of the
final neutrals is always $+1$ for class  a,  and $-1$ for class  b. \\

\item
Class c: ~  It contains the  3 charged boson processes (k=1,3)
\[
G^+G^- ~,~  G^+H^- ~,~  H^+H^- ~.
 \]
The corresponding HC  amplitudes
$F^k_{\pm\mp}$ at high energies may then be expressed as
\bqa
F^k_{\pm\mp} & \to & R^I_{ck} C^I_{\pm\mp}(\theta) +
 R^J_{ck} C^J_{\pm\mp}(\theta) ~~, \label{Class-c1}
\eqa
using  the same angular dependent functions  as in (\ref{Class-a}, \ref{Class-b}).
  The corresponding couplings in (\ref{Class-c1})  are again real and given by
\bqa
R^I_{c1} = m^2_t+m^2_b=R_{a1} & , & R^J_{c1} = m^2_t-m^2_b \simeq R_{a1} ~~,    \nonumber \\
R^I_{c2} =m^2_t\cot\beta-m^2_b\tan\beta=R_{a2} &,&
R^J_{c2} =m^2_t\cot\beta +m^2_b\tan\beta
 ~~,    \nonumber \\
 R^I_{c3} = m^2_t\cot^2\beta+m^2_b\tan^2\beta=R_{a3} &,&
 R^J_{c3} = m^2_t\cot^2\beta -m^2_b\tan^2\beta ~~. \label{Class-c2}
\eqa
Since the relative phase of the two terms
in  (\ref{Class-c1}) is always $\pi/2$,  there is never
any interference between them, in the differential cross sections.\\

\end{itemize}

To recapitulate  on the $gg\to HH'$ processes in MSSM, we note that the
high energy limits of the dominant HC amplitudes $F_{\pm\mp}$  in
  (\ref{Class-a}-\ref{Class-c2}),
are determined  by the quark boxes
in Fig.\ref{ggHH-diag-fig}. As the energy decreases to intermediate values,
the relative magnitudes of the HC amplitudes for the various processes
are changed,  due to   squark contributions that start becoming important.
In addition to it, the HV amplitudes $F_{\pm \pm}$ also
become  important, at intermediate energies.\\

\subsection{Analytical  results for $gg \to VH$ in MSSM}

The helicity structure (shown  in (\ref{independent-F-VH}))  is now richer
than for the $gg\to HH'$ case.
But the HCns rule greatly simplifies its asymptotic structure in MSSM, predicting that
   $F_{\pm\mp 0}$ dominates, while all other  amplitudes must  be  vanishing.

Again, it is possible to   understand analytically many of the  high energy
properties of these amplitudes, by looking at the diagrams of
 Fig.\ref{ggVH-diag-fig} \cite{techpaper}.
 Using the  names for the diagrams indicated in this figure,  we  find that:
\begin{itemize}

\item
The HC amplitudes $F_{\pm\mp 0}$, which satisfy $\mu+\mu'-\tau=0$,
are the only ones that do not vanish
asymptotically, and tend instead to   "constants".
This comes from  combining the contributions of
the various diagrams in  Fig.\ref{ggVH-diag-fig}. The high energy values of these
amplitudes  may most easily be obtained  by
using the equivalence theorem \cite{equiv}, which respectively relates $F_{\pm\mp 0}$  for
\[
gg \to Z_{\tau=0} G^0 ~,~ Z_{\tau=0} A^0 ~,~ Z_{\tau=0} H^0 ~,~ Z_{\tau=0} h^0 ~,~
W^+_{\tau=0}G^- ~,~ W^+_{\tau=0}H^- ~,
\]
to  the $F_{\pm\mp}$ amplitudes for
\[
gg \to G^0 G^0 ~,~ G^0 A^0 ~,~ G^0 H^0 ~,~ G^0 h^0 ~,~
G^+G^- ~,~ G^+H^- ~,
\]
determined in  (\ref{Class-a}-\ref{Class-c2}).

A "constant"  asymptotic   behaviour for  $F_{\pm\mp 0}$ in $gg \to VH$
turns out to be  true in SM also; but in this later case,
  some of the HV amplitudes may also tend  asymptotically to comparable "constants". \\

\item

For amplitudes with  $(\mu=\mu',~ \tau=0)$, we always have   $|\mu+\mu'-\tau|=2$.
In this case, non-vanishing asymptotic contributions  may only come from the diagrams
$F$, $H$ and $A$. Their sum     is always strongly suppressed, though,
forcing  these amplitudes  to vanish quickly at high energies.\\

\item

For amplitudes with $\mu=\mu'=-\tau$, which always satisfy $|\mu+\mu'-\tau|=3$,
non-vanishing asymptotic contributions only come from
the $F$ and $H$ boxes. These boxes are very small in this case,  and
strongly canceling each other. Therefore,  $F_{\pm\pm\mp}$  are
very small and  quickly vanishing at high energies. In fact,
these amplitudes are vanishing at high energies faster, than those of the previous item. \\

\item
We next turn to amplitudes satisfying $|\mu+\mu'-\tau|=1$, for processes
involving  neutral final particles.
These  HV amplitudes receive their asymptotic  contributions from the $F$
and $H$ diagrams of Fig.\ref{ggVH-diag-fig}; with their  sum often behaving   like
  $\sim m/\sqrt{s}$, and thus being strongly suppressed.

Occasionally though, this  suppression is reduced by a $\ln^2(s)$ factor, which
makes their vanishing very slow.
Below we list only these  slowly vanishing  amplitudes, for the relevant MSSM processes.
Their  high energy structure
is determined by\footnote{As already sated above, a color factor $\delta_{ab}$
is always removed from the amplitudes. Moreover $I_3^q$ in (\ref{neutral-slow1})
describe the third isospin component of the $t$ and $b$ quarks.}
\bq
F_{\mu\mu'\tau} \sim \sum_{q=t,b} ~{\alpha_s\alpha
(2I_3^q)m^2_q\sqrt{s}\sin\theta\over8\sqrt{2}s^2_Wc_W\mw} \tilde F_{\mu\mu'\tau} ~~,
\label{neutral-slow1}
\eq
with
\bqa
&& \tilde F_{+++}  \simeq    \left ({1\over t}-
~{1\over u}\right )\ln^2 \left ({-s\over m^2_q} \right )
+ \left ({1\over t}+~{1\over u}\right )
\left [ \ln^2\left ({-t\over m^2_q}\right ) - \ln^2 \left ({-u\over m^2_q}\right )\right ]~~,
 \nonumber \\
&& \tilde F_{+-+} \simeq    - {1\over t}
\left [\ln^2 \left ({-s\over m^2_q}\right ) -\ln^2\left ({-t\over m^2_q}\right )
- \ln^2\left ({-u\over m^2_q} \right) \right ] ~~, \nonumber \\
&& \tilde F_{+--} \simeq    {1\over u} \left [ \ln^2 \left ({-s\over m^2_q} \right )
-\ln^2 \left ({-t\over m^2_q}\right ) - \ln^2\left ({-u\over m^2_q}\right )\right ] ~.
\label{neutral-slow2}
\eqa
Corresponding expressions  for the amplitudes related to them by Bose statistics
and CP-invariance, may be obtained from  (\ref{VH-Bose-CP}) for neutral final bosons.
Note that the non-vanishing contributions in
(\ref{neutral-slow1}, \ref{neutral-slow2}) solely arise from the $t$ and $b$ quarks;
and that they indeed have an
$(m/\sqrt{s})\ln^2 s$-behaviour\footnote{In fact (\ref{neutral-slow1}, \ref{neutral-slow2})
 describe these slowly vanishing amplitudes for  $gg\to ZH$ in SM,
which, as observed in Section 4, "accidentally" also obeys  HCns.}.
Depending on the  neutral final state, the corresponding  slowly vanishing amplitudes
for the various MSSM processes  are:

Process  $gg\to ZH^0$:  The slowly vanishing HV amplitudes are    given
by (\ref{neutral-slow1}, \ref{neutral-slow2}), provided we include the  extra
factors $(\sin\alpha / \sin\beta) $   in  the top contribution,
and $(\cos\alpha / \cos\beta ) $ in  the  bottom contribution.

Process  $gg\to Zh^0$:  The slowly vanishing HV amplitudes are   given
by (\ref{neutral-slow1}, \ref{neutral-slow2}), provided we include the  extra factors
 $( \cos\alpha /\sin\beta )$ in  the  top contribution, and
 $-( \sin\alpha / \cos\beta )$ in the  bottom one.

Process  $gg\to ZA^0$:  The slowly vanishing HV amplitudes may again be obtained from
 (\ref{neutral-slow1}, \ref{neutral-slow2}), provided we include the extra factors
  $-i\cot\beta$ in the   top contribution,  and   $- i\tan\beta$
in  the bottom contribution, and a sign-change is made to  $F_{+-+}$.

Process  $gg\to ZG^0$: The slowly vanishing HV amplitudes are again given by
 (\ref{neutral-slow1}, \ref{neutral-slow2}), provided  an extra
 factor $-i$ is  included  for the top, and $+i$ for the  bottom contributions,
 and an additional sign-change  is made  to $F_{+-+}$.\\

\item
Finally we consider the  amplitudes satisfying $|\mu+\mu'-\tau|=1$,
for the charged final state processes   $gg\to W^+\{H^-, ~G^-\}$.
Their dominant contribution again come from the $F,~H$ boxes.
In this case the constraints from Bose statistics and
CP invariance are different though; see (\ref{VH-Bose-CP}). Now,  $F_{+++}$
and $F_{---}$ receive no logarithmic enhancement and
vanish quickly at high energies. Thus, the only slowly vanishing HV
amplitudes, behaving like $\sim (m/\sqrt{s})\ln^2 s$, are
\bqa
 F_{+-+} & \simeq  &   {\alpha_s\alpha
m^2_t\sqrt{s}\sin\theta\over4\sqrt{2}s^2_W\mw}
\left [\ln^2 \Big ( {-s\over m^2_t} \Big ) -\ln^2 \Big ( {-t\over m^2_t} \Big )
- \ln^2\Big({-u\over m^2_t}\Big) \right ]\frac{\{\cot\beta,~1\}}{t} ~, \nonumber \\
F_{-++} & \simeq & -   {\alpha_s\alpha m^2_t\sqrt{s}\sin\theta\over4\sqrt{2}s^2_W\mw}
\left [ \ln^2 \Big ({-s\over m^2_t} \Big) -\ln^2\Big ({-t\over m^2_t} \Big)
- \ln^2\Big ({-u\over m^2_t } \Big) \right ]\frac{\{\cot\beta,~1\}}{u}~, \nonumber \\
F_{+--} &\simeq & - {\alpha_s\alpha m^2_b\sqrt{s}\sin\theta\over4\sqrt{2}s^2_W\mw}
\left [\ln^2\Big ({-s\over m^2_b}\Big )-\ln^2\Big ({-t\over m^2_b} \Big)
- \ln^2\Big ({-u\over m^2_b }\Big ) \right ] \frac{\{\tan\beta,~-1\}}{u}~~, \nonumber \\
F_{-+-} &\simeq &  {\alpha_s\alpha m^2_b\sqrt{s}\sin\theta\over4\sqrt{2}s^2_W\mw}
\left [\ln^2 \Big({-s\over m^2_b}\Big )-\ln^2({-t\over m^2_b} \Big )
- \ln^2\Big ({-u\over m^2_b } \Big ) \right ] \frac{\{\tan\beta,~-1\}}{t}~~,
\label{charged-slow}
\eqa
for $H^-$ and $G^-$ production respectively. Note that the magnitudes of the
 first two amplitudes in (\ref{charged-slow}) are  determined by the top mass, while
 those of the later two  are determined by the bottom.\\

\end{itemize}

As the energy decreases, squark contributions will also start affecting the
 HC amplitudes $F_{\pm\mp0}$.
In addition to it, other amplitudes will also start contributing to  these  processes;
most notably the purely transverse amplitudes  discussed in
(\ref{neutral-slow1}, \ref{neutral-slow2}, \ref{charged-slow}).\\

\subsection{Asymptotic $R_i $ relations  in MSSM}

We next turn to the so called   $\tilde \sigma$-quantities
\bqa
 \tilde \sigma(gg\to HH')  & \equiv & \frac{512 \pi}{\alpha^2 \alpha_s^2}\,
\frac{s^{3/2}}{p}\, {d\sigma (gg\to HH') \over d\cos\theta}~~~, \nonumber \\
 \tilde \sigma(gg\to VH) & \equiv & \frac{512 \pi}{\alpha^2 \alpha_s^2}\,
\frac{s^{3/2}}{p}\, {d\sigma (gg\to VH) \over d\cos\theta}~~, \label{sigma-tilde}
\eqa
which should be  measurable  at a hadronic collider;
see  (\ref{dsigmaHH1 }, \ref{dsigmaVH1 }).
In MSSM, where HCns is satisfied, the dimensionless $\tilde \sigma$ quantities
behave  asymptotically like angle-dependent "constants",
solely determined by the HC amplitudes; while in SM  some HV amplitudes may also
contribute at high energies. From here on,  all other results  in this section,
are valid in MSSM only.

Using (\ref{sigma-tilde}) and the results in (\ref{Class-a},\ref{Class-b} ),
we get  asymptotically
\bqa
 &R_1 \Rightarrow & \tilde \sigma (gg\to G^0G^0)  \simeq \tilde \sigma (gg\to G^0A^0)
\left (\frac{R_{a1}}{R_{a2}} \right )^2
\simeq \tilde \sigma (gg\to A^0A^0) \left (\frac{R_{a1}}{R_{a3}} \right )^2 \nonumber \\
&& \simeq \tilde \sigma (gg\to H^0H^0) \left (\frac{R_{a1}}{R_{a4}} \right )^2
\simeq \tilde \sigma (gg\to h^0h^0) \left (\frac{R_{a1}}{R_{a5}} \right )^2
\simeq \tilde \sigma (gg\to H^0h^0) \left (\frac{R_{a1}}{R_{a6}} \right )^2 \nonumber \\
 && \simeq \tilde \sigma (gg\to Z^0G^0)  \simeq \tilde \sigma (gg\to Z^0A^0)
\left (\frac{R_{a1}}{R_{a2}} \right )^2 ~~,  \label{R1-rel}
\eqa
and
\bqa
 &R_2 \Rightarrow & \tilde \sigma (gg\to G^0H^0)  \simeq \tilde \sigma (gg\to G^0h^0)
\left (\frac{R_{b1}}{R_{b2}} \right )^2
 \simeq \tilde \sigma (gg\to A^0H^0)\left (\frac{R_{b1}}{R_{b3}} \right )^2 \nonumber \\
 &&  \simeq \tilde \sigma (gg\to A^0h^0)\left (\frac{R_{b1}}{R_{b4}} \right )^2 \nonumber \\
&& \simeq \tilde \sigma (gg\to ZH^0)  \simeq \tilde \sigma (gg\to Zh^0)
\left (\frac{R_{b1}}{R_{b2}} \right )^2 ~~.  \label{R2-rel}
\eqa
Note that the last lines in  (\ref{R1-rel}, \ref{R2-rel}) receive at non-asymptotic energies
also  contributions from
the slowly vanishing amplitudes involving transverse final vector bosons; see the
discussion around (\ref{neutral-slow1}, \ref{neutral-slow2}).\\

We can also relate the cross sections of the charged sector,
to those of the neutral sector; classes a,b,c above.
Thus,  combining
(\ref{Class-a}, \ref{Class-b}, \ref{Class-c1}-\ref{Class-c2}, \ref{sigma-tilde}), we obtain
\bqa
 &R_3 \Rightarrow& \tilde \sigma (gg\to G^+G^-)  \simeq \tilde \sigma (gg\to G^0G^0)
+\left (\frac{R^J_{c1}}{R_{b2}}\right )^2 \tilde \sigma (gg\to G^0h^0) ~~,
  \label{R3-rel} \\
&R_4 \Rightarrow&  \tilde \sigma (gg\to G^+H^-)  \simeq
\left ( \frac{R^I_{c2}}{R_{a1}} \right )^2
\tilde \sigma (gg\to G^0G^0) +\left (\frac{R^J_{c2}}{R_{b2}}\right )^2
\tilde \sigma (gg\to G^0h^0)~,  \label{R4-rel} \\
&R_5 \Rightarrow&  \tilde \sigma (gg\to H^+H^-)  \simeq
\left ( \frac{R^I_{c3}}{R_{a1}} \right )^2 \tilde \sigma (gg\to G^0G^0)
+ \left ( \frac{R^J_{c3}}{R_{b2}} \right )^2 \tilde \sigma (gg\to G^0h^0)
~,  \label{R5-rel}
\eqa
connecting charged and neutral final states.

Eliminating the neutral channels  from
(\ref{R3-rel},\ref{R4-rel}, \ref{R5-rel}),  we obtain
\bqa
 && \tilde \sigma (gg\to H^+H^-)  \simeq {1\over \left ({m^4_t\over\tan^2\beta}+m^4_b \right )}
 \Bigg \{ \left ({m^4_t\over\tan^4\beta}+m^4_b \tan^2\beta \right )
 \tilde \sigma (gg\to G^+G^-)\nonumber\\
 && +\left ({m^4_t\over\tan^4\beta}-m^4_b \right )(1-\tan^2\beta)
 \tilde \sigma (gg\to G^+H^-)\Bigg \}
 ~~,  \label{sigma-charged}
\eqa
which in fact is a relation among $R_3,~R_4, ~R_5$, that
could have also been obtained directly from (\ref{Class-c1}-\ref{Class-c2}).\\

Concerning $gg \to VH$,  with charged final sates, we get two more relations,
\bqa
 &R_6 \Rightarrow&  \tilde \sigma (gg\to G^+G^-)   \simeq  \tilde \sigma (gg\to W^+G^-) ~,
 \label{R6-rel} \\
&R_7 \Rightarrow&  \tilde \sigma (gg\to G^+H^-)  \simeq  \tilde \sigma (gg\to W^+H^-) ~,
\label{R7-rel}
\eqa
using  (\ref{sigma-tilde}). In deriving these relations,
the high energy equivalences theorem was used, and
the slowly vanishing transverse amplitudes discussed   in
(\ref{charged-slow}) were neglected. Since these relations constrain
the $W$ production processes,
 they should be considered in conjunction with the last two parts of
(\ref{R1-rel}, \ref{R2-rel}), affecting  corresponding $Z$
 cross sections.\\

The relations $R_i$ of (\ref{R1-rel}-\ref{R7-rel}) are  analogous, in spirit, to those
concerning    the cross sections for  $ug\to dW$  and $ug\to \tilde d_L\tchi^+_i$, derived in
\cite{ugdW-ugsdWino}.  At asymptotic energies,
they   should be very  accurate,   depending only on the  MSSM angles   $\beta$ and $\alpha$;
   see  (\ref{Class-a}, \ref{Class-b}, \ref{Class-c2}).
 Provided the SUSY particles
 are sufficiently light, or the hadronic collider sufficiently energetic,
  $\beta$ and $\alpha$ could be  determined from such relations.

As the energy decreases to intermediate values, deviations appear in $R_i$,
which are due to 2 types of contributions. The first  comes from  the sub-dominant HV amplitudes
which are slowly vanishing, like $m/\sqrt{s}$ times logarithmic terms.
The second one, comes  from  the squark boxes.
Thus, at intermediate energies, further model dependence is introduced,
whose investigation should offer  a deeper  insight to the MSSM picture.
At such energies, we also expect on general grounds,  that $R_i$ become better
in the central angular region, away from the forward and backward angles \cite{techpaper}.\\

\section{Numerical results.}

As already said, the helicity amplitudes for all the
gluon-gluon fusion
 processes in (\ref{ggHH-proc2}, \ref{ggVH-proc2}) are
calculated  in terms of Passarino-Veltman (PV) functions
\cite{Veltman}, using   \cite{looptools} and the  FORTRAN codes
gghhcode and ggvhcode   \cite{code}. The resulting  helicity
amplitudes are  expressed as functions of the c.m. energy and
angle,
 in either   the   SM or the  MSSM models. Input  couplings and masses
 are always assumed to be real and at  the electroweak scale, while   the quark masses of
the first two generations are neglected. The output files
generated after running the various codes,  are  specified as
".dat" for the $gg\to HH'$ case; and as ".dat1, .dat2" for the
$gg\to VH$ case.
 An accompanying Readme, fully explains the compilation  of the codes.\\

In the figures presented here, we can only give examples
of the helicity  amplitudes for the various processes.
Thus, for $gg\to HH'$, we just  plot  the two independent amplitudes $F_{++}, F_{+-}$;
see (\ref{ggHH-proc2}, \ref{HH-Bose-CP}, \ref{independent-F-HH}).
Correspondingly,  for $gg\to VH$, the figures contain
  the six independent amplitudes $F_{+++}$, $F_{++-}$ , $F_{++0}$, $F_{+-+}$,
$F_{+--}$, $F_{+-0}$, for neutral final particles; while in the charged case,
the amplitudes    $F_{---}$, $F_{--0}$ and $F_{--+}$ are  also included;
see (\ref{ggVH-proc2}, \ref{VH-Bose-CP}, \ref{independent-F-VH}).

As a first example,  Fig.\ref{h0h0-SPA-SM-fig} shows the HC and HV amplitudes
for $gg\to h^0 h^0$, in both, the MSSM benchmark   $SPS1a'$ \cite{SPA},
and in the SM cases. Panel (a)   addresses the MSSM amplitudes  $(F_{+-}, F_{++})$,
 in a sufficiently  high energy region  elucidating the asymptotic behavior;
 while panel  (b) is restricted  to  a  more LHC-type energy range.
 Panels (c, d) give the corresponding amplitudes for $gg\to HH$ in SM.

As shown in  Fig.\ref{h0h0-SPA-SM-fig}, above 6 TeV, the HC amplitude for this process strongly
dominates in  MSSM, but not in SM.

\begin{figure}[p]
\vspace*{-1cm}
\[
\epsfig{file=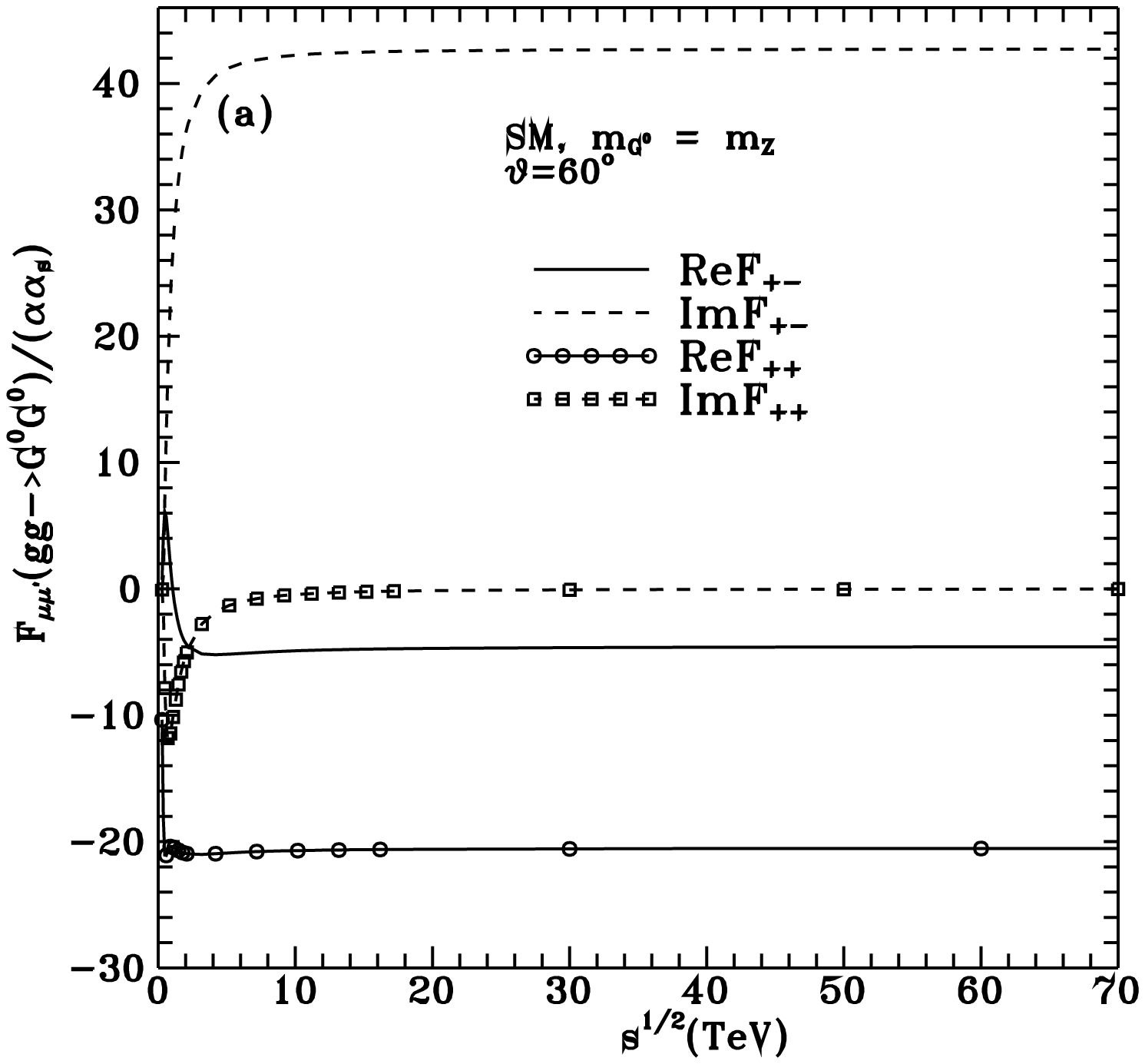, height=7.cm}\hspace{1.cm}
\epsfig{file=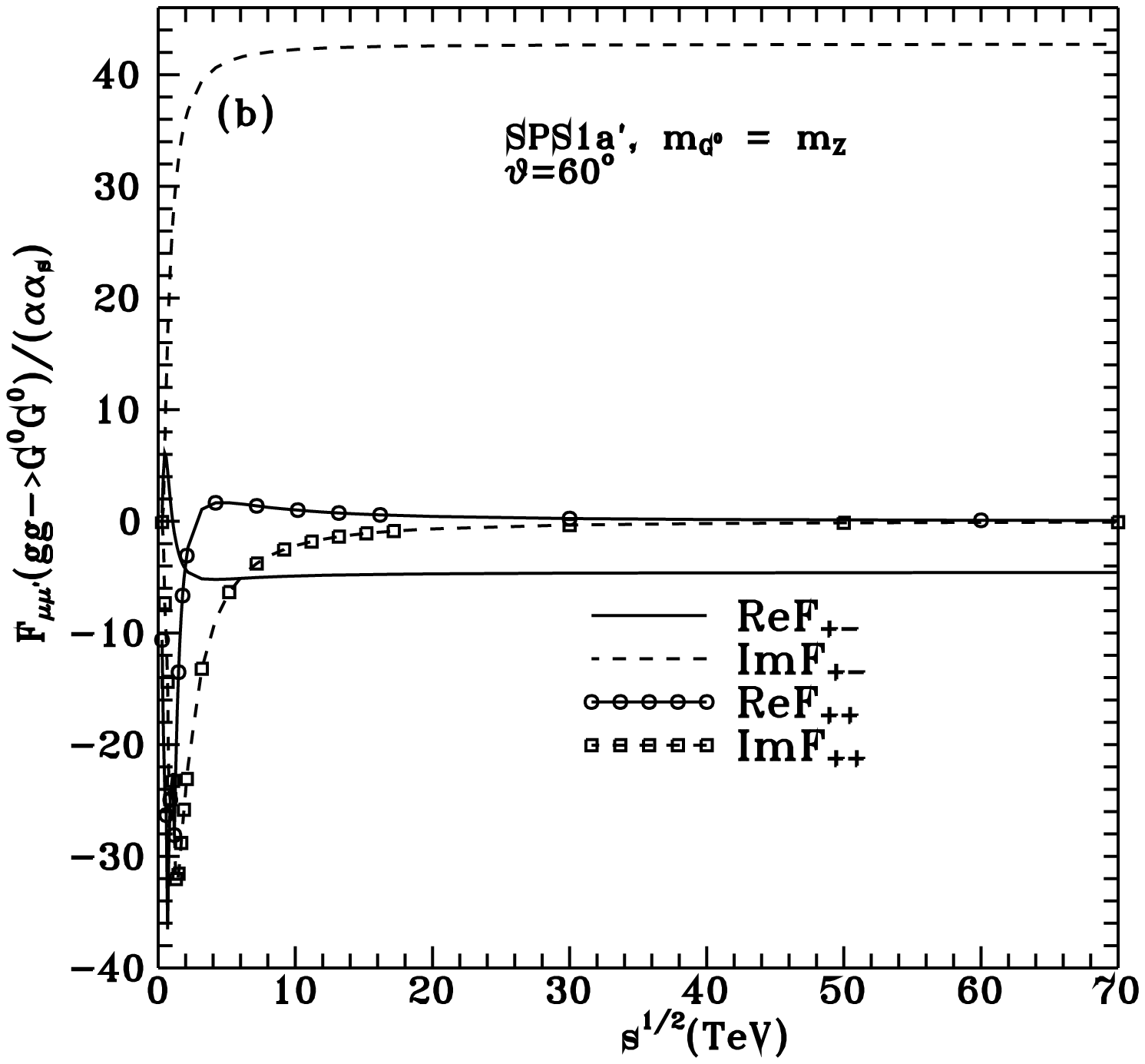, height=7.cm}
\]
\caption[1]{Amplitudes for  $gg \to G^0 G^0 $  in SM (a) and in $SPS1a'$ (b). }
\label{G0G0-SPA-SM-fig}
\end{figure}

\begin{figure}[t]
\vspace*{-1cm}
\[
\epsfig{file=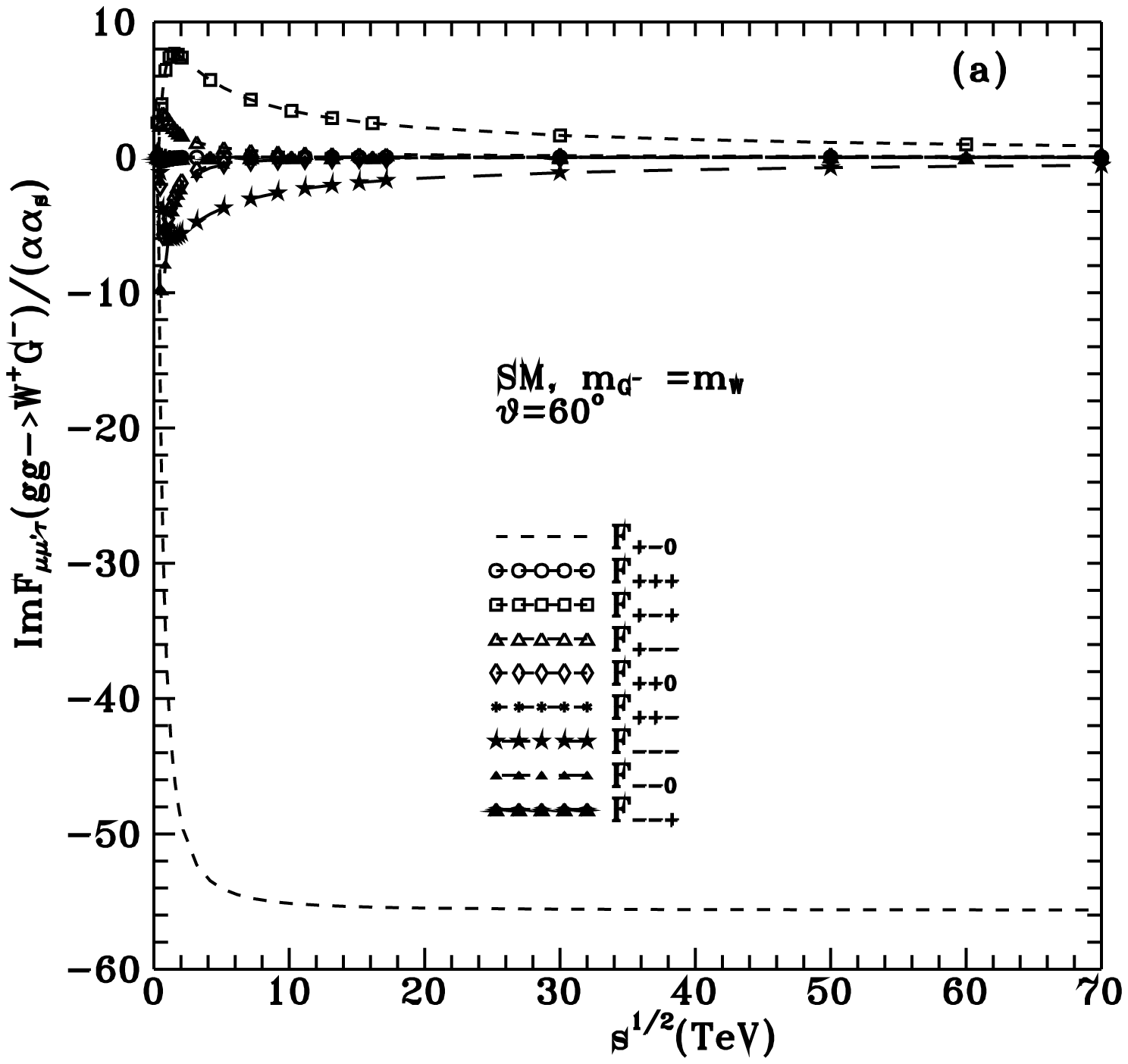, height=7.cm}\hspace{1.cm}
\epsfig{file=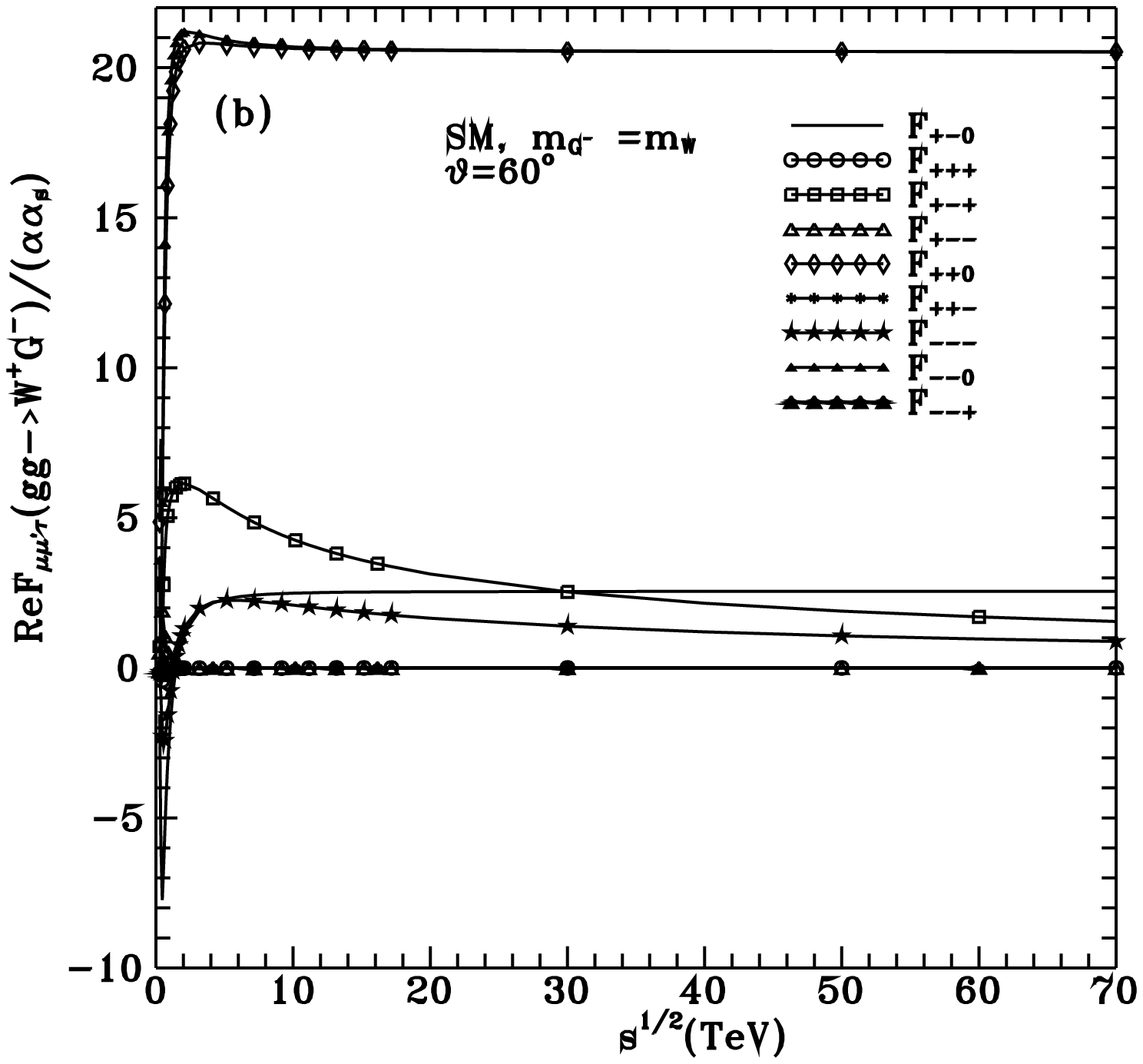, height=7.cm}
\]
\[
%\hspace{-0.5cm}
\epsfig{file=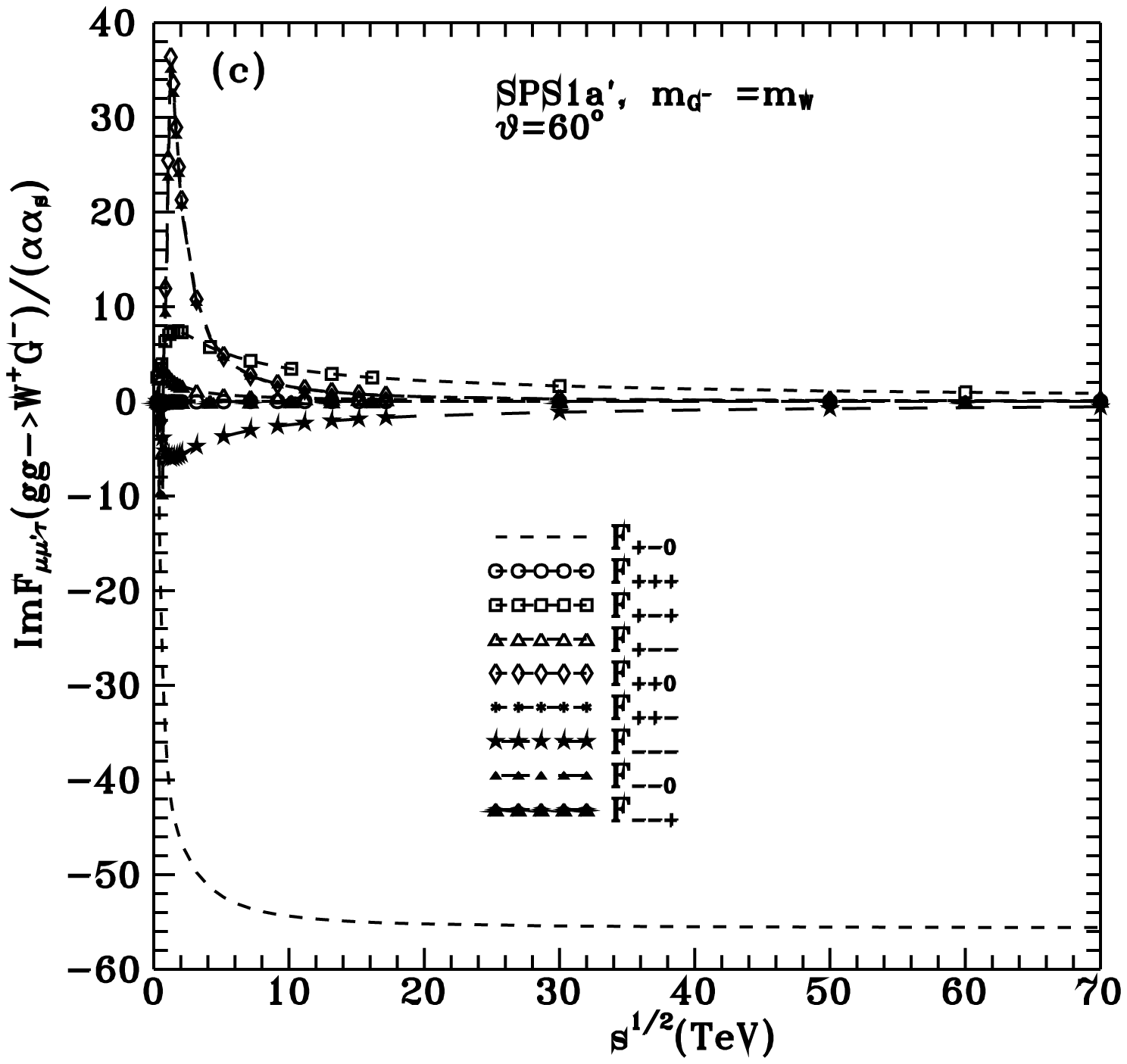, height=7.cm}\hspace{1.5cm}
\epsfig{file=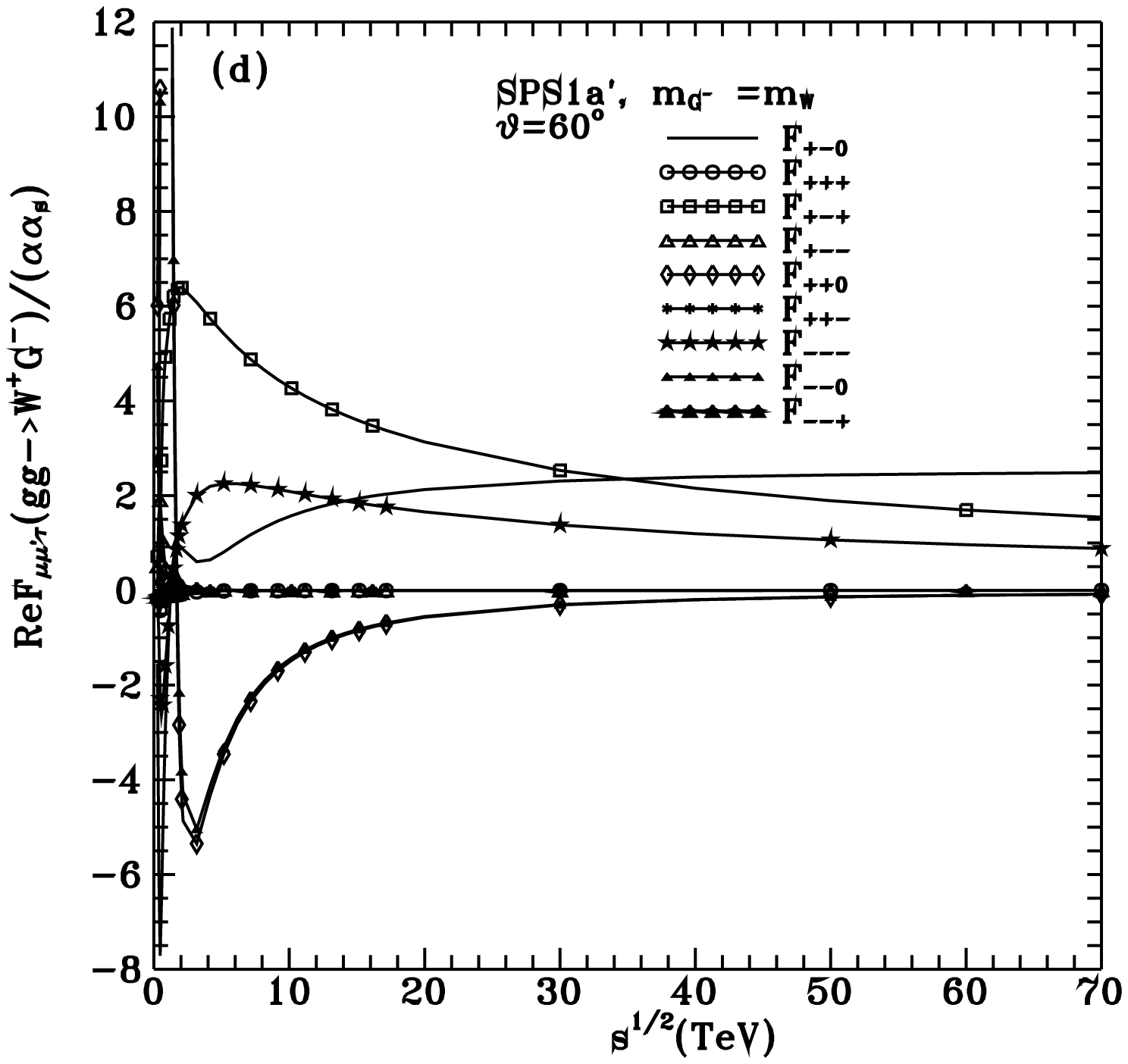, height=7.cm}
\]
\caption[1]{Amplitudes for  $gg \to W^+G^- $  in SM (a,b), and  $SPS1a'$ (c,d). }
\label{WpGm-SPA-SM-fig}
\end{figure}

In fact, the SM process  $gg \to HH $, constitutes an   example
where HCns is   strongly violated in SM.
Similar  violations of  HCns  in SM,
 may also been seen  for  $gg\to G^0 G^0$
in Fig.\ref{G0G0-SPA-SM-fig}a; and for $gg\to W^+G^-$ in Fig.\ref{WpGm-SPA-SM-fig}a,b.
Because of these and the equivalence theorem, a clear  violation of HCns
 for the SM processes $gg\to Z G^0$ and $gg\to G^+G^-$ is also true.
These are the only known examples where HCns is not even approximately
obeyed in SM.

Contrary to them, the corresponding MSSM results in Figs.\ref{G0G0-SPA-SM-fig}b
and \ref{WpGm-SPA-SM-fig}c,d satisfy helicity conservation.\\

\begin{figure}[p]
\vspace*{-1cm}
\[
\epsfig{file=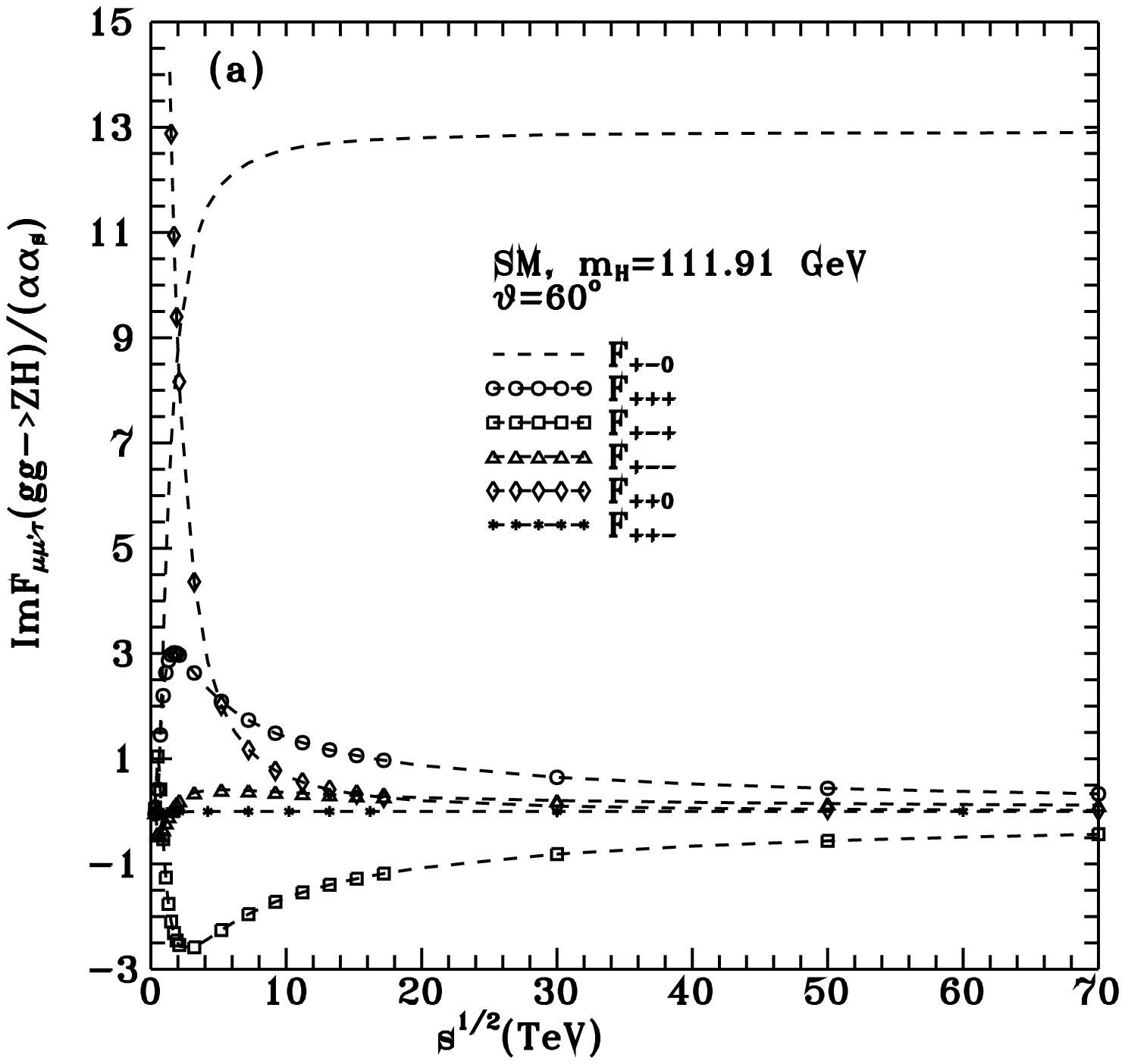, height=7.cm}\hspace{1.cm}
\epsfig{file=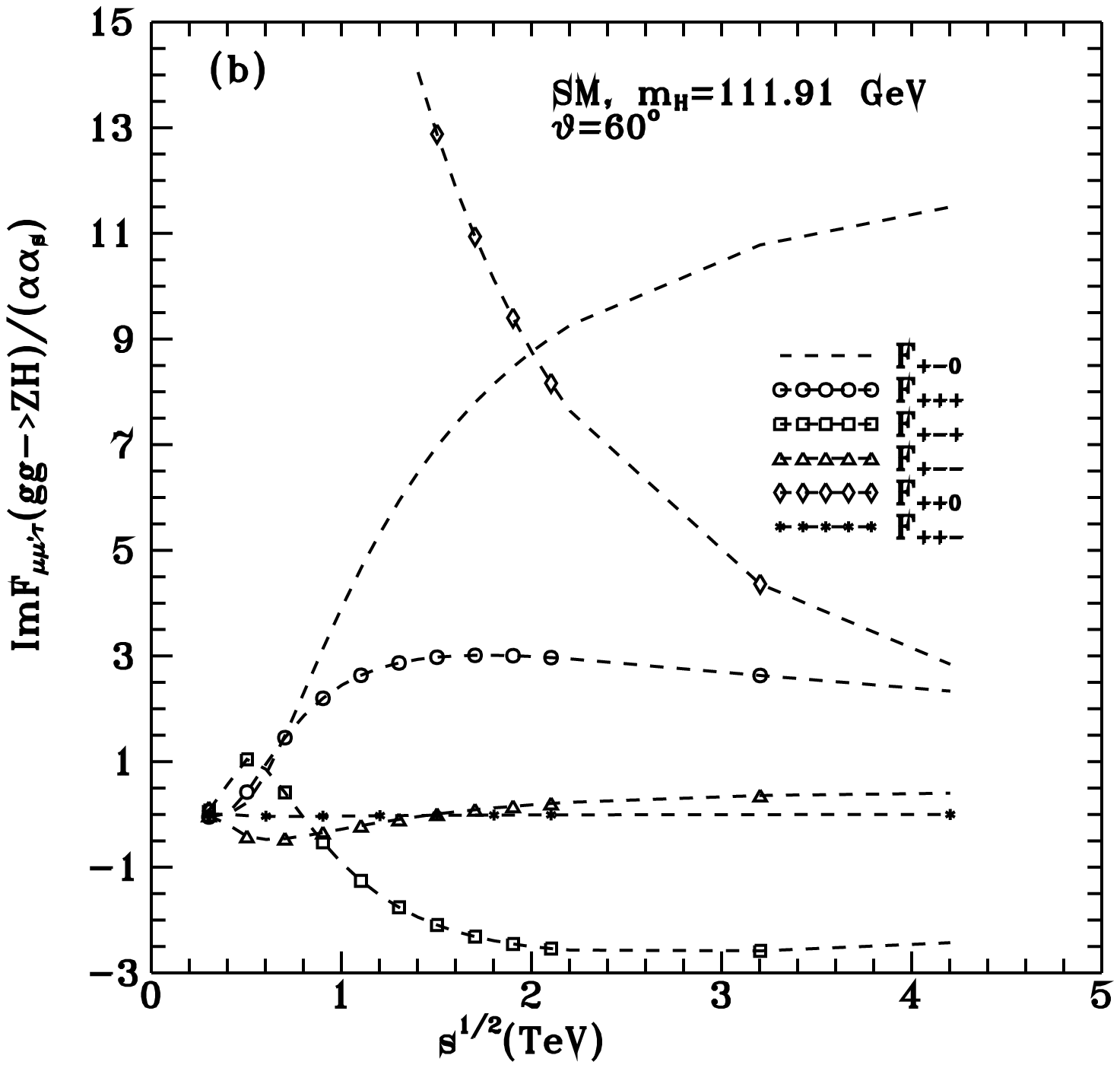,height=7.cm}
\]
\[
\epsfig{file=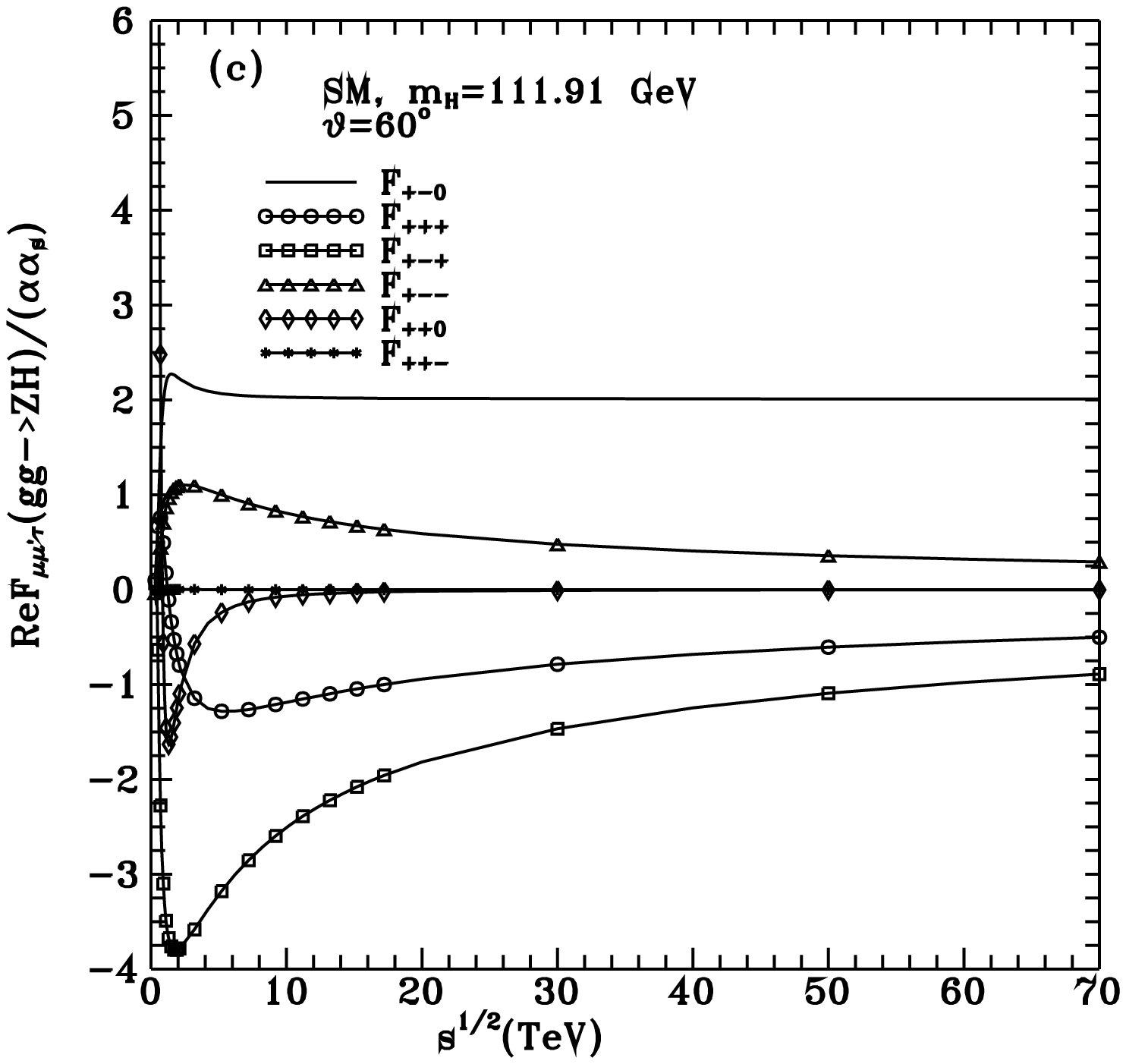, height=7.cm}\hspace{1.cm}
\epsfig{file=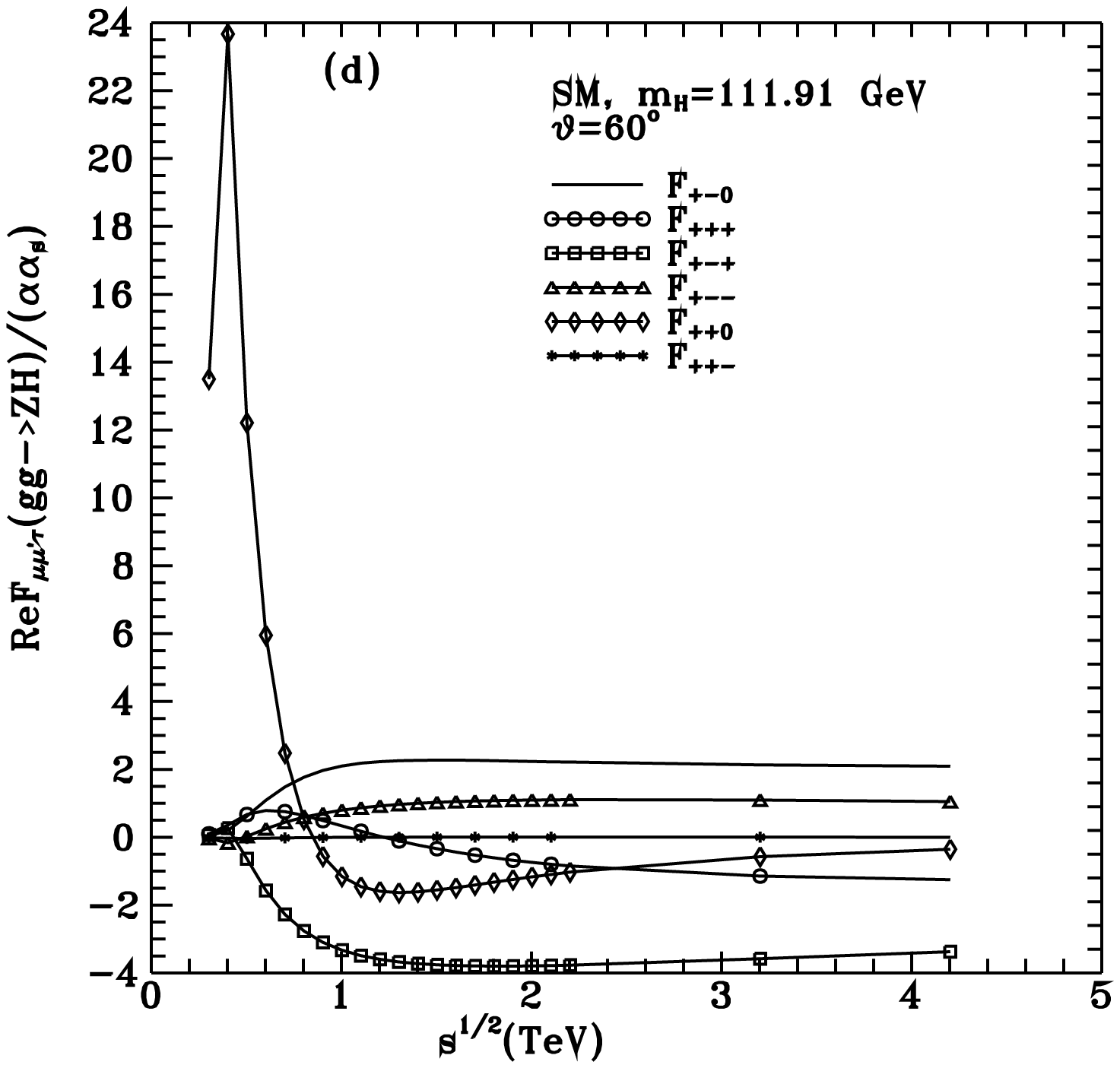,height=7.cm}
\]
\caption[1]{Amplitudes  for  $gg \to Z H $  in SM;  (a,c) describe the high energy behaviour,
while  (b,d)  emphasize the  LHC range.}
\label{Zh0-SM-fig}
\end{figure}

\begin{figure}[p]
\vspace*{-1cm}
\[
\epsfig{file=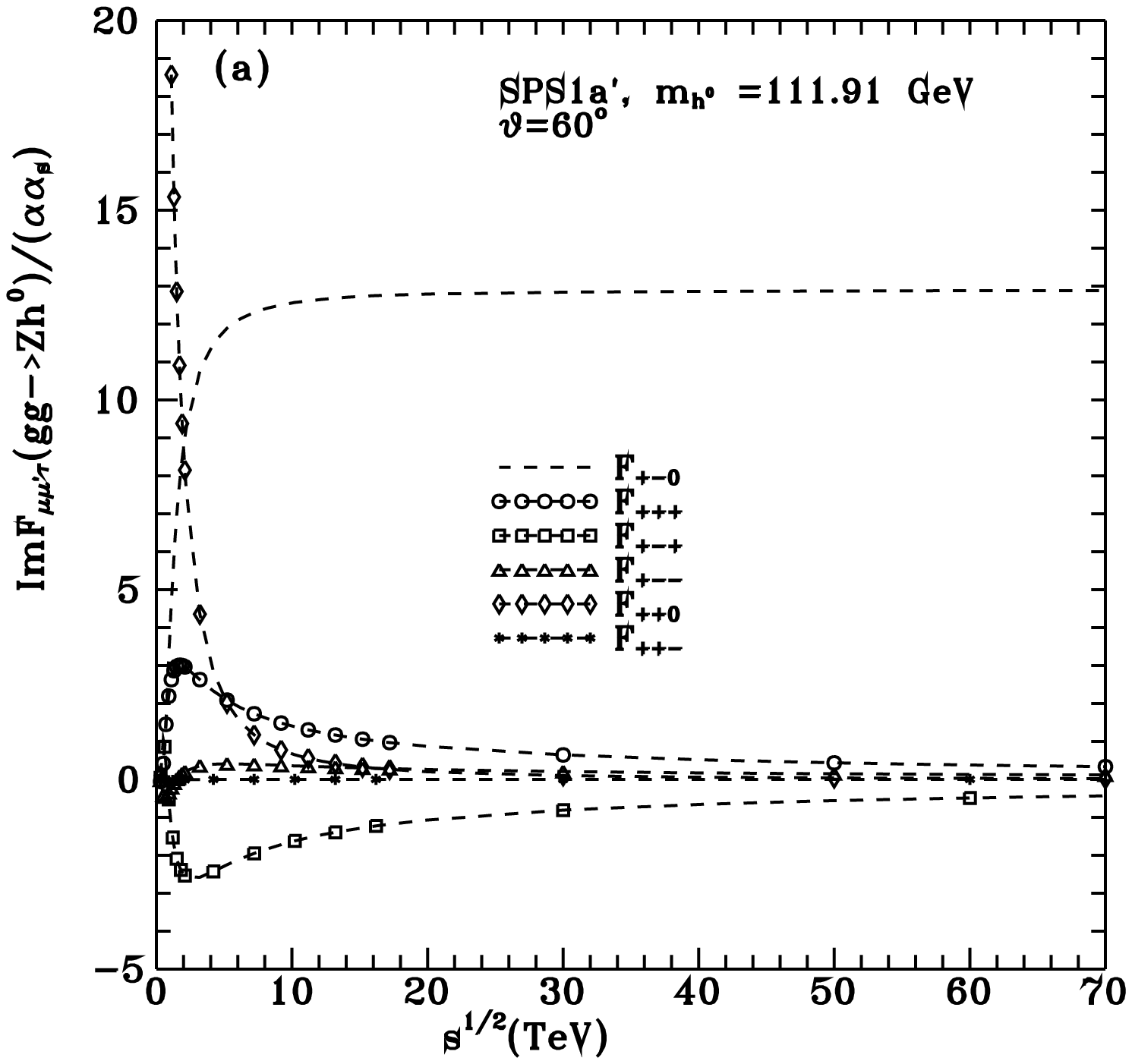, height=7.cm}\hspace{1.cm}
\epsfig{file=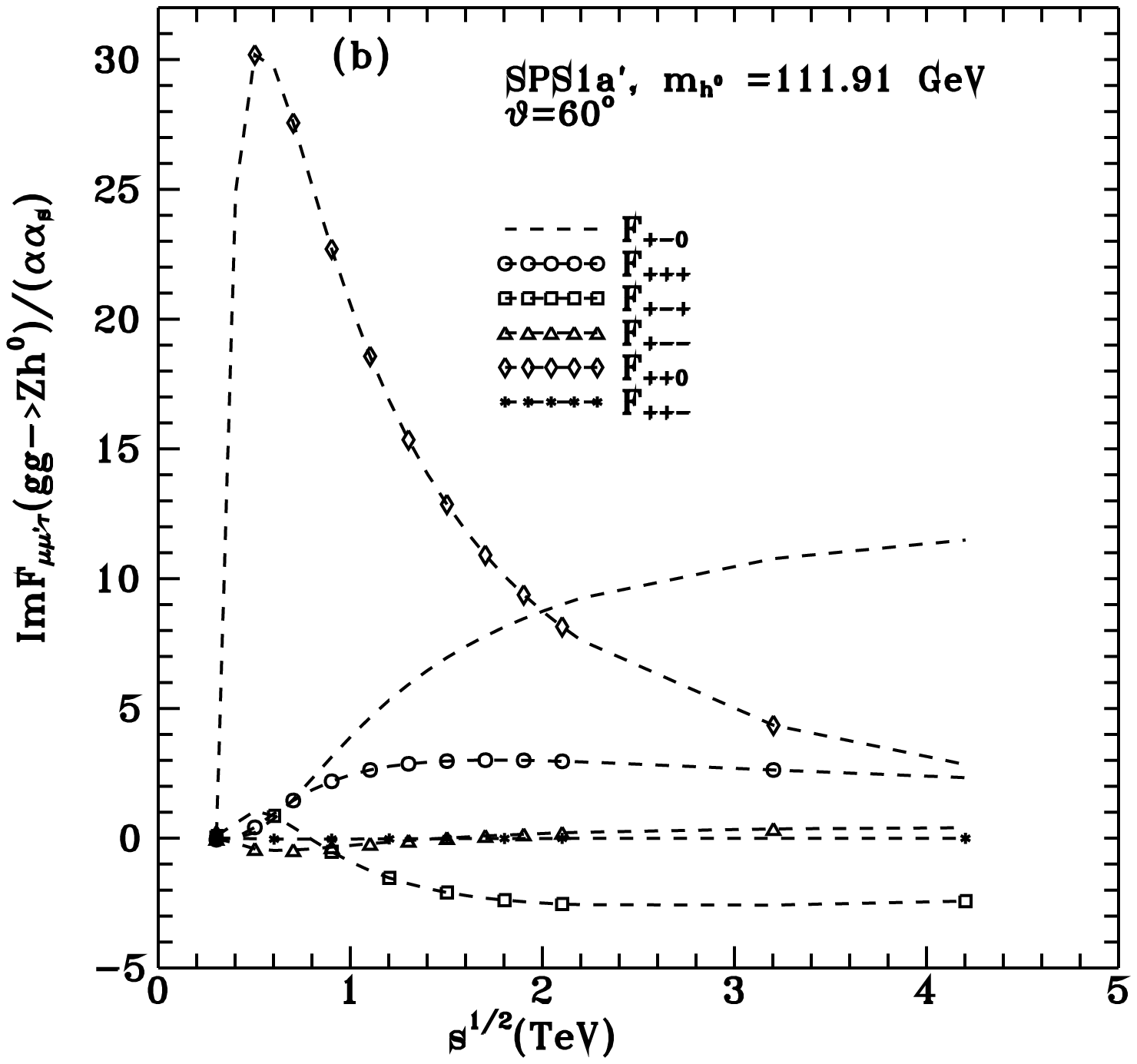,height=7.cm}
\]
\[
\epsfig{file=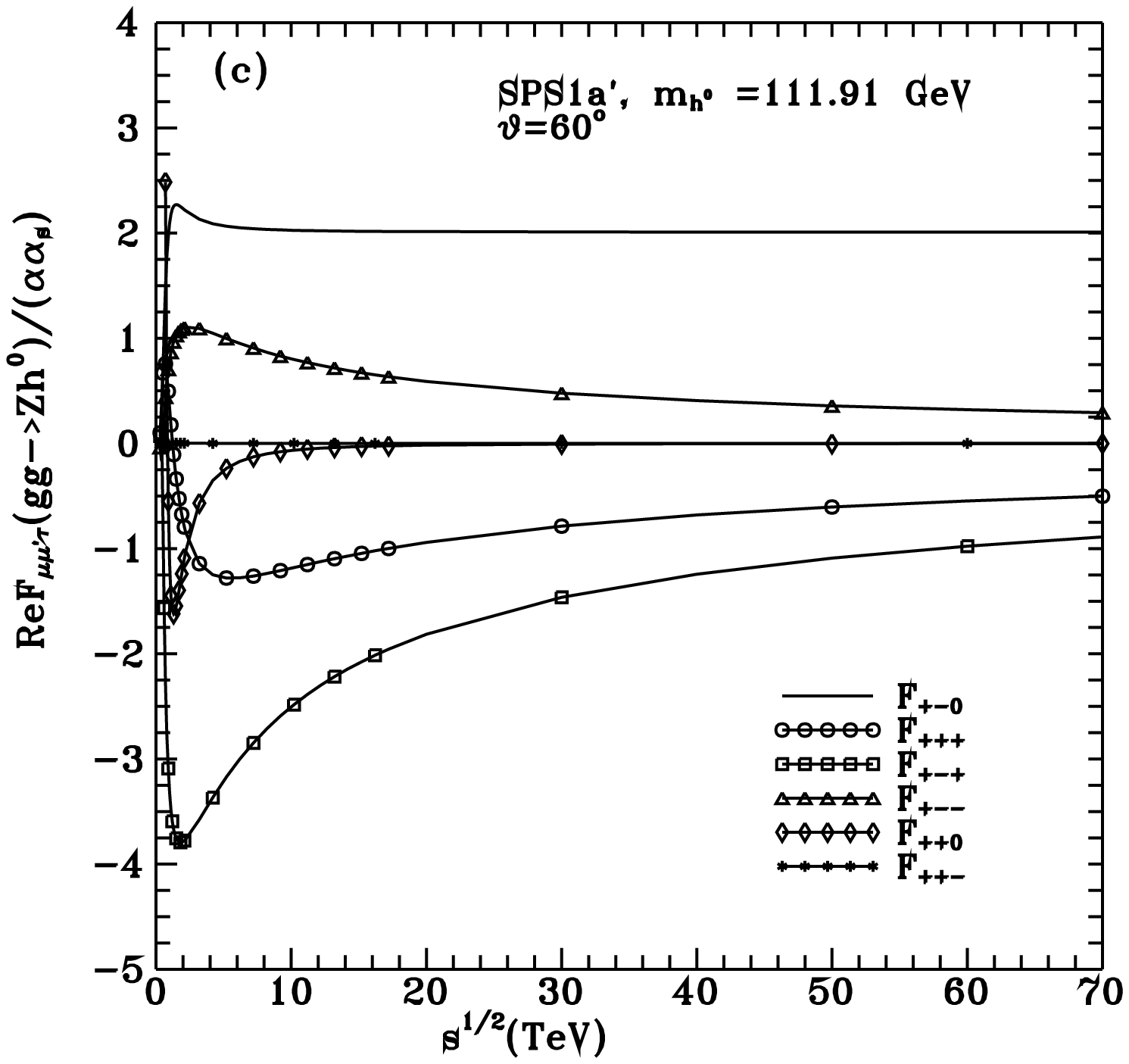, height=7.cm}\hspace{1.cm}
\epsfig{file=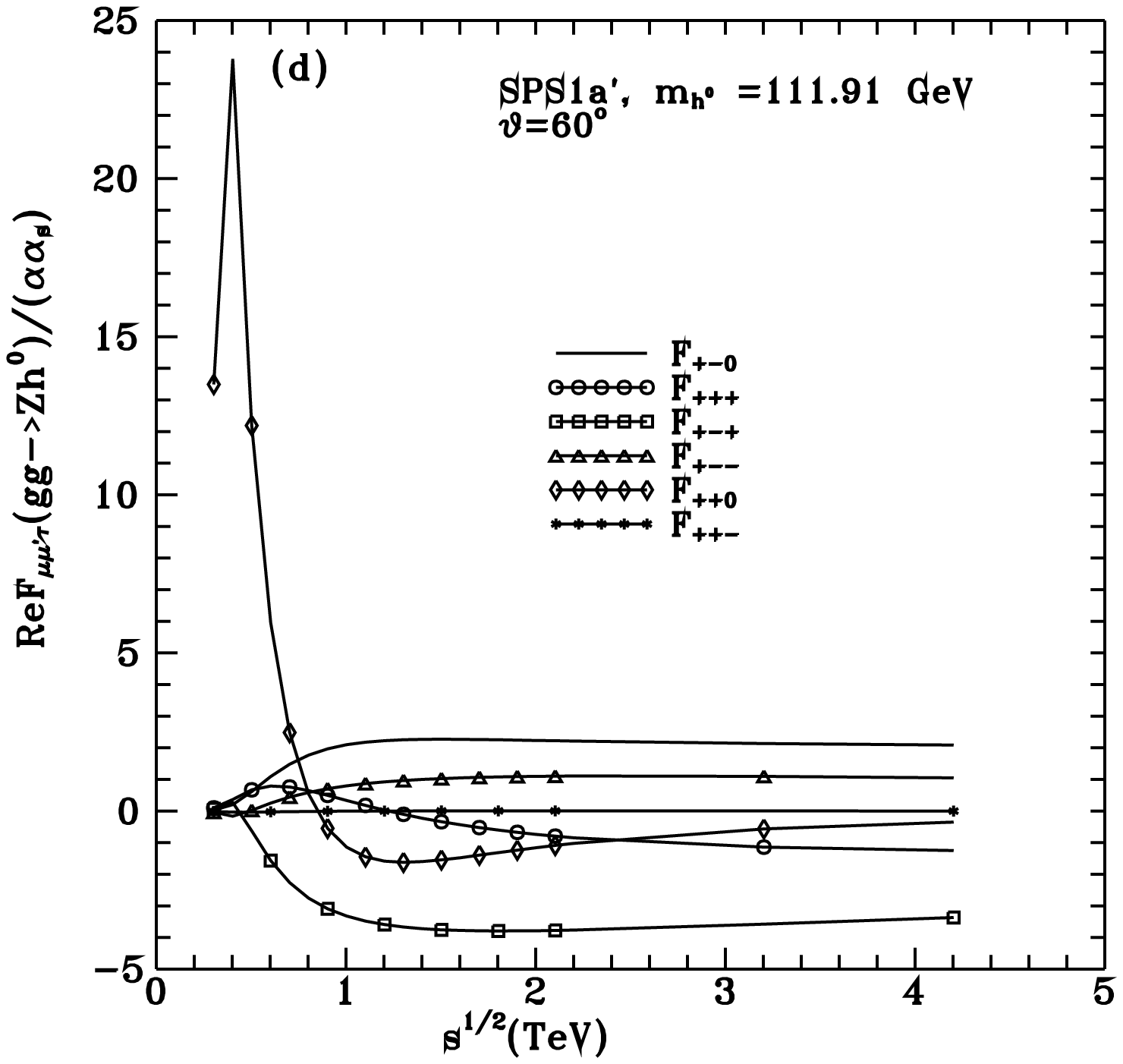,height=7.cm}
\]
\caption[1]{Amplitudes for  $gg \to Z h^0 $  in $SPS1a'$;
(a,c) describe the high energy behaviour while  (b,d)  emphasize the  LHC range. }
\label{Zh0-SPA-fig}
\end{figure}

A peculiarity arises for  the SM process  $gg \to Z H$ presented in
Fig.\ref{Zh0-SM-fig}, and the corresponding MSSM process $gg \to Z h^0$
 shown in Fig.\ref{Zh0-SPA-fig}. The validity of  HCns in both   cases seems equally good.
 A similar situation arises also
 for the SM process $gg\to G^0H$ and the MSSM process $gg \to G^0h^0$,
 related to the previous ones by the equivalence theorem.
 Such an  "accidental" validity of helicity conservation for e.g. $gg\to G^0H$ in SM,
 must be related to the  absence of a  squark contribution
 in the corresponding  MSSM process $gg \to G^0 h^0$,  which  makes   the SM and MSSM  amplitudes
 very similar. \\

\begin{figure}[p]
\vspace*{-1cm}
\[
\epsfig{file=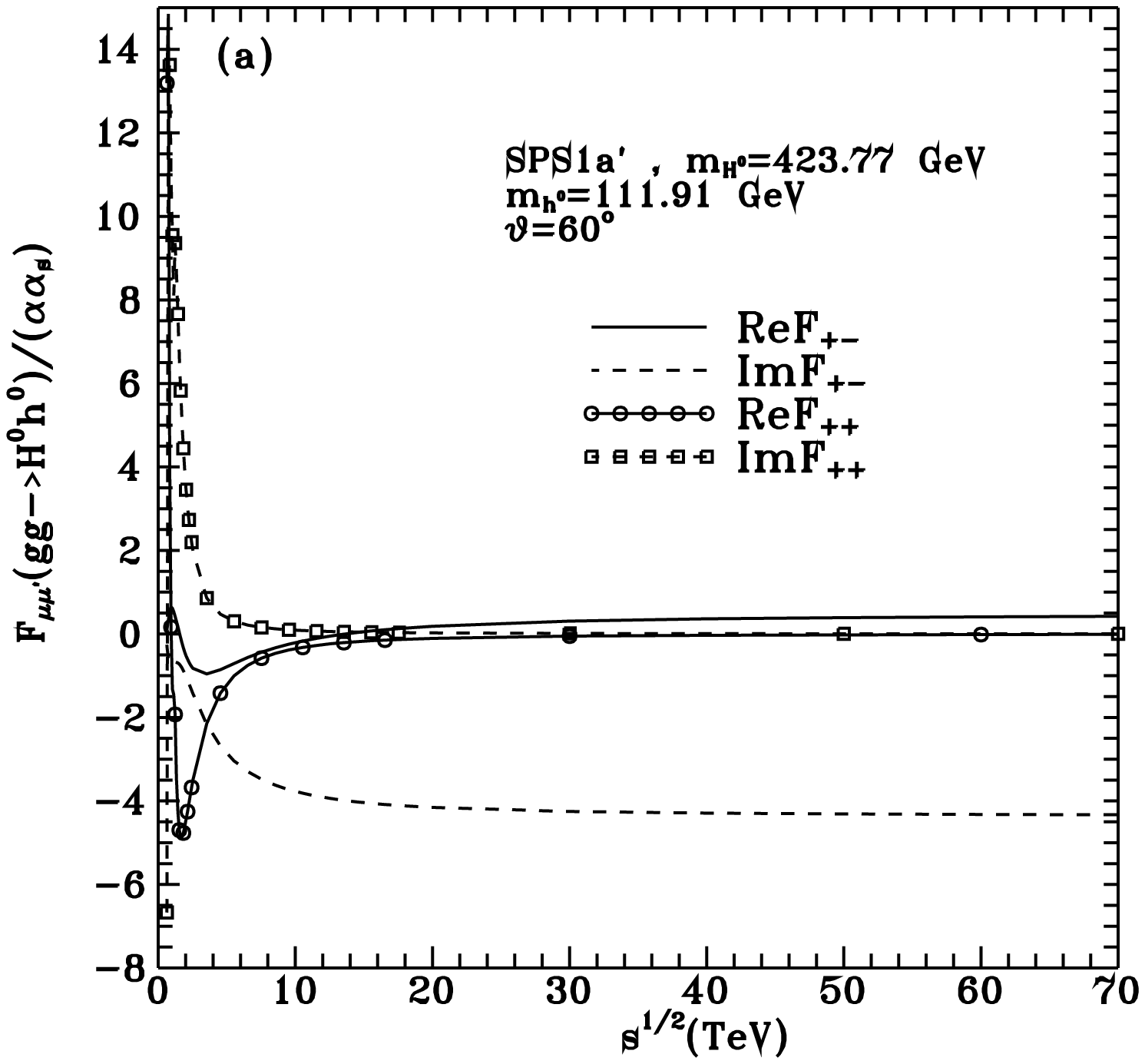, height=6.cm}\hspace{1.cm}
\epsfig{file=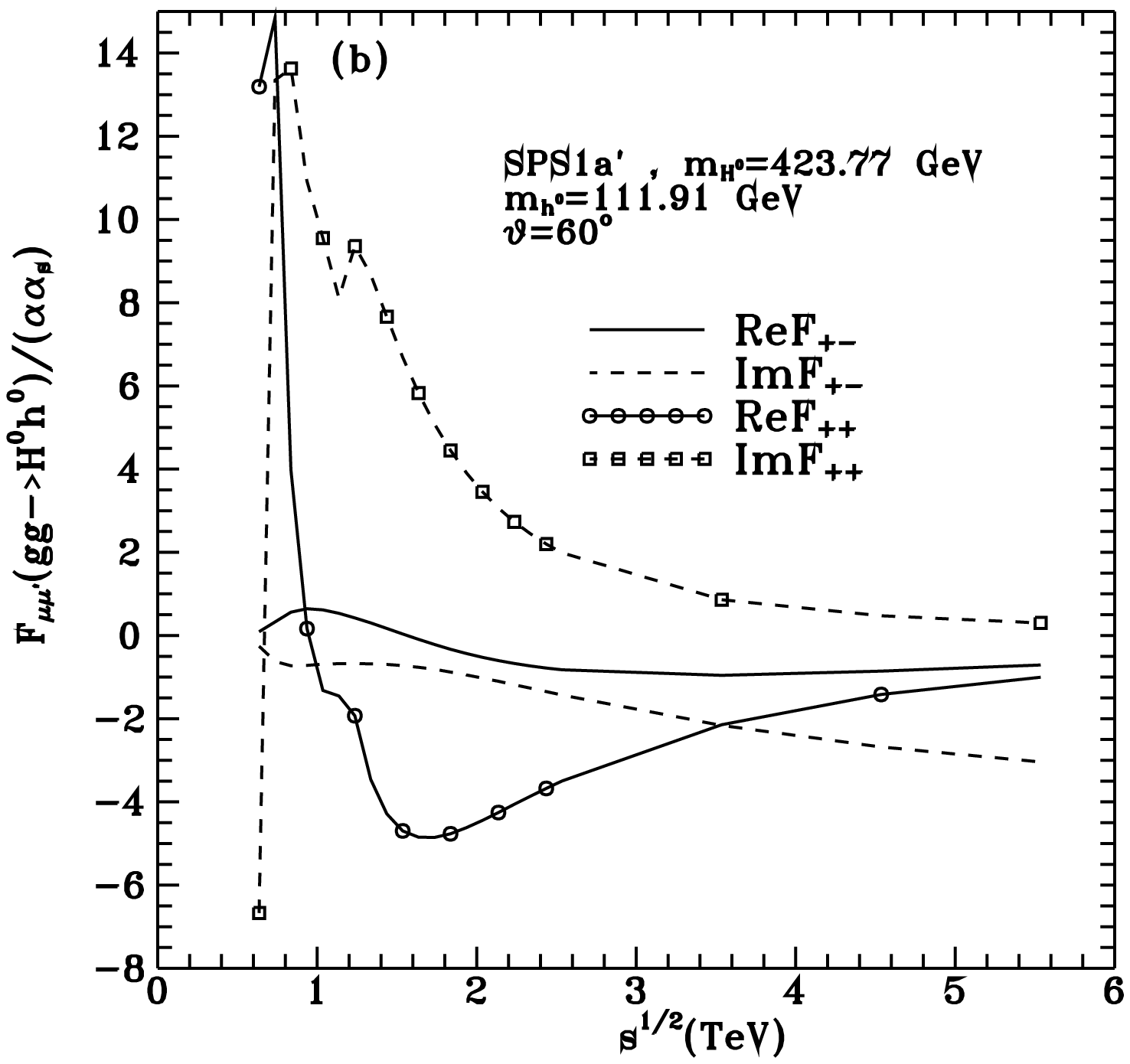,height=6.cm}
\]
\[
\epsfig{file=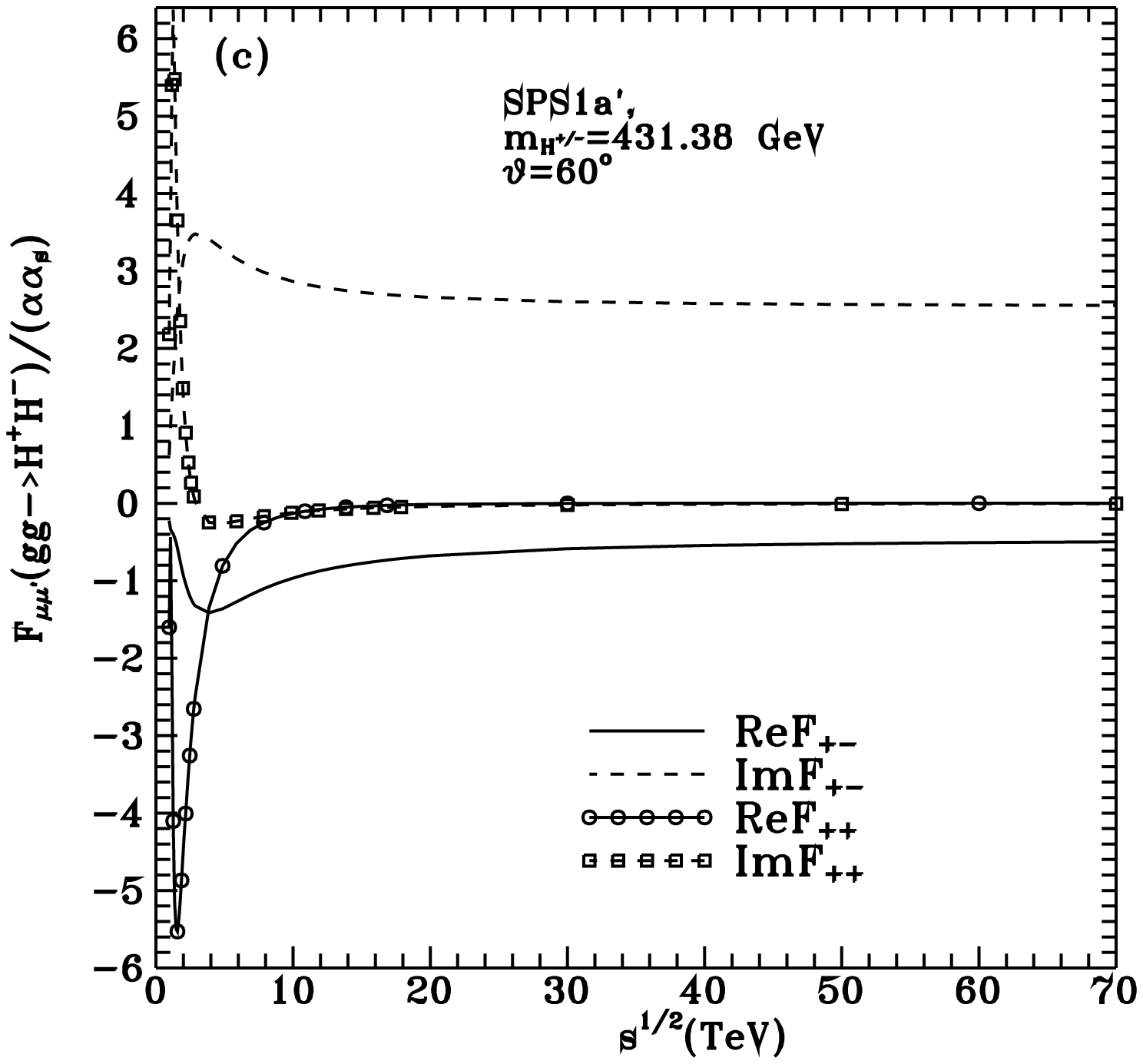, height=6.cm}\hspace{1.cm}
\epsfig{file=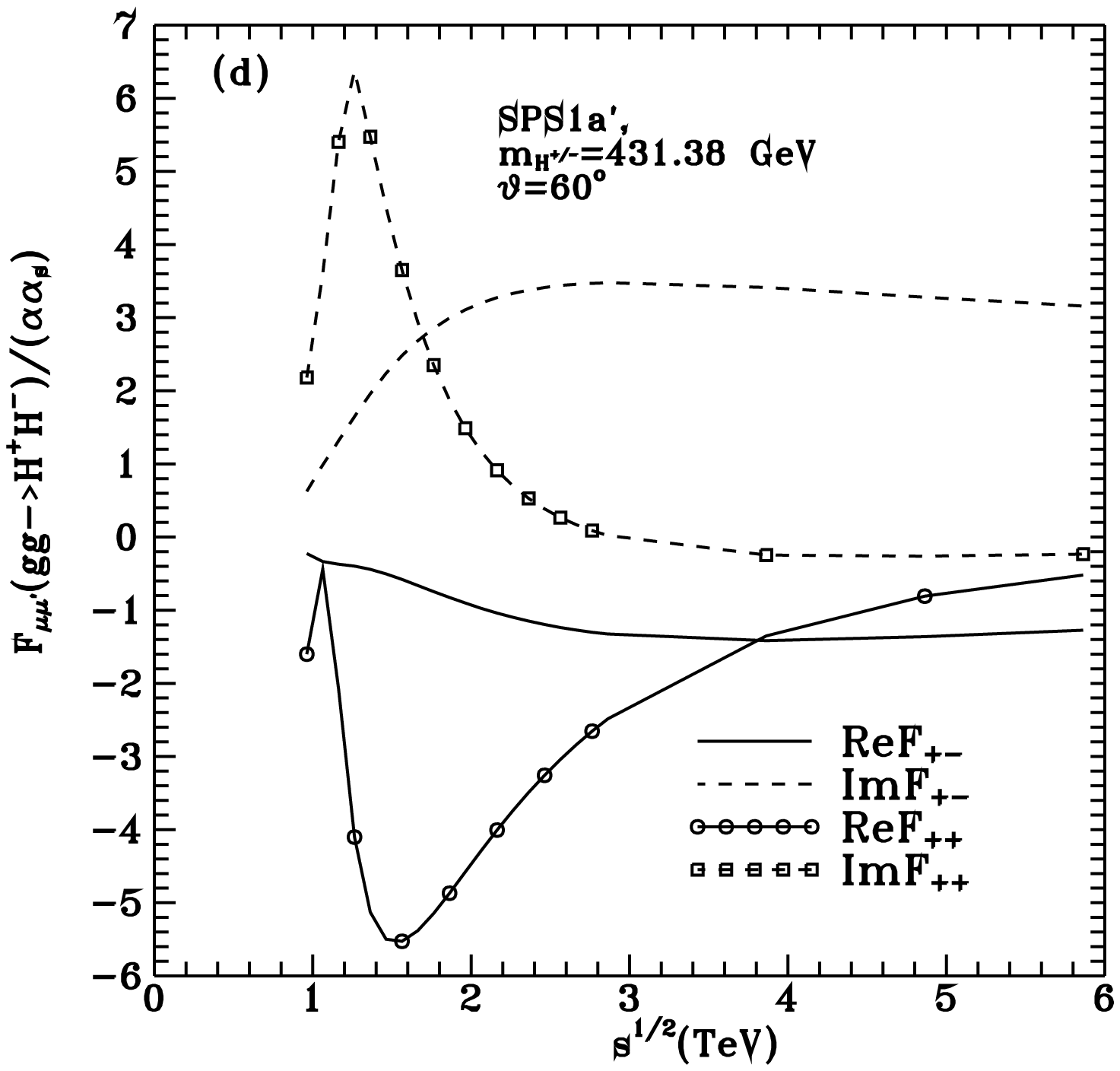,height=6.cm}
\]
\[
\epsfig{file=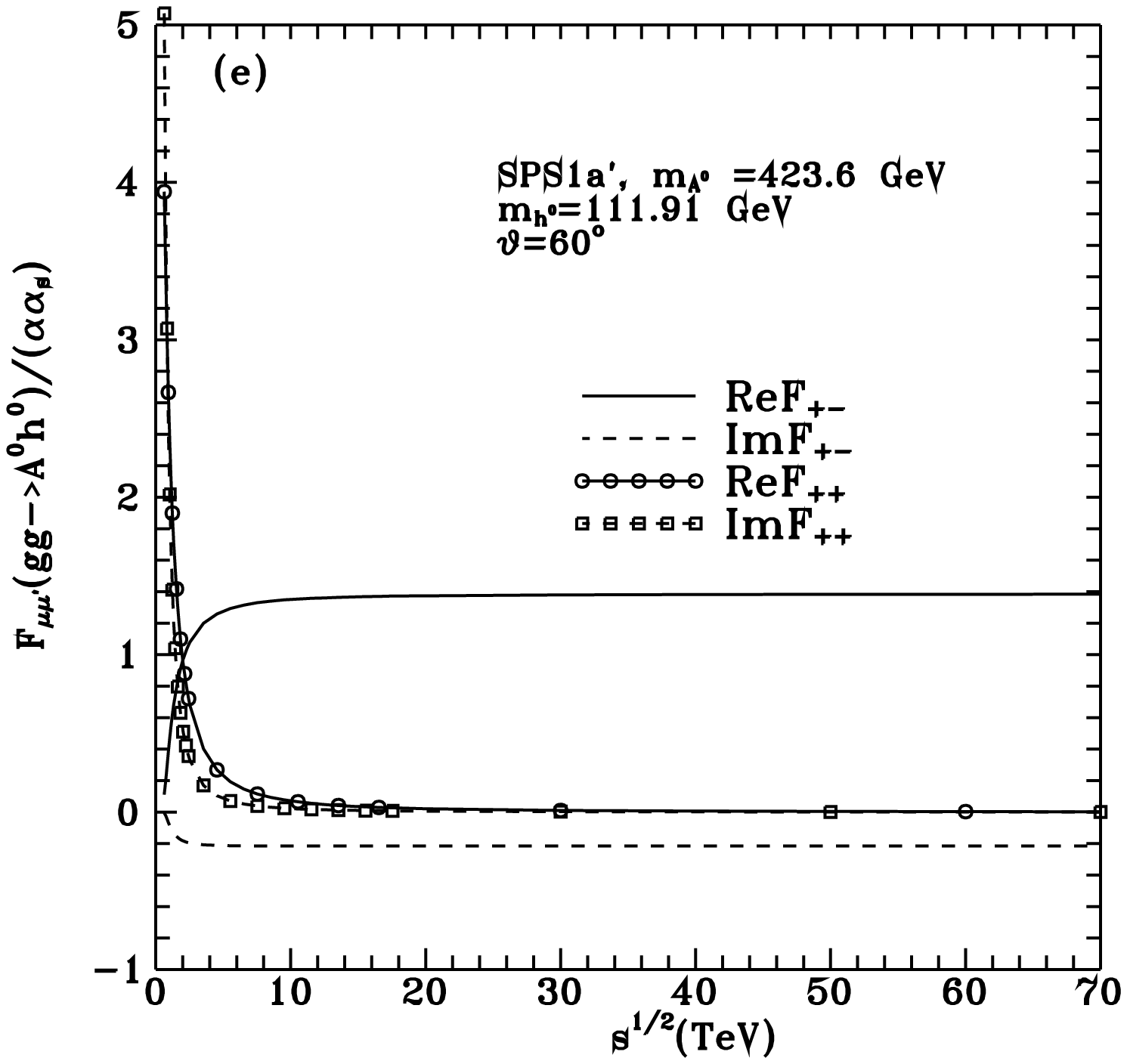, height=6.cm}\hspace{1.cm}
\epsfig{file=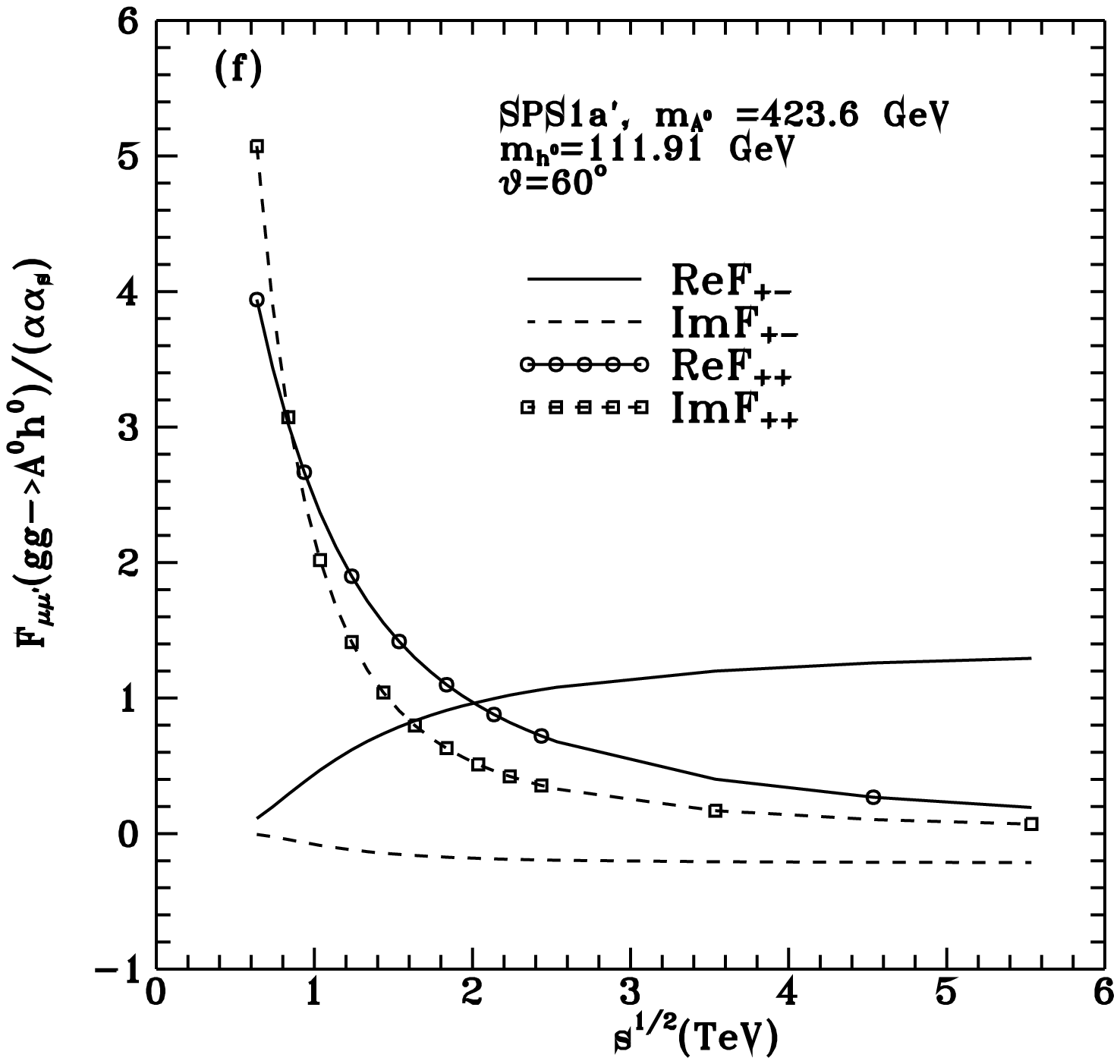,height=6.cm}
\]
\caption[1]{Amplitudes describing the high-energy and the LHC-type-energies
(see previous caption) for  $gg \to H^0 h^0 $  (a,b),   $gg \to H^+ H^- $ (c,d),
and $gg \to A^0 h^0 $  (e,f)  in $SPS1a'$. }
\label{Hbh0-HpHm-A0h0-SPA-fig}
\end{figure}

  We next focus on the MSSM  helicity amplitudes,
  always using the $SPS1a'$ benchmark \cite{SPA}.
  As it can been seen from Fig.\ref{h0h0-SPA-SM-fig}b,
   the HV amplitude for  $gg\to h^0h^0$   vanishes very quickly with energy.
   But for $gg\to H^0h^0, ~H^+H^-, ~  A^0h^0$, this   vanishing
 seems  slower, apparently due to a larger squark contribution;
 see Figs.\ref{Hbh0-HpHm-A0h0-SPA-fig}, which
 suggest that  a minimum energy of $\sim 10$ TeV is required, for HCns  to be
 approximately realized.

\begin{figure}[p]
\vspace*{-1cm}
\[
\epsfig{file=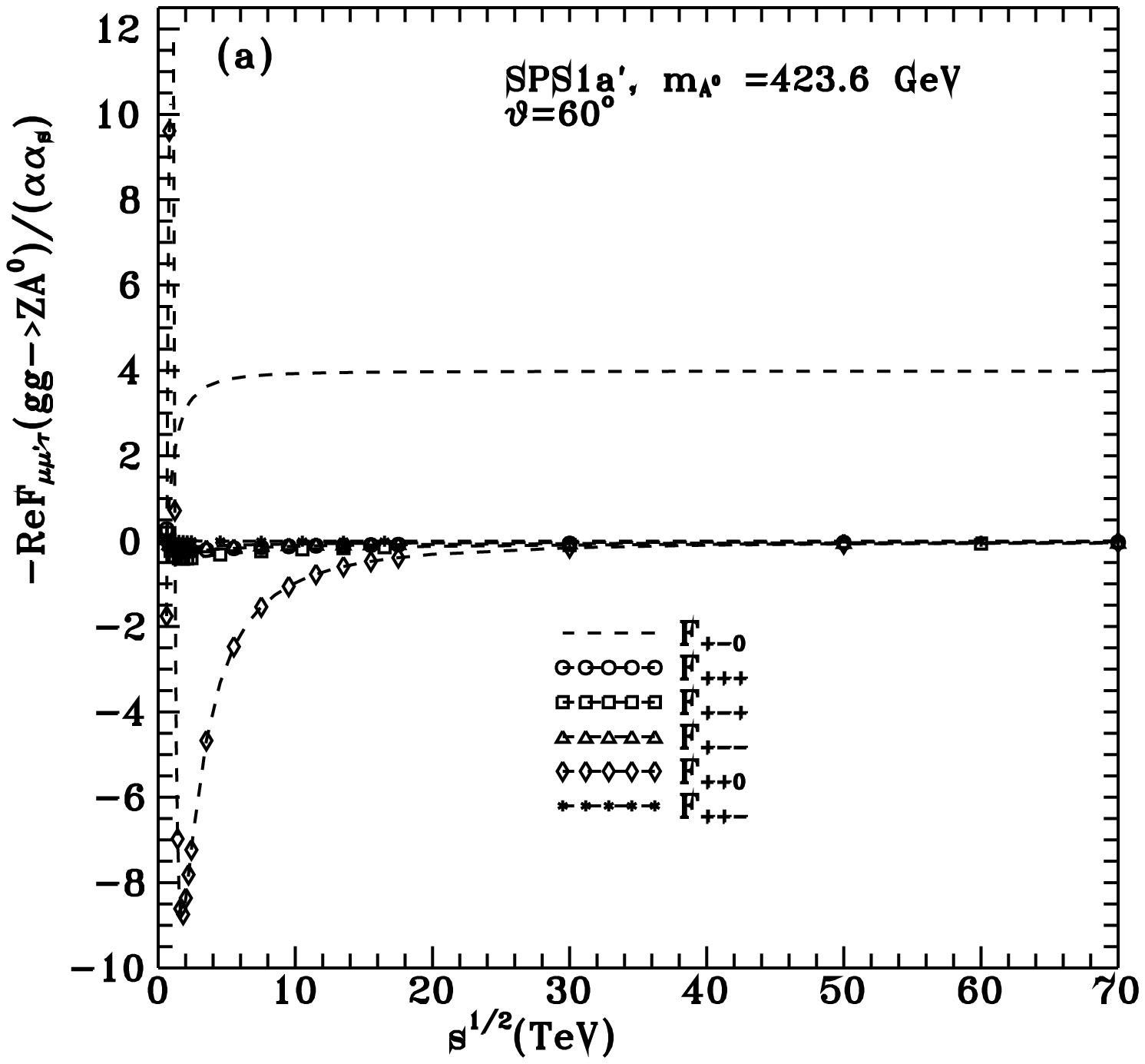, height=7.cm}\hspace{1.cm}
\epsfig{file=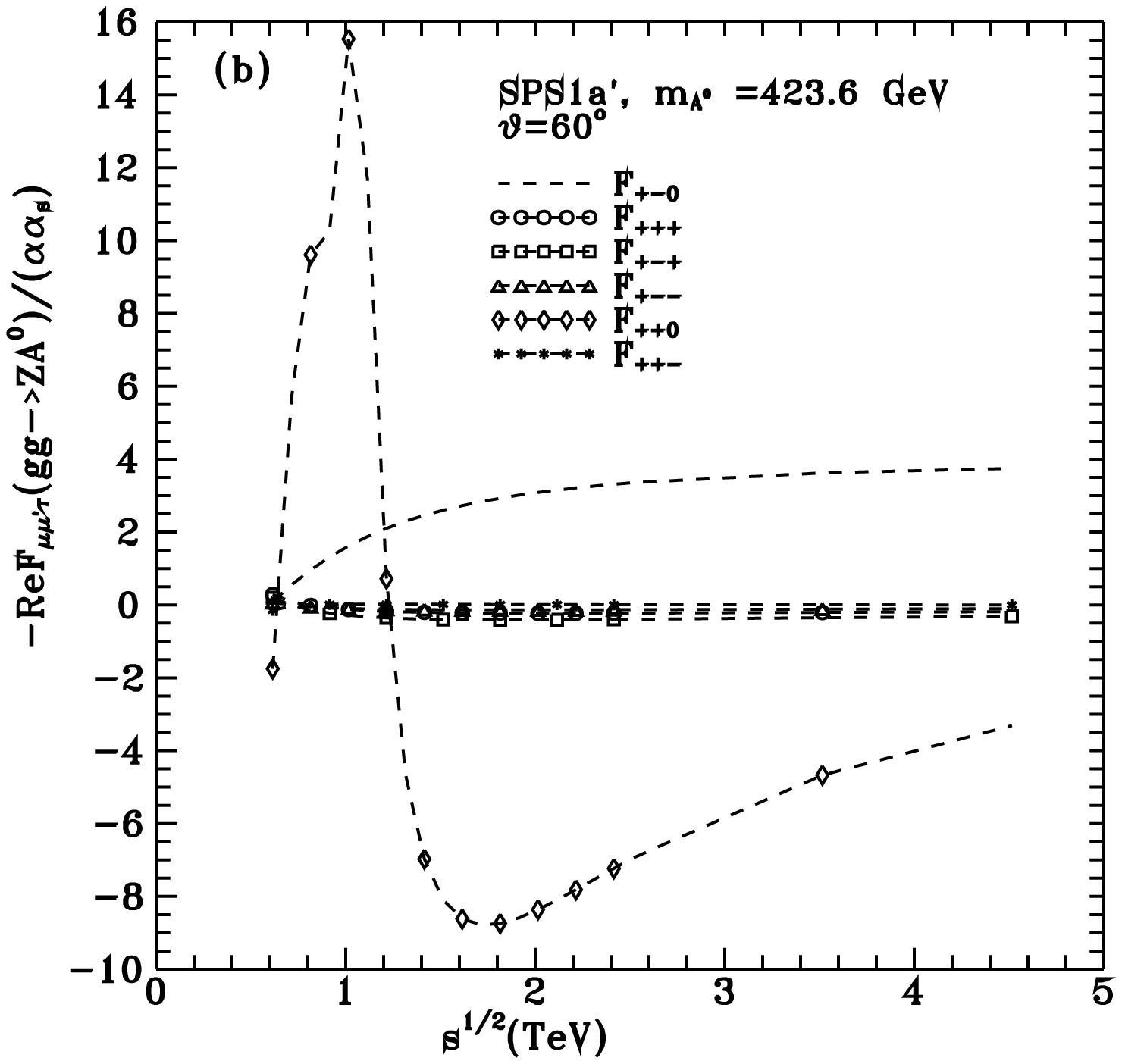,height=7.cm}
\]
\[
\epsfig{file=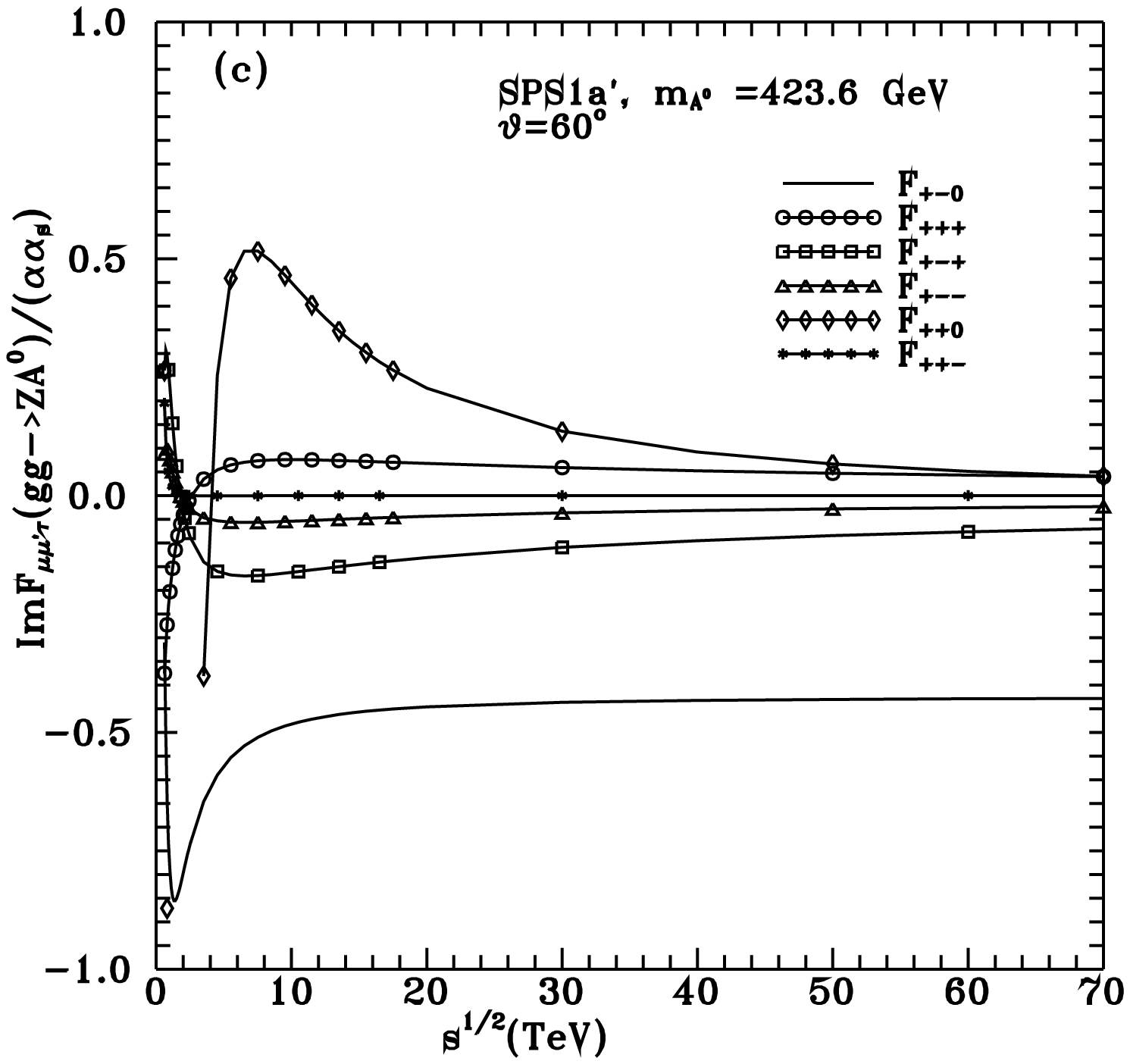, height=7.cm}\hspace{1.cm}
\epsfig{file=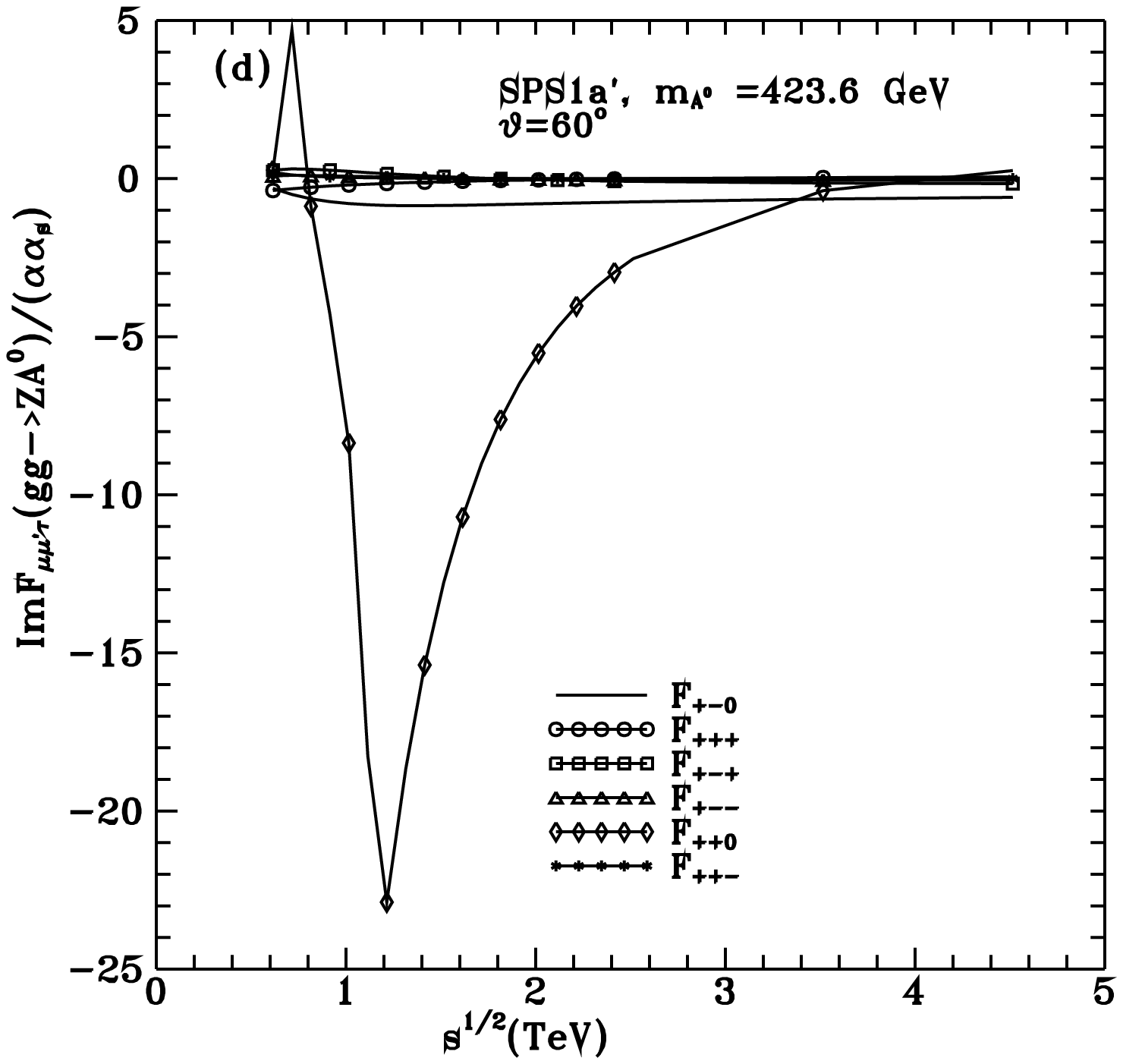,height=7.cm}
\]
\caption[1]{Amplitudes for  $gg \to Z A^0 $  in $SPS1a'$; (a,c) describe
the high energy behaviour while  (b,d)  emphasize the  LHC range. }
\label{ZA0-SPA-fig}
\end{figure}

\begin{figure}[p]
\vspace*{-1cm}
\[
\epsfig{file=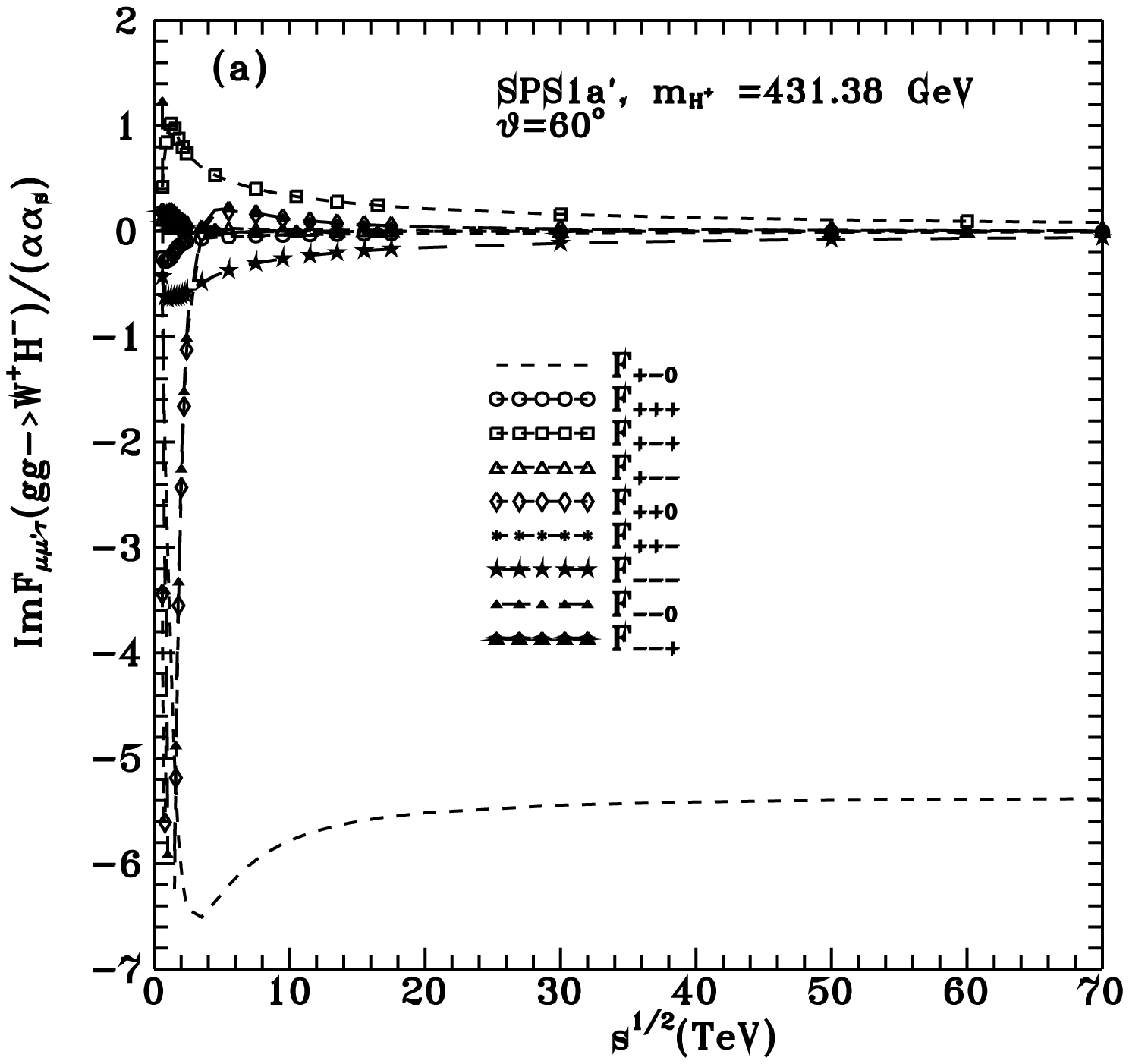, height=7.cm}\hspace{1.cm}
\epsfig{file=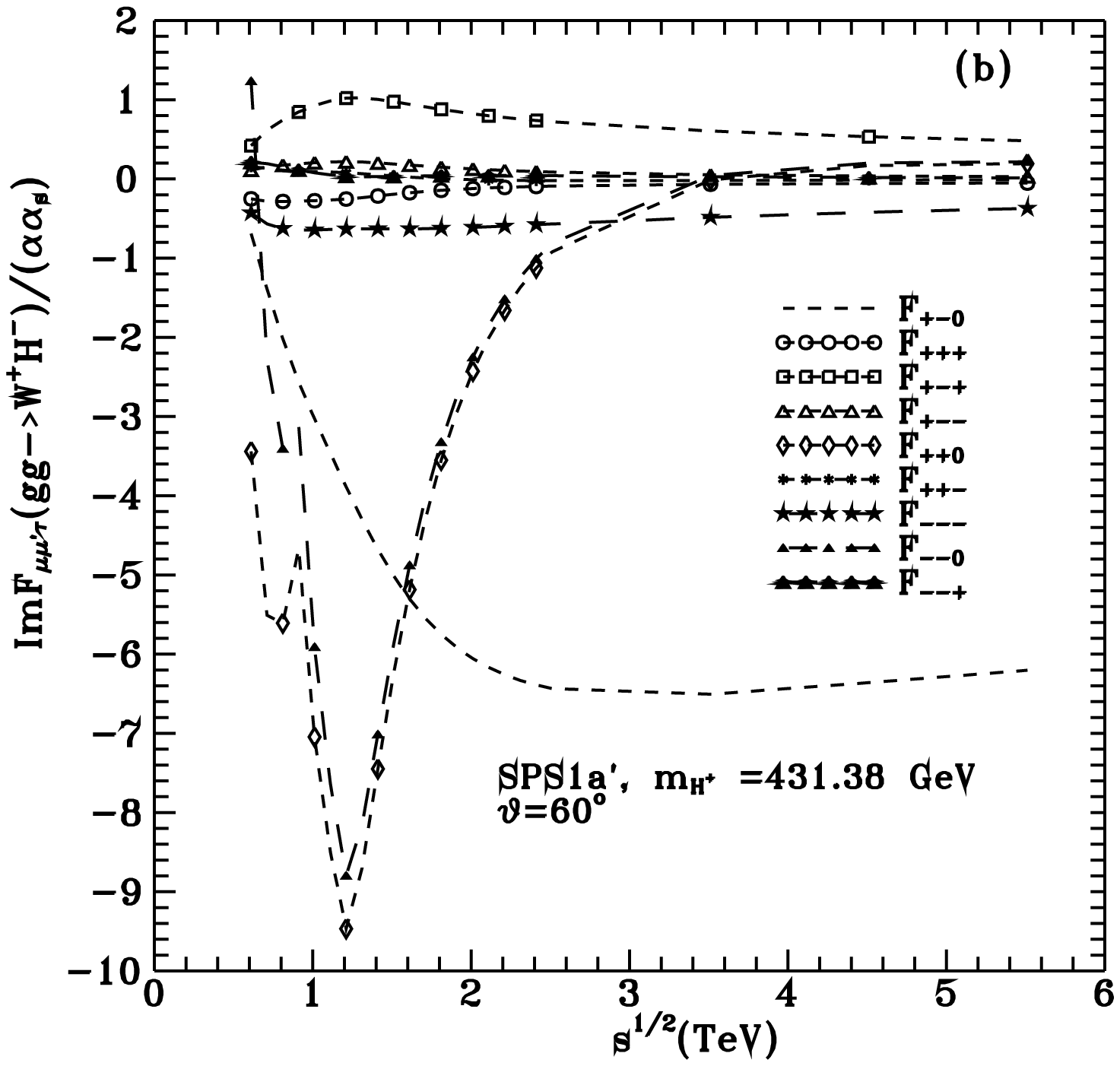,height=7.cm}
\]
\[
\epsfig{file=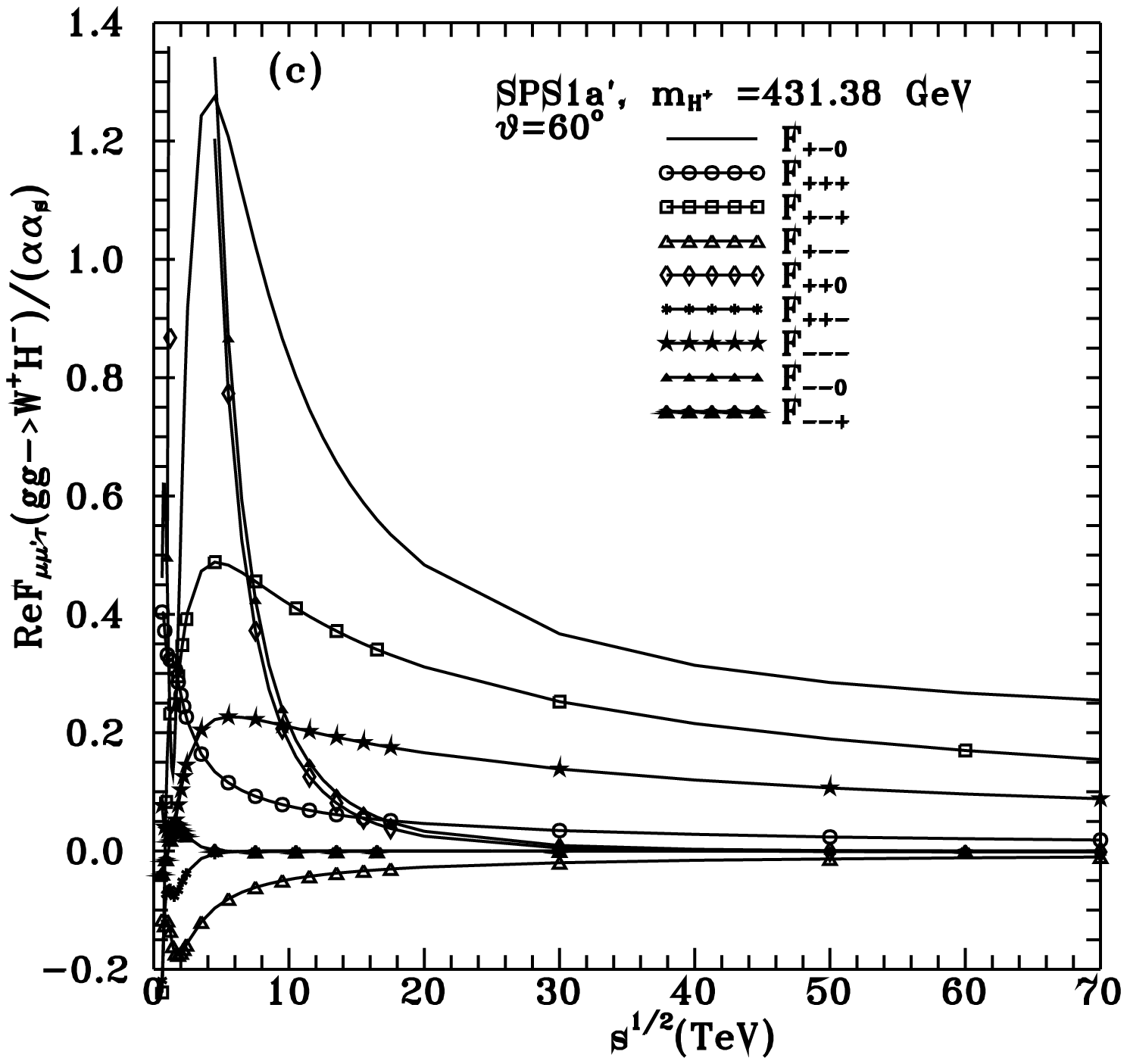, height=7.cm}\hspace{1.cm}
\epsfig{file=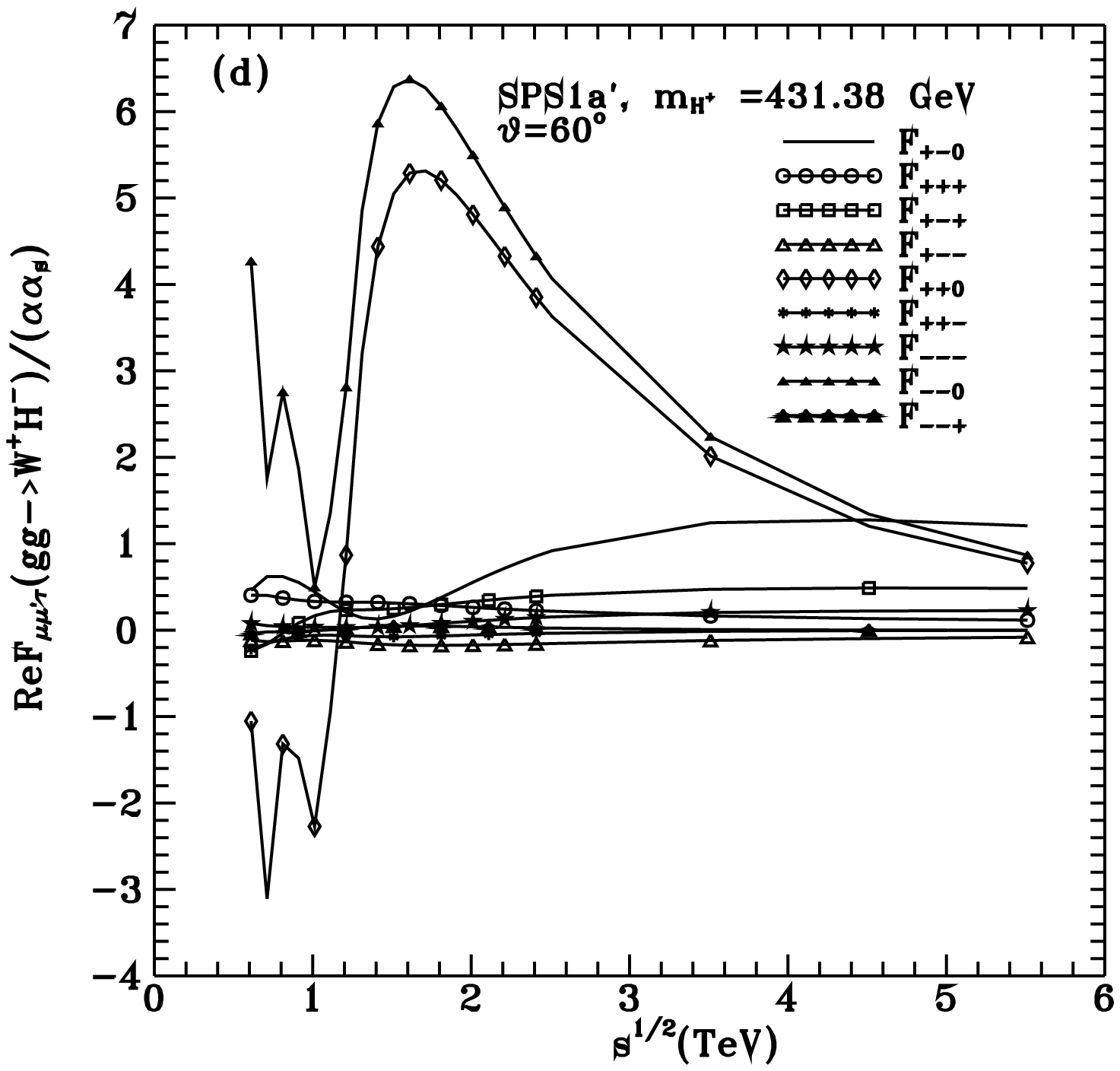,height=7.cm}
\]
\caption[1]{Amplitudes  for  $gg \to W^+H^- $  in $SPS1a'$; (a,c) describe the high
energy behaviour while  (b,d)  emphasize the  LHC range. }
\label{WpHm-SPA-fig}
\end{figure}

  In  Figs.\ref{ZA0-SPA-fig} and\ref{WpHm-SPA-fig} we
 show the amplitudes for  $gg\to Z A^0$ and $gg\to W^+H^-$.
 These results, together with those for $gg \to Zh^0$
  (see   Figs.\ref{Zh0-SPA-fig}), indicate that the high energy vanishing of the HV amplitudes
  in the $gg\to VH $ cases is generally slower, than in the $gg\to HH'$ cases.
  Particularly for $W^+H^-$, center of mass  energies of $\gsim 20 {\rm TeV}$
  are required in $SPS1a'$, for helicity   conservation to approximately  establish itself.
  Such a slow approach to the HCns regime, should be partly  due to  the slow vanishing
  of the transverse amplitudes in
  (\ref{charged-slow}, \ref{neutral-slow1}, \ref{neutral-slow2} ).\\

Finally, in Figs.\ref{R1-fig}-\ref{R6-R7-fig}, we compare
the energy- and angle-dependence of the various parts of the cross section relations $R_i$,
defined in (\ref{R1-rel}-\ref{R7-rel}). These $R_i$-parts should always become identical
at high energies; while their deviations at intermediate energies give a measure of the
violations in $R_i$.

In the  $SPS1a'$ benchmark we are using, which belongs to the so called
 decoupling MSSM regime,  the  $h^0$ self-couplings, as well as its couplings to
 the quarks,  leptons and the gauge bosons, are very close to the SM ones,
implying     $\alpha \simeq \beta -\pi/2$  \cite{Djouadi}.
Through (\ref{Class-a}-\ref{Class-c2}), this leads to
 \bqa
&& R_{a4} \simeq R_{a3}  ~,~ R_{a5} \simeq R_{a1} ~,~ R_{a6} \simeq  -R_{a2} ~,~ \nonumber \\
&& R_{b1} \simeq -R_{b4} \simeq  -R^J_{c2} ~,~ R_{b2} \simeq R^J_{c1} ~,~
R_{b3} \simeq -R^J_{c3} ~, \nonumber
\eqa
which explain   many features   in Figs.\ref{R1-fig}-\ref{R6-R7-fig}.\\

Concentrating on  $R_1$ defined in (\ref{R1-rel}), we  compare
in Figs.\ref{R1-fig} the magnitudes of its various parts;
here panels (a,b) describe  the energy dependencies  at  $(\theta=30^o, 60^o)$,
 while (c) gives the angular dependence at a c.m. energy $\sqrt{s}=8~{\rm TeV}$.
 As seen from (a,b),  five of the $R_1$ parts  reach
 their common asymptotic value already at $\sim 6 ~{\rm TeV}$, for the above  angles.
 Deviations  persist only for the  $gg\to H^0H^0, ~H^0h^0$ parts, which seem to come
 from squark boxes\footnote{For $gg \to G^0G^0, ~G^0A^0, ~A^0A^0$,
 the approach to asymptopia is faster because the squark plus antisquark box
contributions vanish identically.};
 and for the $gg\to ZG^0$ part (related through the equivalence theorem to
 $gg\to Z Z_{\rm longitudinal}$), which is due to the slowly vanishing contributions
 discussed  in (\ref{neutral-slow1},\ref{neutral-slow2}). Energies of $\gsim 20 ~{\rm TeV}$ are
 needed for all these $R_1$  deviations to fall below the  10\% level.

These deviations are also reflected in Fig.\ref{R1-fig}c, presenting the angular
distributions of the various $R_1$-parts   at 8 TeV.  \\

In Figs.\ref{R2-fig}, the corresponding results for $R_2$ defined in (\ref{R2-rel}),
are presented. At an energy of $\sim 8 ~ {\rm TeV}$ and angles in the central region,
$R_2$ is much better satisfied than $R_1$.\\

In Figs.\ref{R3-R4-R5-fig}, the left and right parts of $(R_3,~R_4,~R_5)$ defined in
(\ref{R3-rel}-\ref{R5-rel}) are compared; panels (a,b) gives the energy- and angle-dependence
for $R_3$, (c,d) correspondingly for $R_4$, and (e,f) for $R_5$.
The agreement between the left and right parts in the central region is  rather poor,
at an energy of $\sim 8 ~$TeV.  In fact for $R_4$, even the shapes of the two parts
are different at 8 TeV. As the energy increases, these relations  gradually improve;
the deviations reducing to  the 20\% level at 20 TeV, and to the permille level at 400 TeV.\\

Finally in Figs.\ref{R6-R7-fig},  we compare the left and right parts of $R_6,~R_7$
defined in (\ref{R6-rel}, \ref{R7-rel}); again (a,b) give the energy- and angle-dependencies
for $R_6$, and (c,d) the corresponding results for  $R_7$.
At an energy scale of  $\gsim 12~ {\rm TeV}$, these relations are satisfied
for angles in the central region.
It appears, that the squark boxes and the HV amplitudes
for the $WG,~ WH $ channels,  are  responsible for most of the deviations
at  $\lsim 12~ {\rm TeV}$.\\

At least, as far as the contribution from the $gg\to VH $ processes is concerned,
we could, in principle extend the validity of the  relations $R_1,~R_2,~R_6,~R_7$
to lower energies, by
subtracting the contributions from the slowly vanishing transverse amplitudes discussed
in Sections  3.2. This, will of course make these relations considerably more complicated.\\

 Of course, the appearance of helicity conservation, as well as the validity of the
 asymptotic $R_i$  relations discussed above,   will be further delayed, if the SUSY masses
 are higher than those at $SPS1a'$.

 In any case, using the codes  \cite{code} which give the exact 1loop EW predictions
 for the above amplitudes, the exact values of all separate parts of the
 $R_i$ relations in (\ref{R1-rel}-\ref{R7-rel}) may be calculated,
 and its validity checked, for any MSSM model at any energy.\\

\section{Summary and outlook}

The helicity conservation theorem, proved to all orders,  for
any 2-to-2 process at asymptotic energies and fixed angles, is
a really impressive property of  any supersymmetric extension of SM \cite{heli}.

It not only greatly simplifies the structure of the asymptotic
amplitudes, but it may also have important implications at realistic LHC energies,
provided the SUSY scale is not too high.
This has been realized for a considerable range of MSSM benchmarks,
by studying the complete 1loop EW contributions to  $ug\to  dW$  \cite{ugdW};
as well as by constructing  asymptotic  cross section relations between
$ug\to dW$ and $ug\to \tilde d_L \tchi^+_j$, which were seen to remain approximate correct
even close to the  LHC range \cite{ugdW-ugsdWino}.

In all examples studied previously, which were all done at the 1loop EW order in MSSM,
the dominant helicity conserving  amplitudes
were always increasing logarithmically, whilst the helicity violating ones were tending to zero.

Comparing to  corresponding 1loop  SM results for processes
with standard external particles, it appeared that HCns is
approximately correct in SM also; in the sense that the
helicity conserving amplitudes
 were again found to increase  logarithmically with energy, while the HV  ones were
 going to much smaller  "constants".
 In fact, in such  cases,   the dominance of the logarithmically increasing  HC  amplitudes
 is  often so overwhelming,  that it should be   impossible to experimentally  discriminate
 a "constant" asymptotic value of an HV SM amplitude, from a strictly vanishing one;
 see e.g. the example of  $\gamma \gamma \to \gamma \gamma, ~Z\gamma, ~ZZ$ in \cite{gamgamV10V20}. \\

In the present work we looked at processes where the dominant HC amplitudes cannot be very large,
so that to obtain  a more stringent view of the way HCns is realized.
Thus, we looked  at the 1loop EW predictions for  processes
where there are no gauge (or gaugino)  contributions within 1loop; and thus, no
 large logarithmic enhancements.
In this spirit,  we have studied  the 13 processes $gg\to HH'$
and the 6 processes $gg\to VH $, within any CP conserving MSSM framework;
see (\ref{ggHH-proc2}, \ref{ggVH-proc2}).

Correspondingly in SM, we have calculated the 1loop EW predictions for the
 4 $gg\to HH'$ processes,
and the 3 $gg\to VH $ processes; see again (\ref{ggHH-proc2}, \ref{ggVH-proc2}).
And  for the first time, we indeed saw
examples where  helicity conservation  is \underline{strongly} violated in SM.

In MSSM, of course, HCns is always  obeyed.
These detail examples  confirm that helicity conservation indeed is a genuine SUSY property;
in accordance with the general, rather formal, all order proof   in   \cite{heli}.

\begin{figure}[t]
\vspace*{-1cm}
\[
\epsfig{file=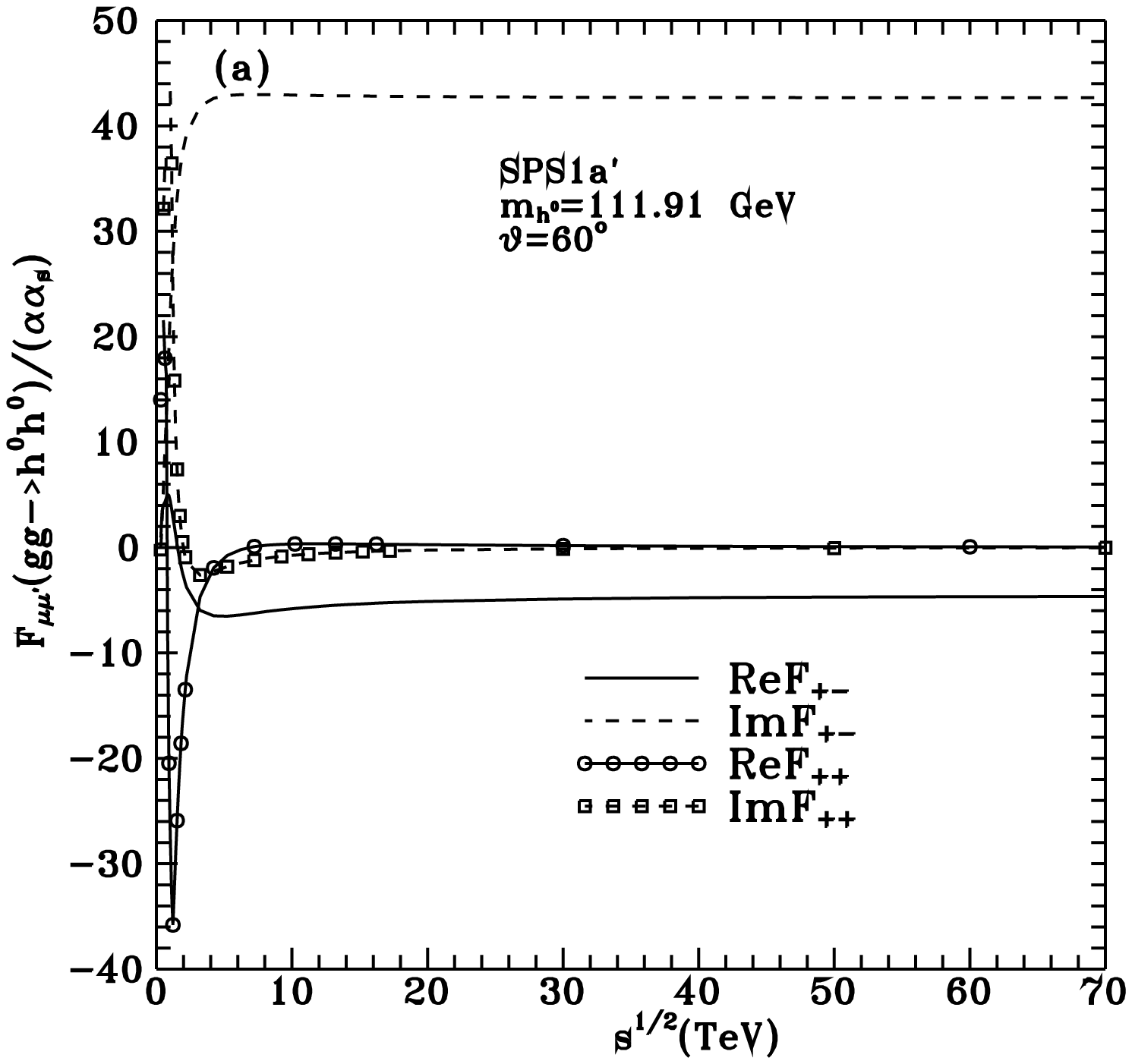, height=7.cm}\hspace{1.cm}
\epsfig{file=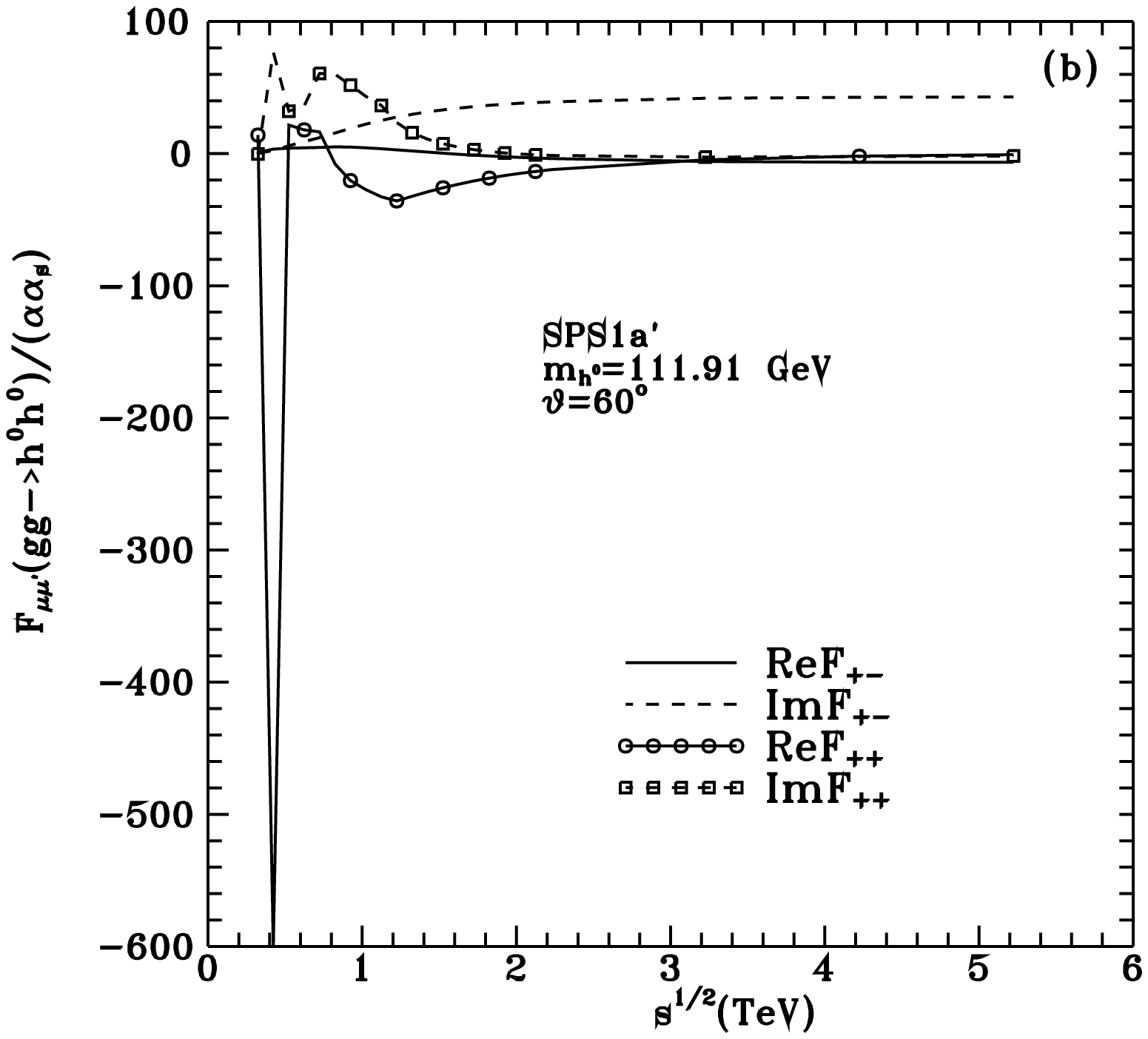,height=7.cm}
\]
\[
\epsfig{file=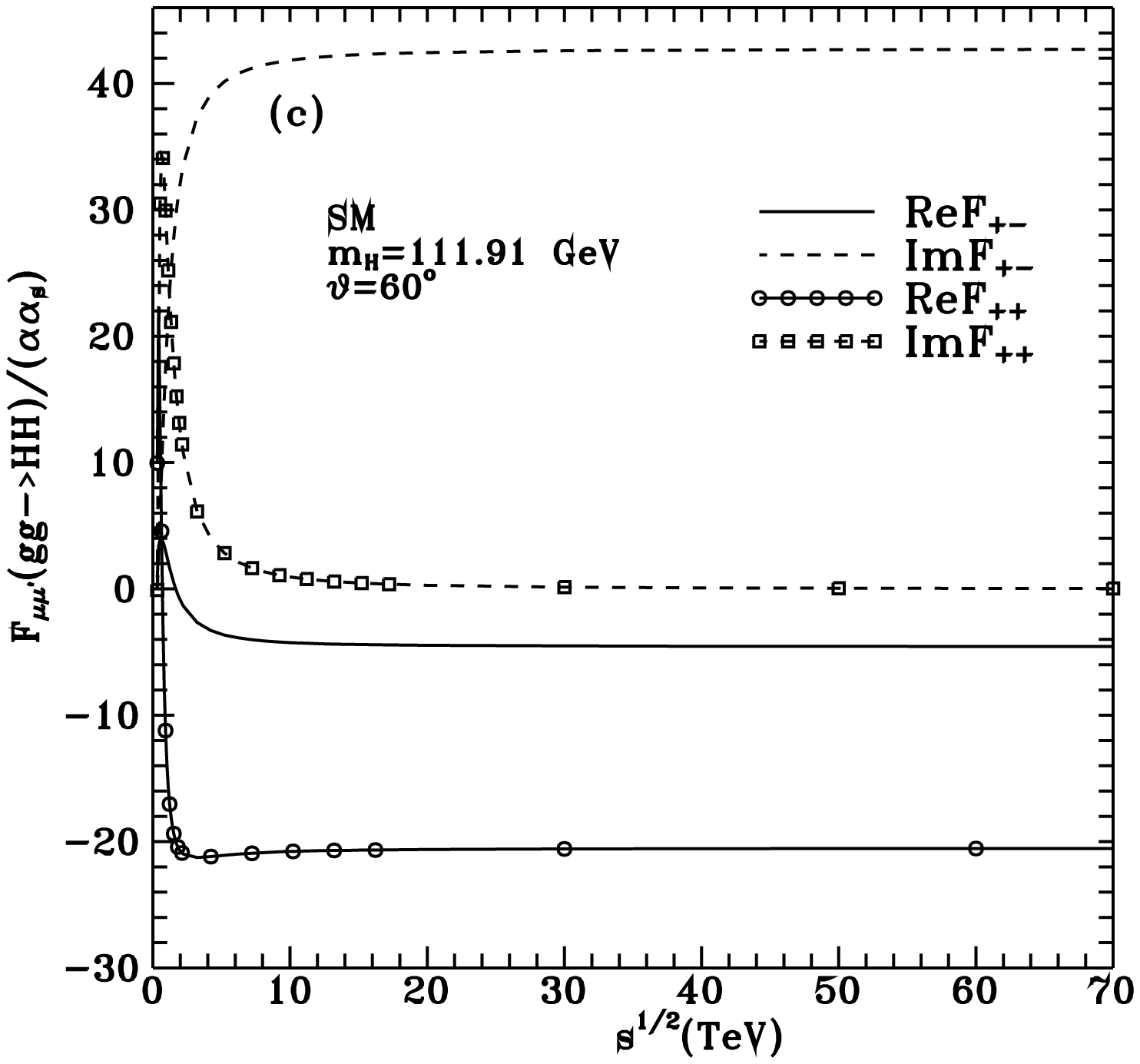, height=7.cm}\hspace{1.cm}
\epsfig{file=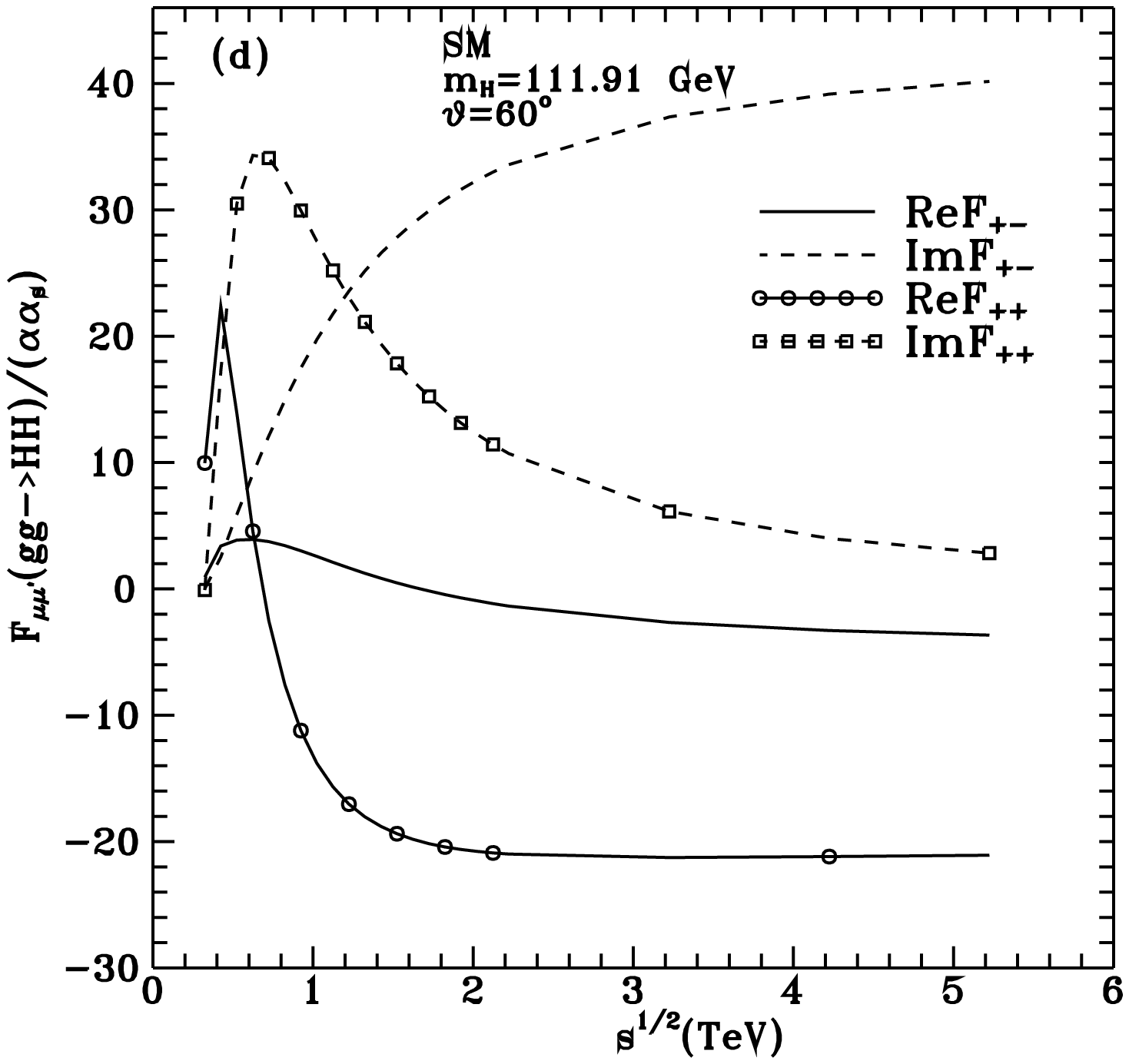,height=7.cm}
\]
\caption[1]{Amplitudes  for  $gg \to h^0 h^0 $  in $SPS1a'$  (a,b) \cite{SPA},
and  for   $gg \to HH $ in SM (c,d). }
\label{h0h0-SPA-SM-fig}
\end{figure}

The most striking example of difference between SM and MSSM is in
the process $gg\to h^0h^0$, as one can see in Fig.\ref{h0h0-SPA-SM-fig}.
In SM, the HV  amplitude for this process tends to a constant value, which is about half
the value of the helicity conserving one. Contrary to it,  in MSSM an opposite contribution
from the  squark loop arises, which exactly cancels the HV amplitude at energies
much larger than the squark masses. The unpolarized cross sections in the two cases
should then differ  by about 20\% at sufficiently high energies, which  could
be observable, particularly if squark candidates are also observed in the TeV range.\\

FORTRAN codes calculating the helicity amplitudes of all processes in
(\ref{gluon-HH-VH}, \ref{ggHH-proc2}, \ref{ggVH-proc2}), as functions of the center of mass
energy and angle, are released in  \cite{code}.
The input parameters in these codes are always  at the electroweak
scale, while the quark masses of the first two generations are neglected. \\

In this work, we have also derived the $R_1-R_7$ asymptotic relations among various cross
sections, within  the  MSSM framework.
Strictly speaking these relations should
be exact (to the 1loop EW order of course)  at asymptotic energies and fixed angles.
The only MSSM parameters they depend on,   are the  $\alpha$ and $\beta$  MSSM angles.
Testing  such  relations (at sufficiently high energies),
 would  constitute a  genuine check of the  MSSM structure.\\

\begin{figure}[p]
\vspace*{-1cm}
\[
\epsfig{file=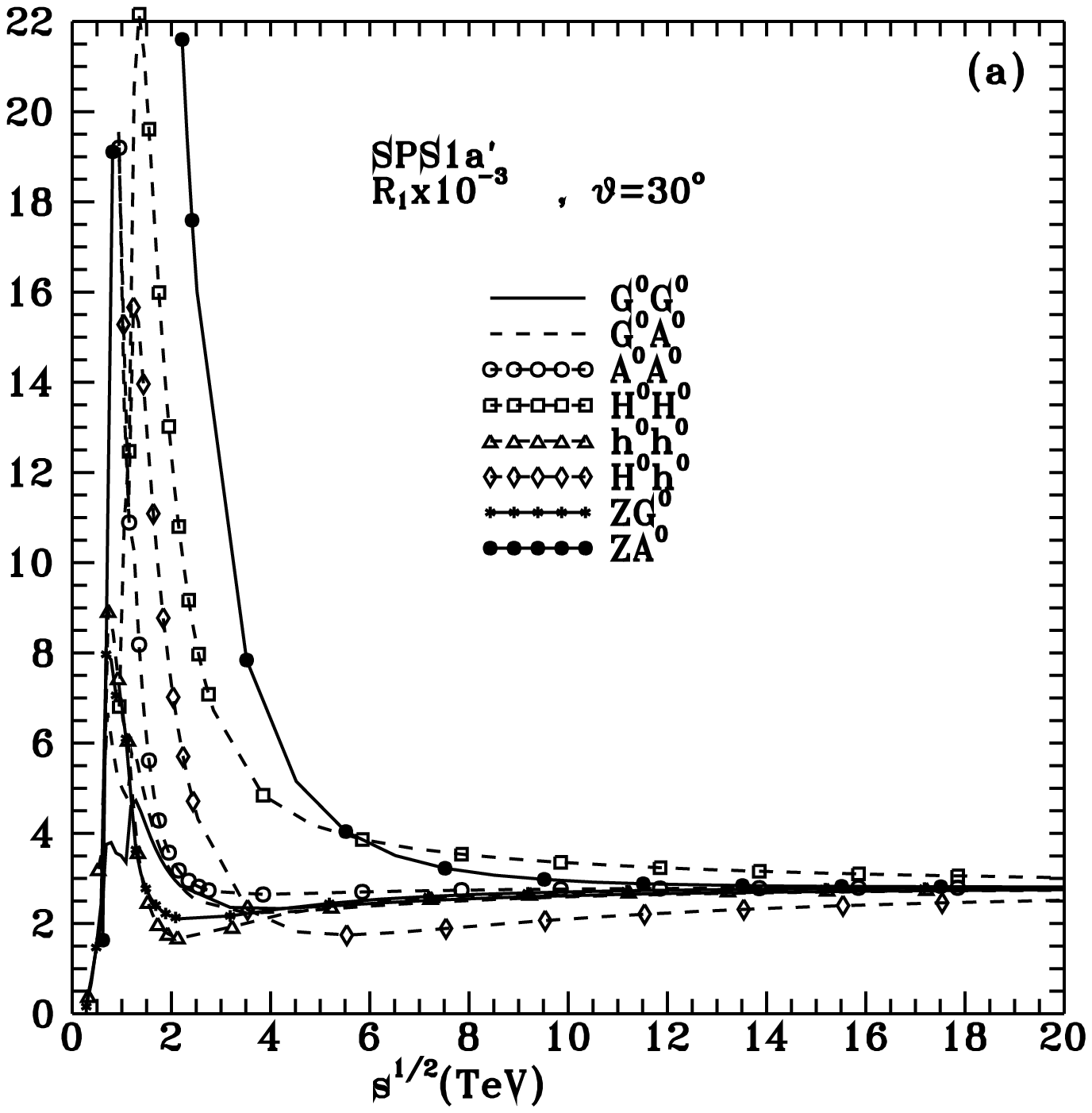, height=7.cm}\hspace{1.cm}
\epsfig{file=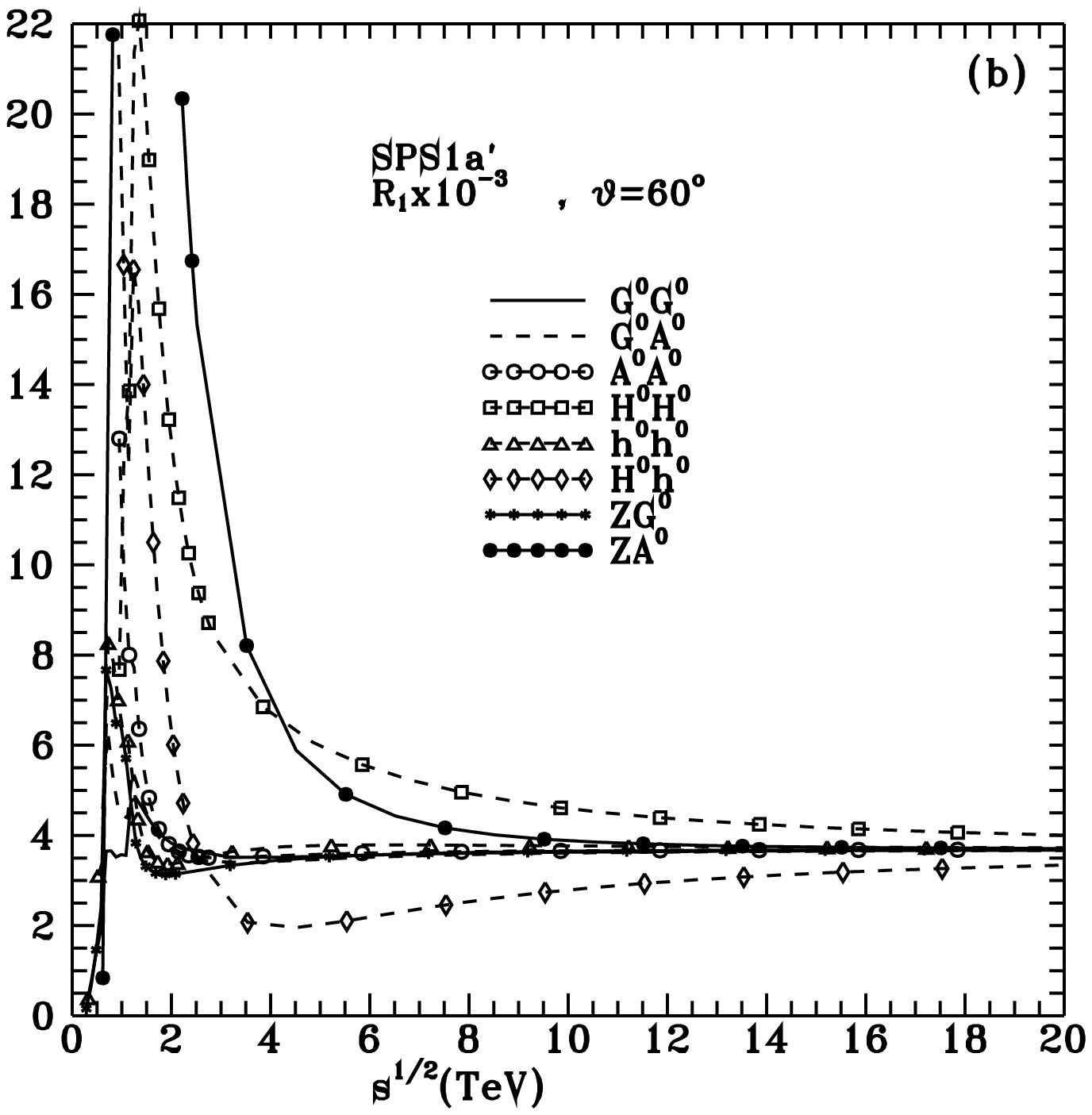,height=7.cm}
\]
\[
\epsfig{file=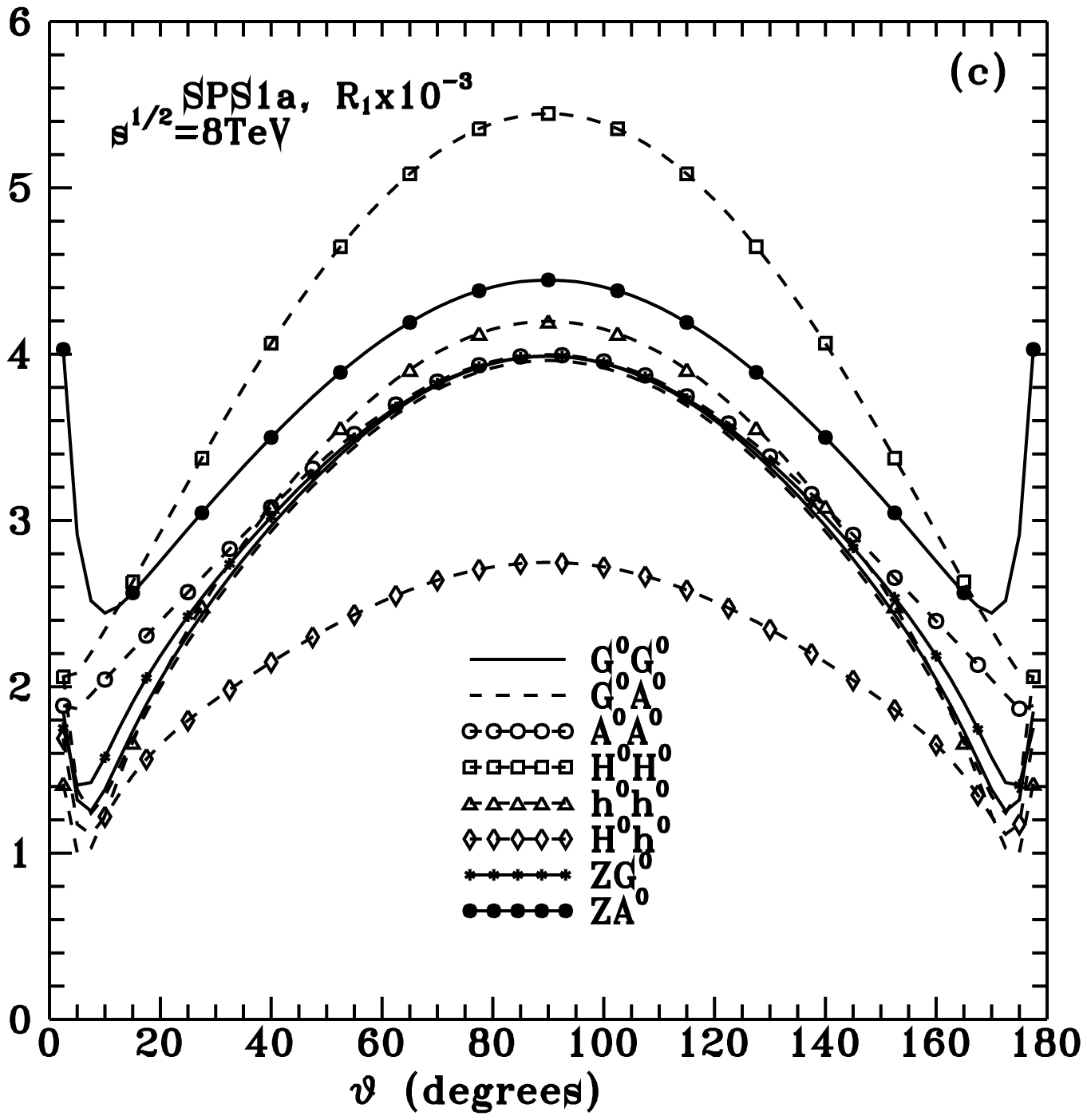, height=7.cm}
\]
\caption[1]{Magnitudes of the various  parts of the asymptotic relation $R_1$
defined in (\ref{R1-rel}); (a,b) describe  the energy dependence at  $\theta=30^o$
and $\theta=60^o$
respectively;  while (c) gives the angular dependence at $\sqrt{s}=8~{\rm TeV}$.}
\label{R1-fig}
\end{figure}

\begin{figure}[p]
\vspace*{-1cm}
\[
\epsfig{file=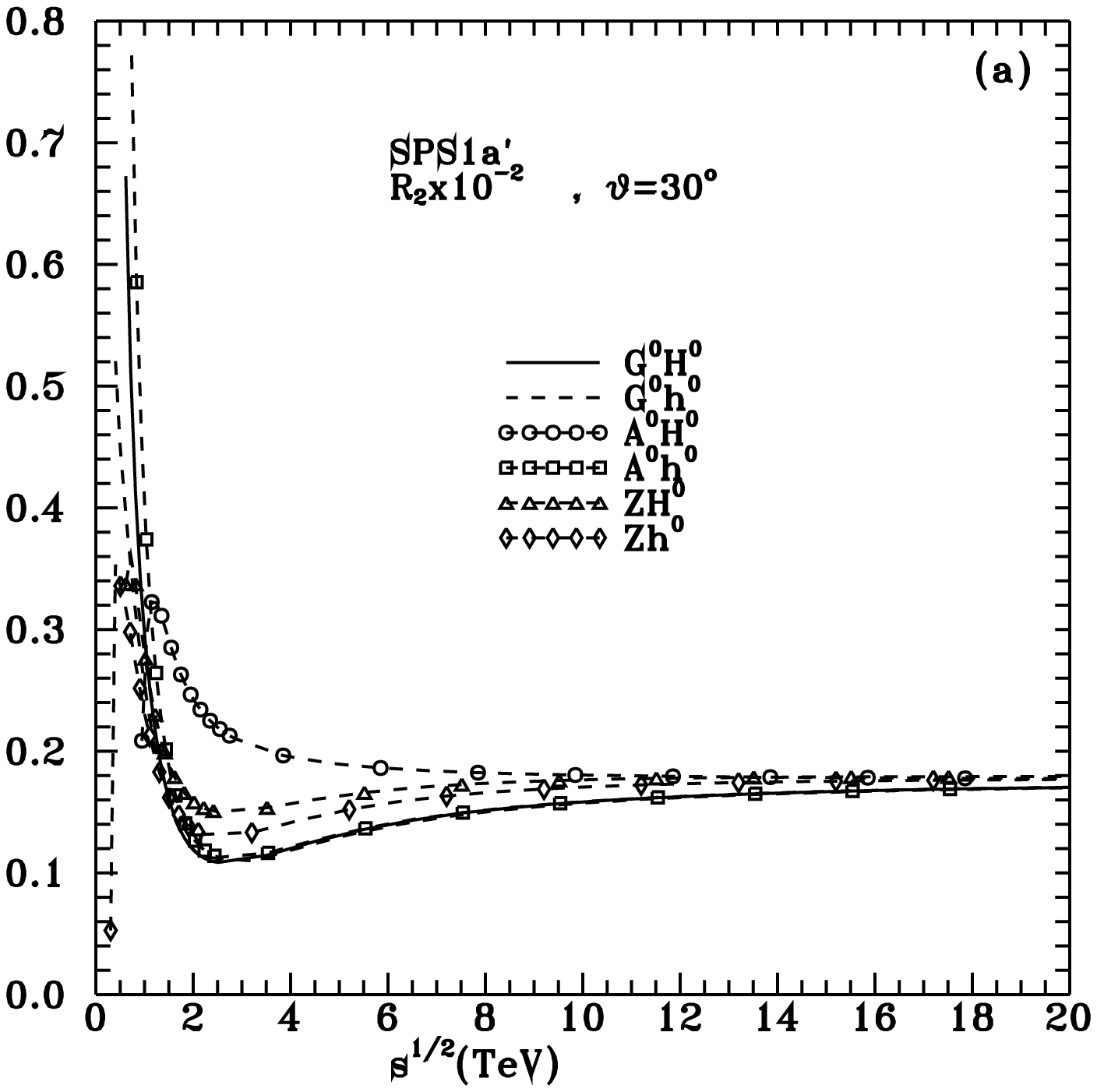, height=7.cm}\hspace{1.cm}
\epsfig{file=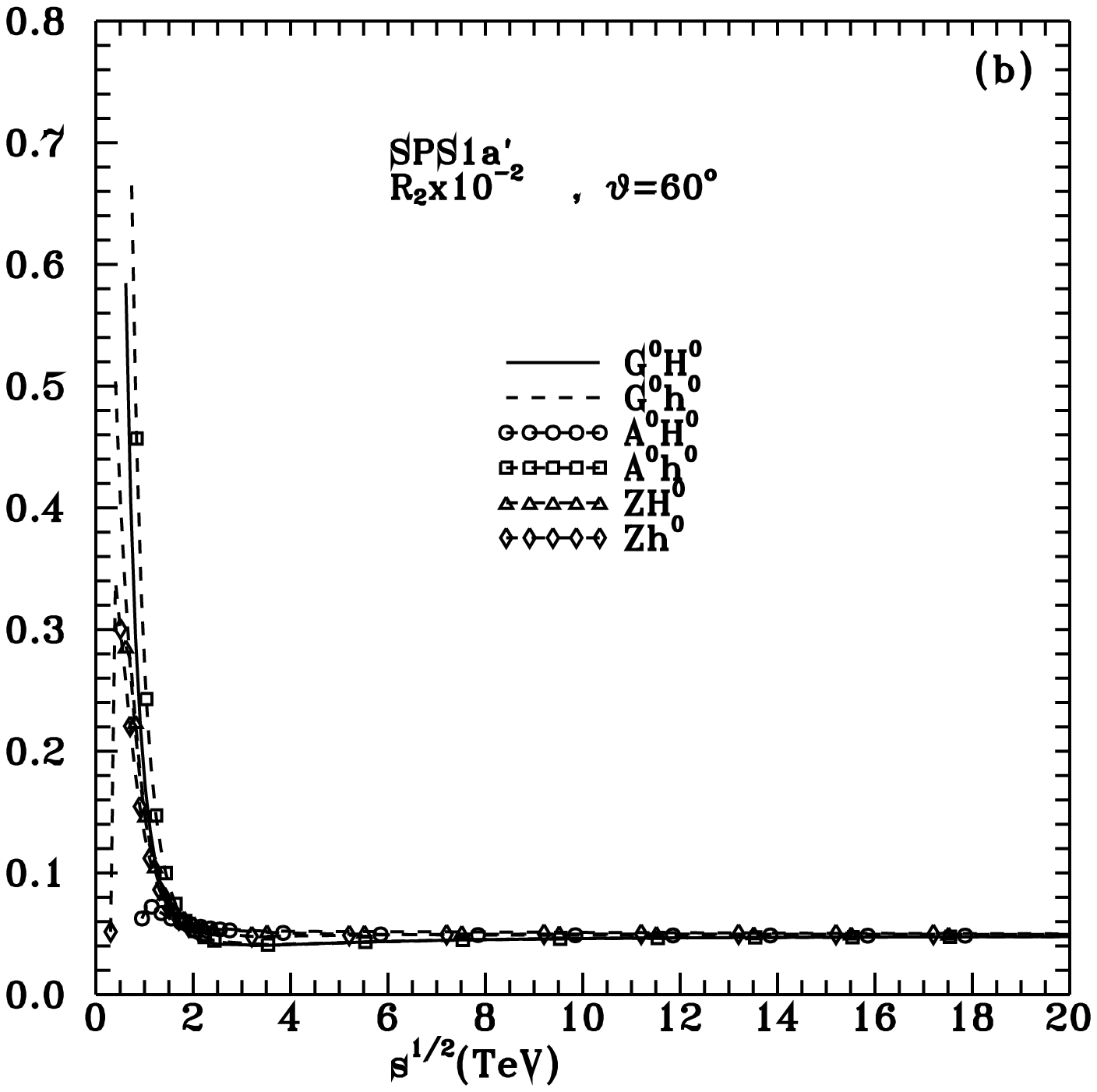,height=7.cm}
\]
\[
\epsfig{file=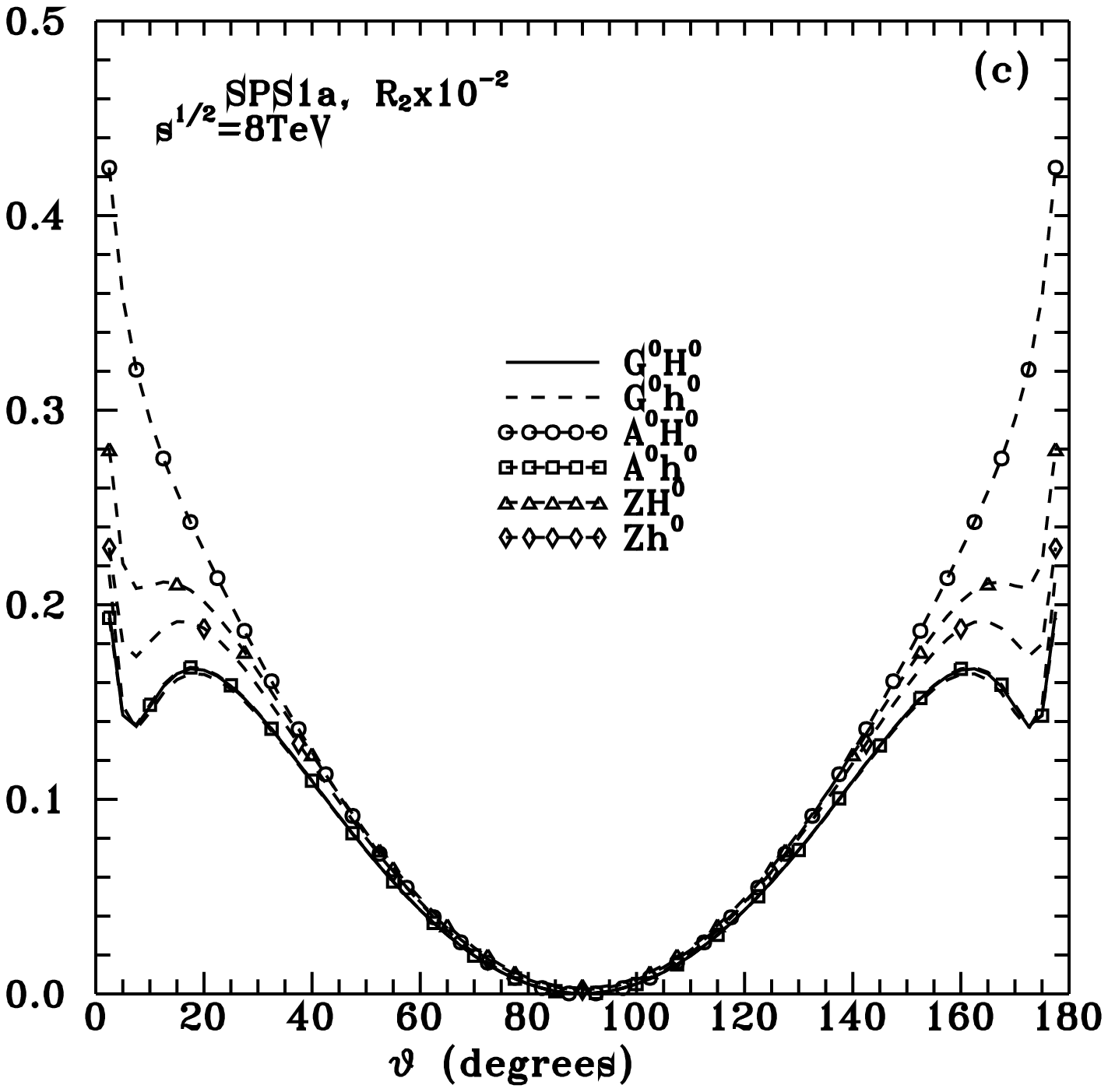, height=7.cm}
\]
\caption[1]{Magnitudes of the various  parts of the asymptotic relation
$R_2$ defined in (\ref{R2-rel}); (a,b) describe  the energy dependence at  $\theta=30^o$
and $\theta=60^o$
respectively;  while (c) gives the angular dependence at $\sqrt{s}=8~{\rm TeV}$.}
\label{R2-fig}
\end{figure}

\begin{figure}[p]
\vspace*{-1cm}
\[
\epsfig{file=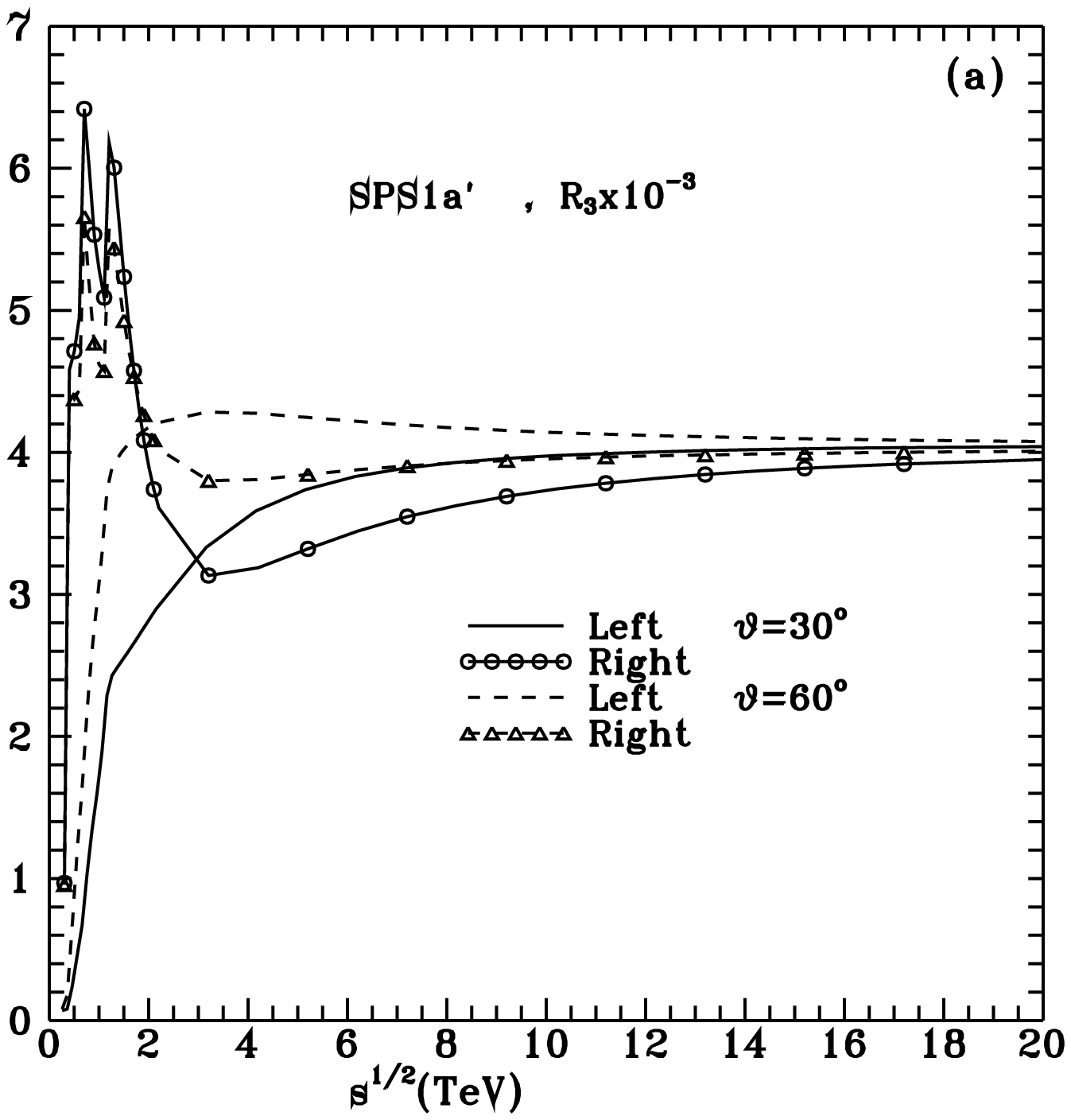, height=6.5cm}\hspace{1.cm}
\epsfig{file=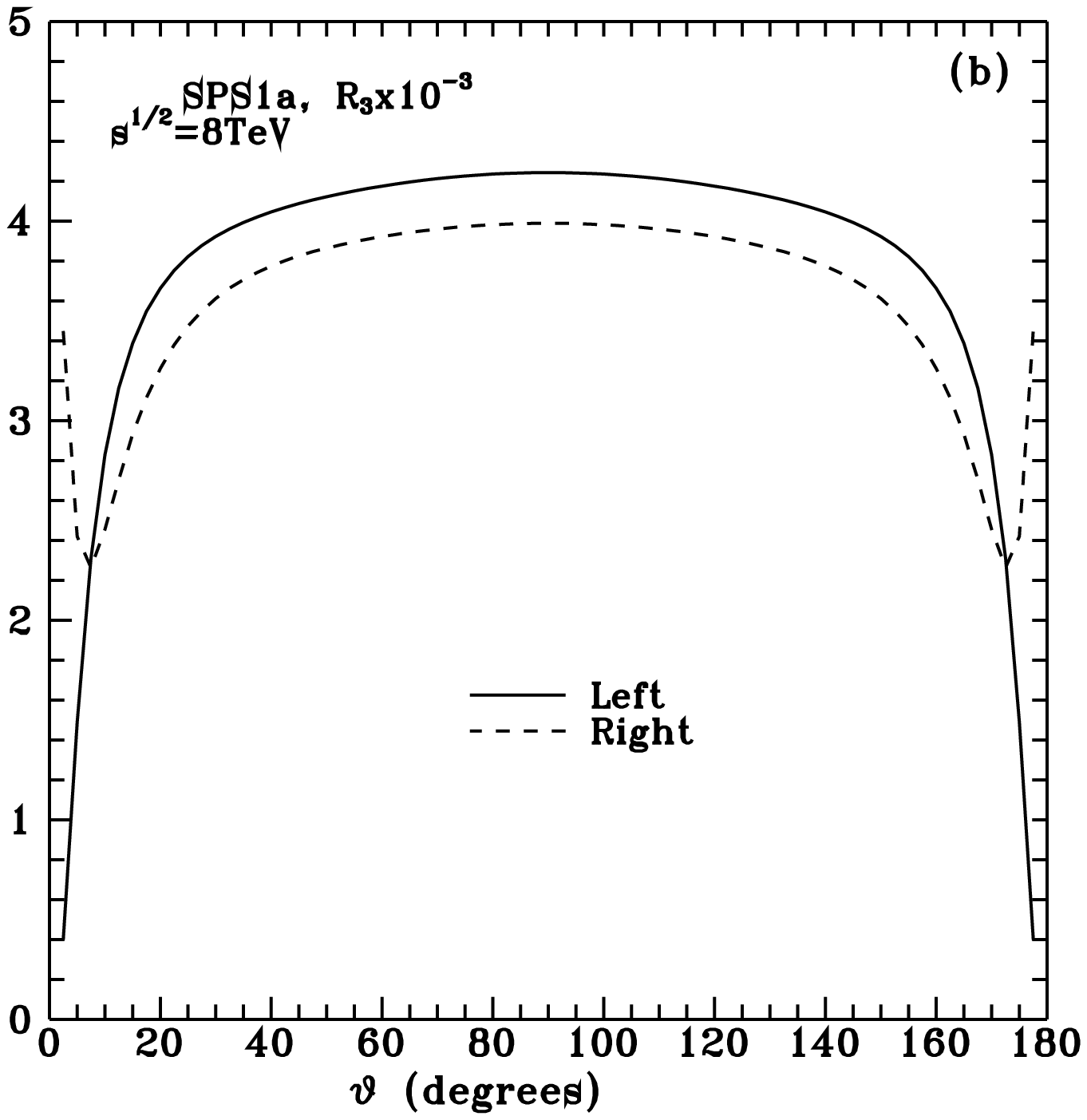,height=6.5cm}
\]
\[
\epsfig{file=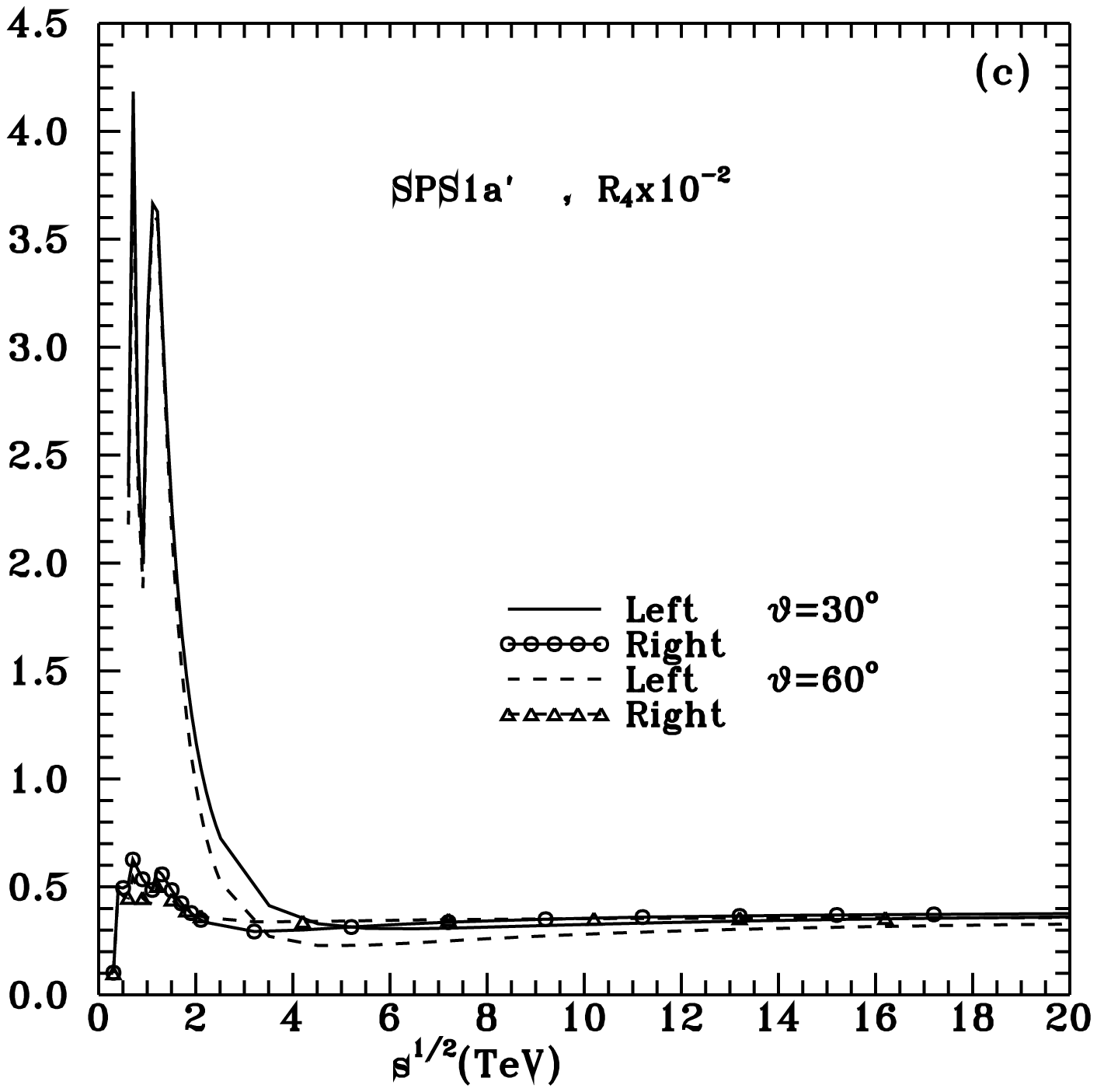, height=6.5cm}\hspace{1.cm}
\epsfig{file=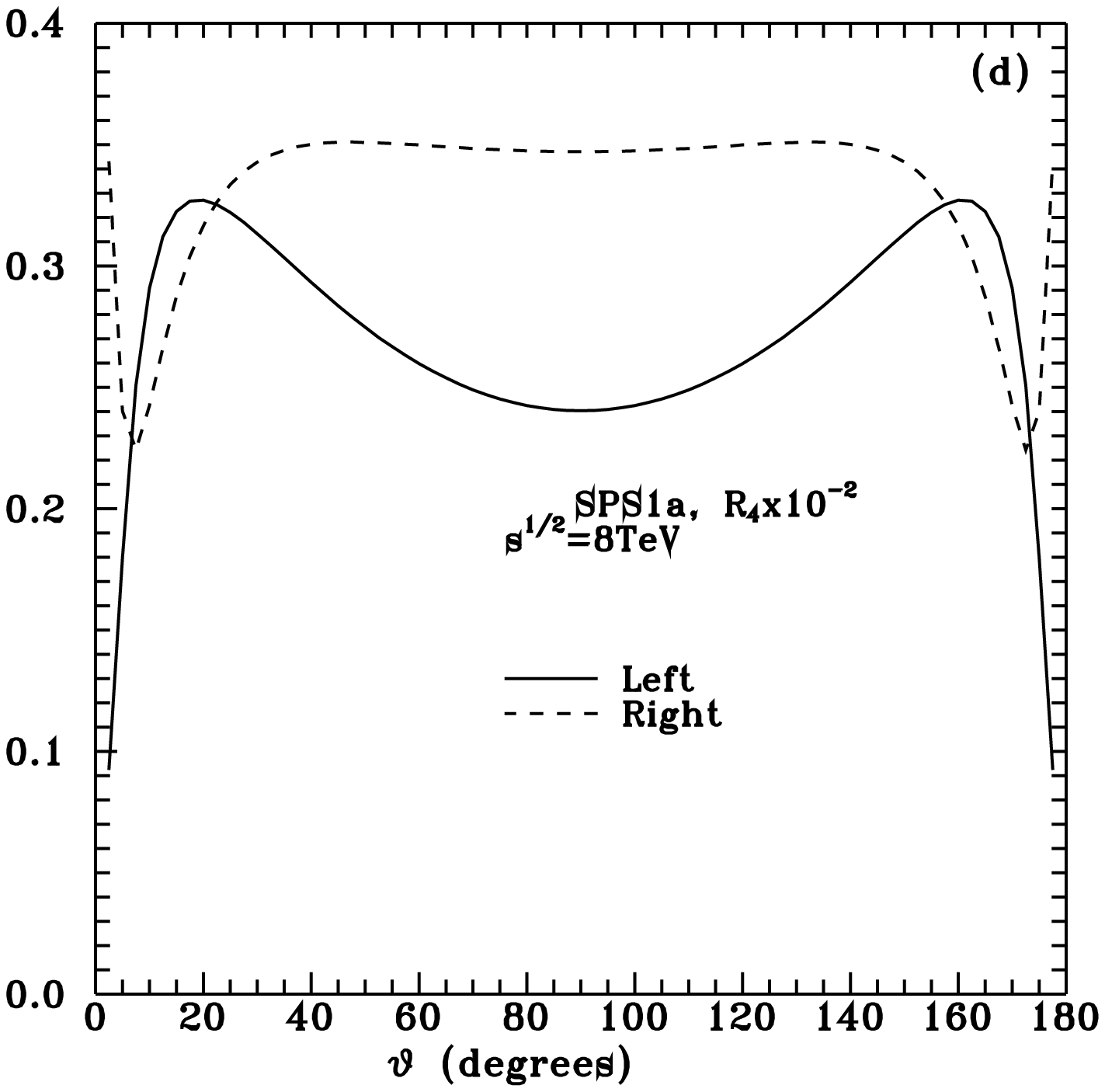,height=6.5cm}
\]
\[
\epsfig{file=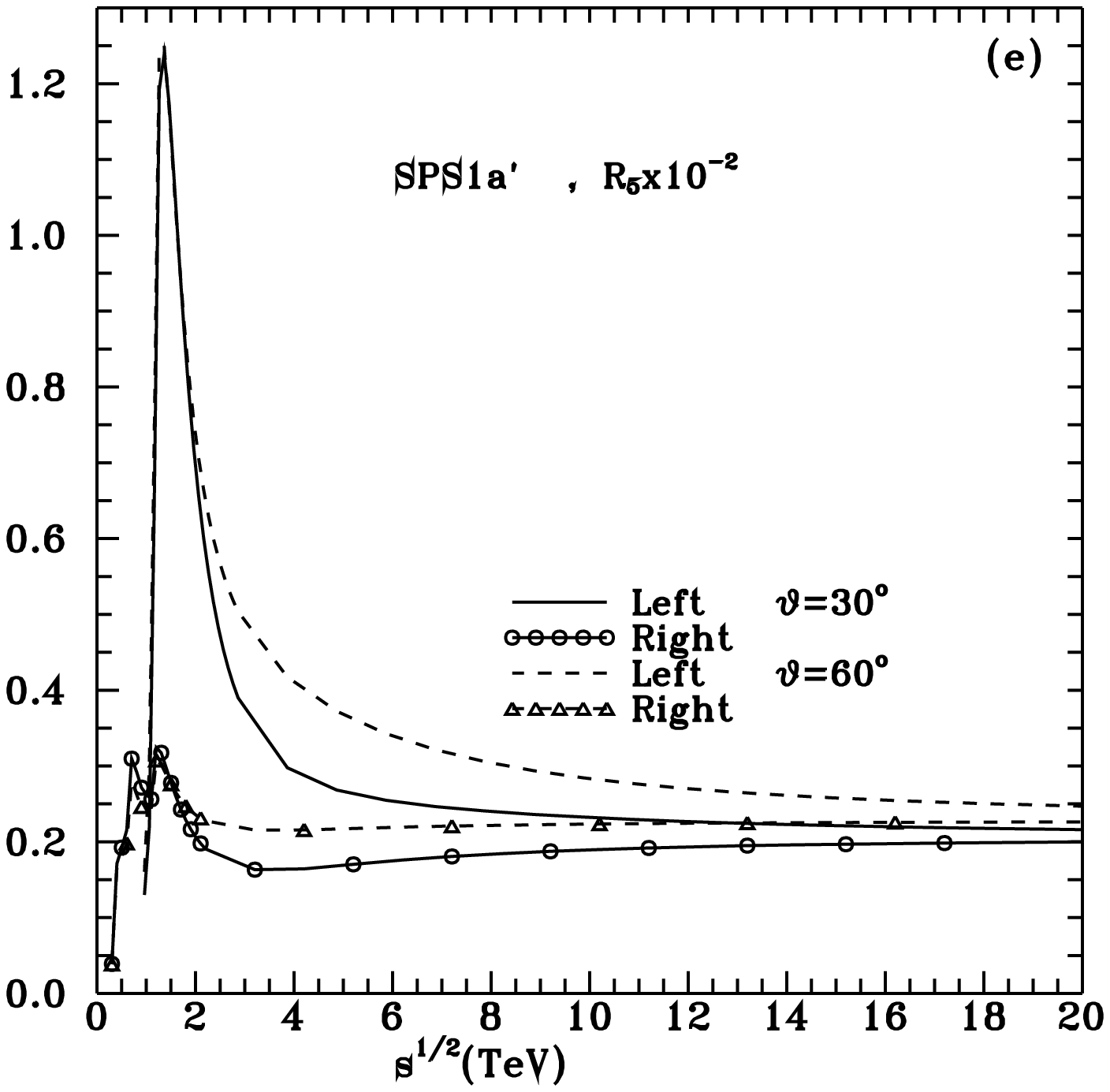, height=6.5cm}\hspace{1.cm}
\epsfig{file=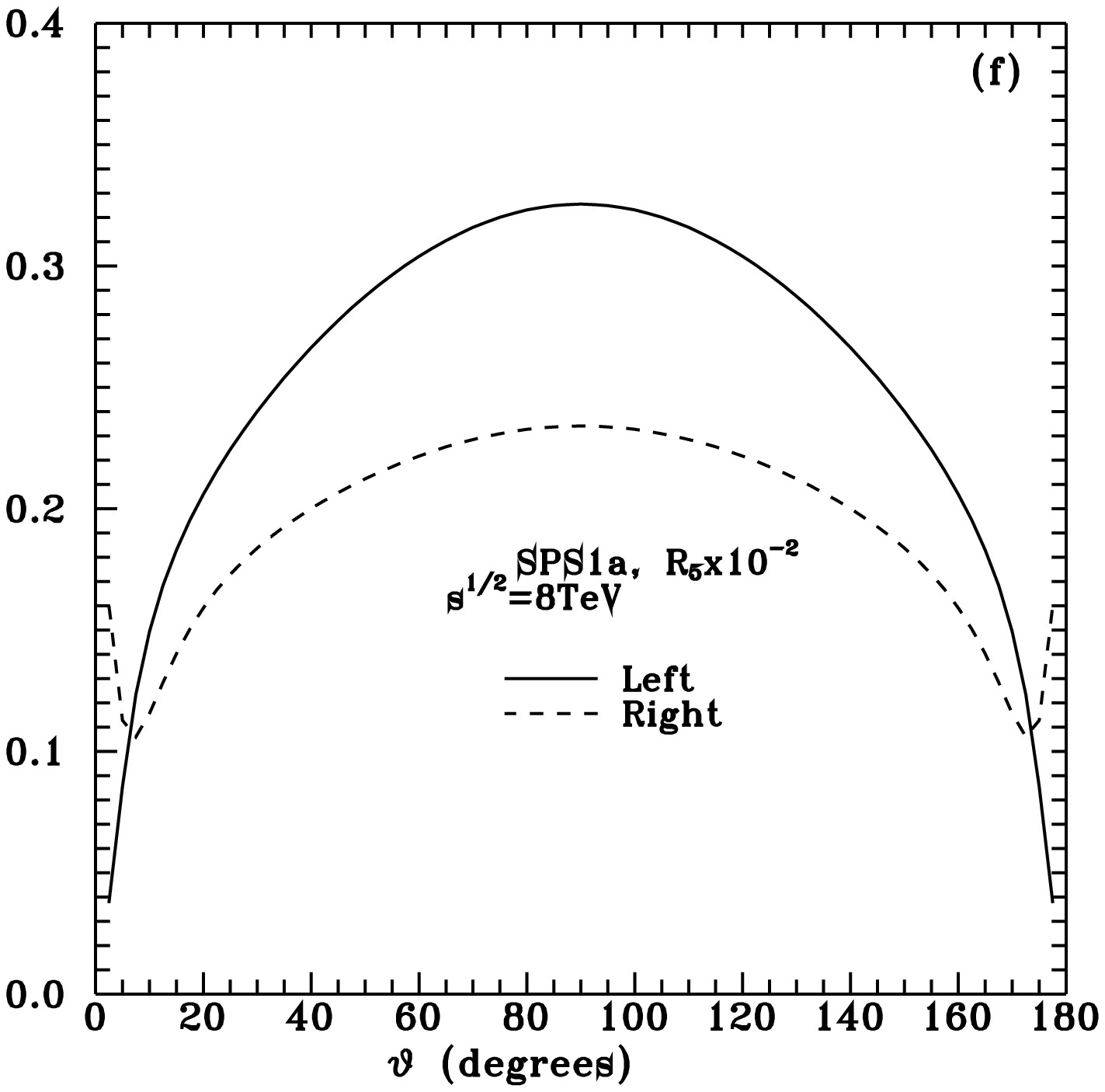,height=6.5cm}
\]
\caption[1]{Magnitudes of the Left and Right  parts of the
asymptotic relation $R_3,~R_4,~R_5$ defined in (\ref{R3-rel}, \ref{R4-rel}, \ref{R5-rel});
(a,c,e) describe  the energy dependencies,  while (b,d, f) give the angular dependencies
at $\sqrt{s}=8~{\rm TeV}$.}
\label{R3-R4-R5-fig}
\end{figure}

\begin{figure}[p]
\vspace*{-1cm}
\[
\epsfig{file=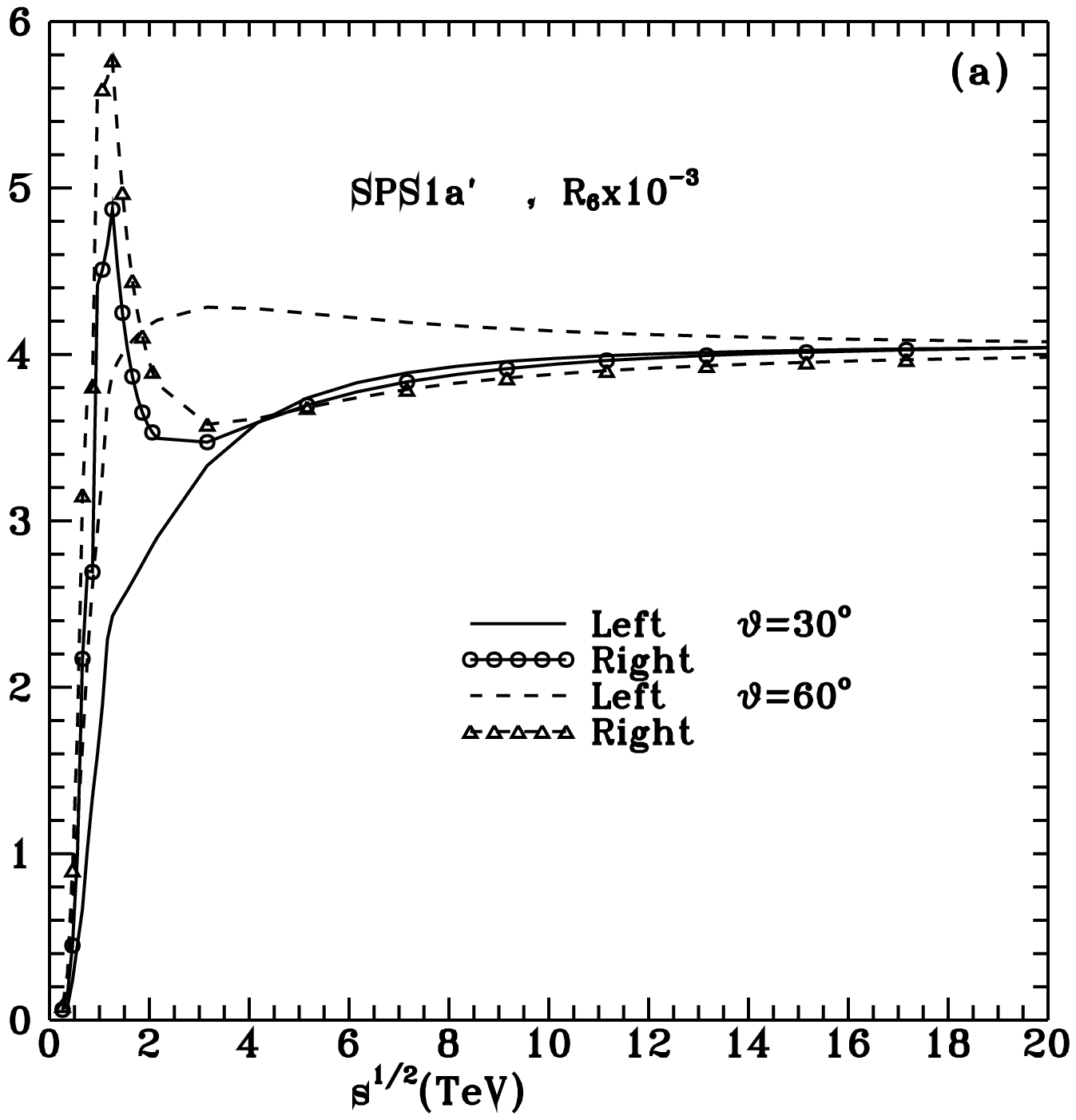, height=6.5cm}\hspace{1.cm}
\epsfig{file=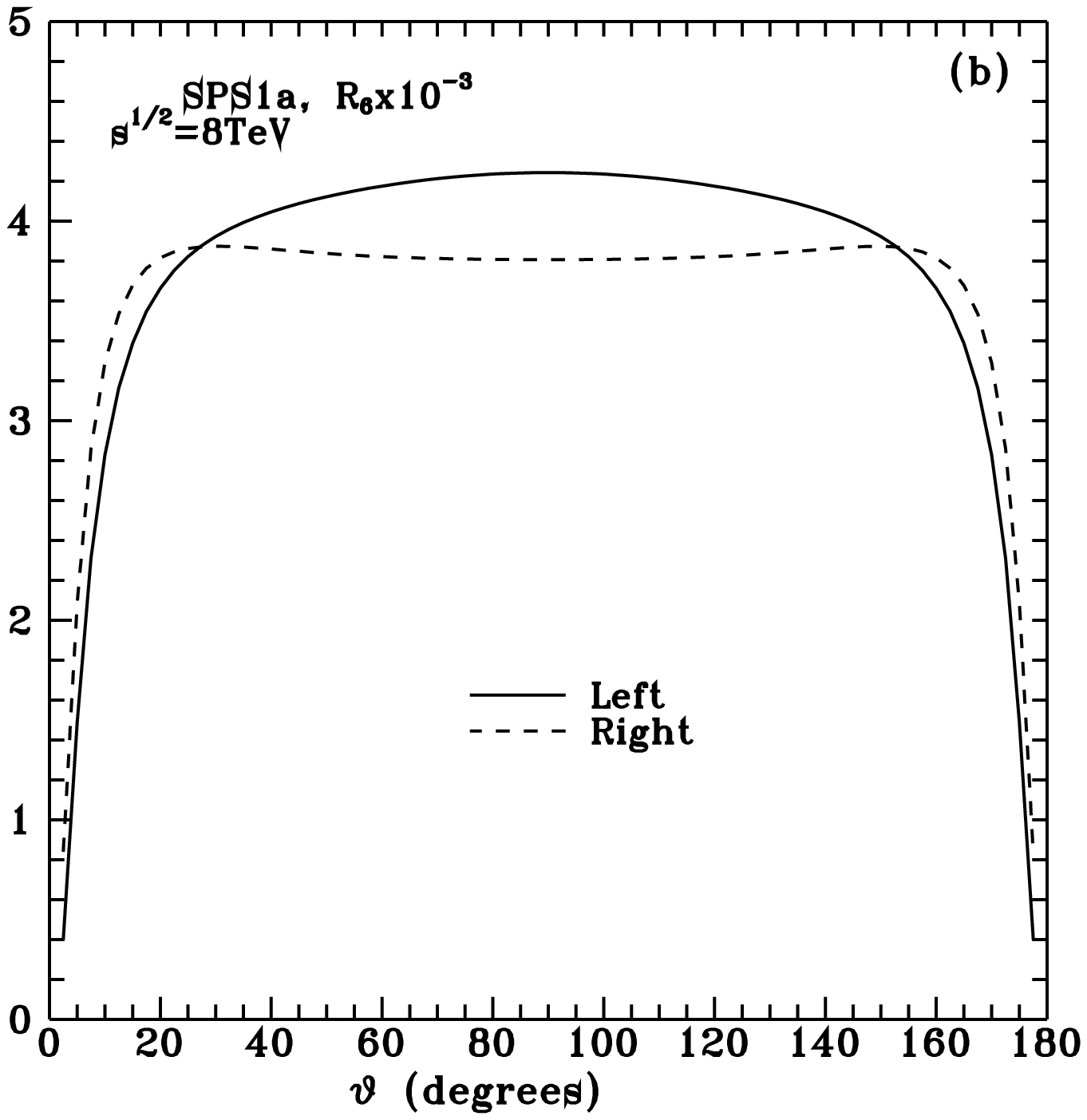,height=6.5cm}
\]
\[
\epsfig{file=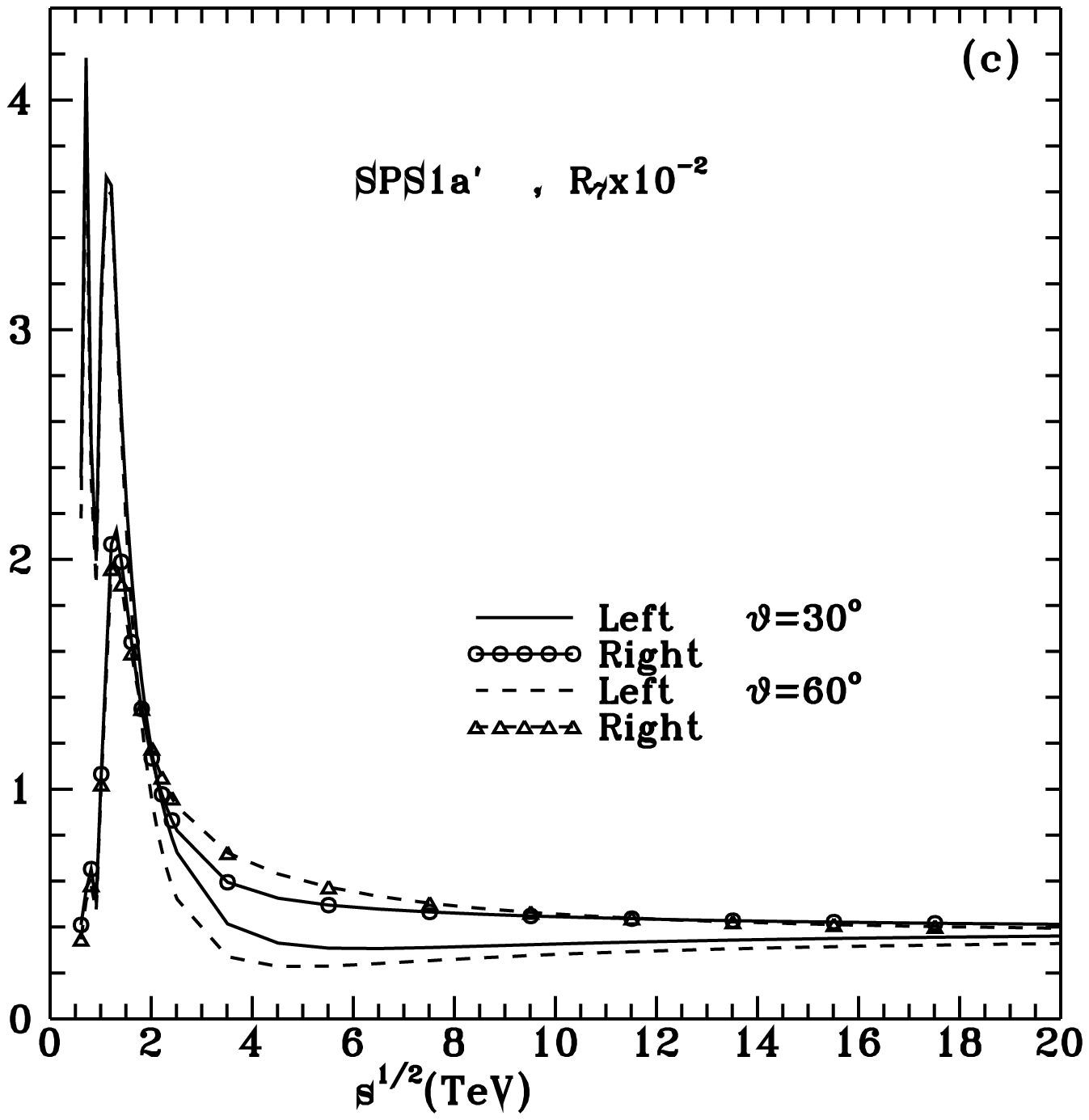, height=6.5cm}\hspace{1.cm}
\epsfig{file=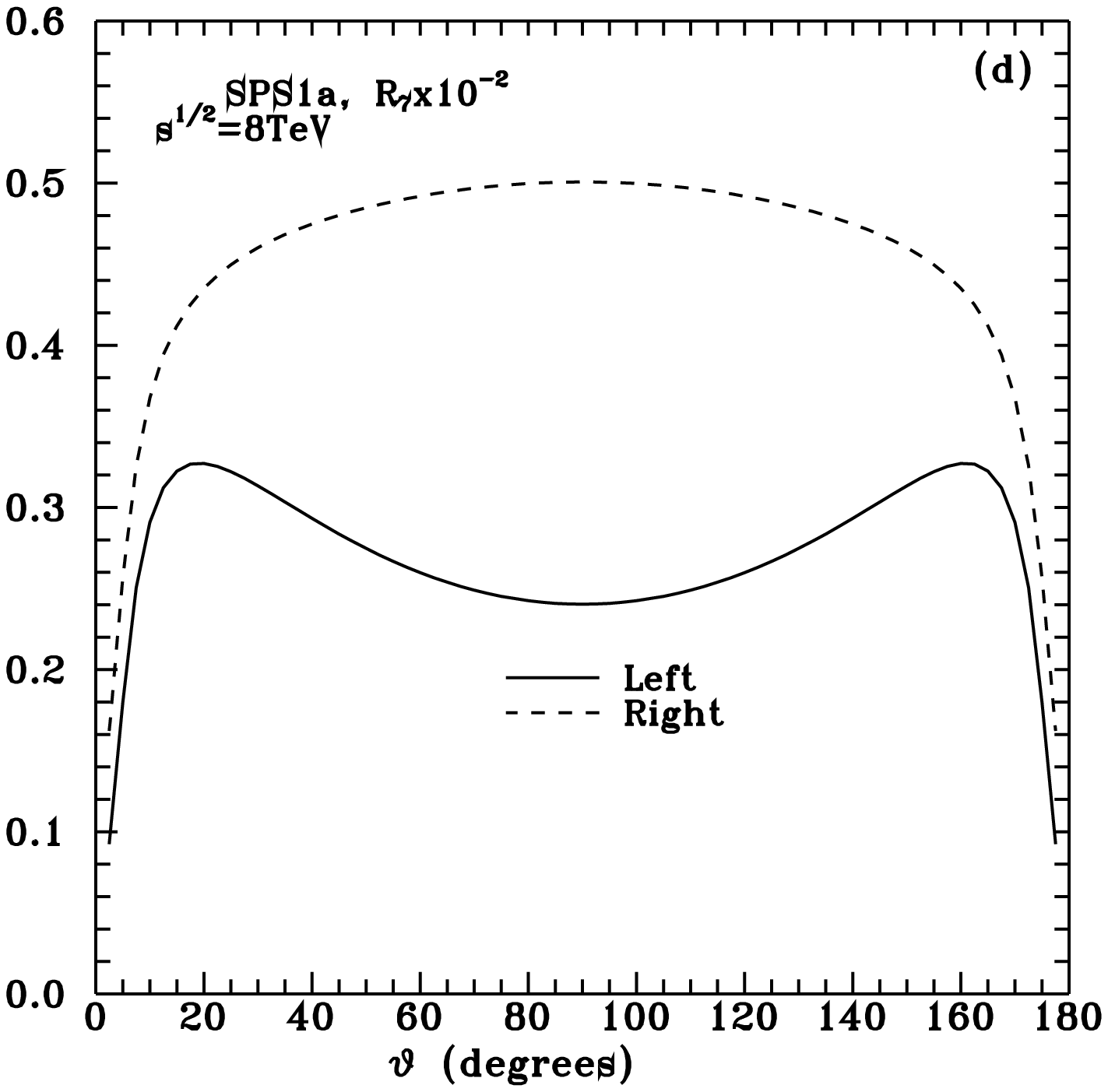,height=6.5cm}
\]
\caption[1]{Magnitudes of the Left and Right  parts of the
asymptotic relation $R_6,~R_7,$ defined in (\ref{R6-rel}, \ref{R7-rel});
(a,c) describe  the energy dependencies,  while (b,d) give the angular dependencies
at $\sqrt{s}=8~{\rm TeV}$.}
\label{R6-R7-fig}
\end{figure}

As the energy decreases though, deviations in the $R_1-R_7$ relations   appear,
like those shown in
Figs.\ref{R1-fig}-\ref{R6-R7-fig} for $SPS1a'$ \cite{SPA}. We have studied in detail
these relations  in $SPS1a'$; and as a general statement we could say that at   an energy of
$\gsim  8~{\rm TeV}$, they are   satisfied to  an accuracy  of $\sim 50\%$ or better.
Of course the accuracy of these relations would become better or worse, depending
on whether the SUSY scale is lower or higher than in this benchmark.

The energies needed for  $R_i$ to acquire a certain accuracy,
are generally larger than those  required  for the relation connecting
the $ug\to dW$ and $ug \to d_L\tchi^+_i$ cross sections \cite{ugdW-ugsdWino}.
This is probably due to the presence of  important Born  contributions to these
later processes, which makes them  less sensitive to higher scale-effects,
than the  purely  1loop processes entering the $R_i$'s.

We also   note that the departures from the asymptotic
 predictions $R_1-R_7$  arise from  global  SUSY-scale  effects.
Measuring such effects   could define a strategy of SUSY analysis, starting from the high
energy range where the basic SUSY properties can be established, and
then going down in energy,   progressively becoming more sensitive
to  specific SUSY  masses.
Such a strategy is to be opposed
to the usual one starting from the low energy with more than 100 free
parameters in MSSM, and then going up in energy.
If the SUSY scale is not too high, such a strategy may be feasible.\\

In the near future we hope  to  look at the 1loop EW predictions for the
gluon-gluon fusion producing   two vector bosons, or  two
charginos or neutralinos \cite{ggVV-ggchichi}. We expect that the combination of these
processes, with those studied here, will supply many more  asymptotic
relations among various, in principle measurable, cross sections .\\

In conclusion, we dare to say that the SUSY best motivated candidacy for describing
the physics beyond SM, is not only due to its  smooth ultraviolet properties,
its inclusion of  dark matter candidates, and its invitation to unification.
Its exact  helicity conservation property for any 2-to-2 process, which
so strongly simplifies its asymptotic amplitudes, also deserves
to be added to this list.

\vspace*{1cm}
\noindent
{\large\bf{Acknowledgements}}\\
\noindent
G.J.G. gratefully acknowledges the support by the European Union  contracts
MRTN-CT-2004-503369 and   HEPTOOLS,  MRTN-CT-2006-035505.

\clearpage

\clearpage

\clearpage

\clearpage

\clearpage

\clearpage

\clearpage

\clearpage

\clearpage

\clearpage

\clearpage

\end{document}